\documentclass{jpp}
\usepackage{graphicx}
\usepackage{marginnote}
\usepackage{setspace}

\usepackage{hyperref}
\usepackage[utf8]{inputenc}
\usepackage[T1]{fontenc}
\usepackage{amsmath}
\usepackage{colortbl}
\usepackage{enumitem}
\usepackage{xcolor}
\usepackage{subcaption}

\newtheorem{lemma}{Lemma}

\usepackage{xparse}
\ExplSyntaxOn
\NewDocumentCommand{\mymatrix}{m}
{
  \begin{pmatrix}
    \tl_set:Nn \l_tmpa_tl { #1 }
    \tl_replace_all:Nnn \l_tmpa_tl { ; } { \\ }
    \tl_replace_all:Nnn \l_tmpa_tl { , } { & }
    \tl_use:N \l_tmpa_tl
  \end{pmatrix}
}
\ExplSyntaxOff

\renewcommand{\t}{\hat{\pmb t}}
\newcommand{\n}{\hat{\pmb \kappa}}
\renewcommand{\b}{\hat{\pmb \tau}}

\shorttitle{Near-axis QI database}
\shortauthor{E. Rodriguez, G. G. Plunk
}

\title{Near-axis quasi-isodynamic database}

\author{E. Rodríguez\aff{1}\corresp{\email{eduardo.rodriguez@ipp.mpg.de}}, G. G. Plunk\aff{1}}

\affiliation{
\aff{1} Max Planck Institute for Plasma Physics, 17491 Greifswald, Germany
}

\begin{document}

\maketitle

\begin{abstract}
In this work, we investigate the landscape of quasi-isodynamic stellarators using the near-axis expansion of the magnetic field.  Building on recent theoretical developments, we construct a database of more than 800,000 stable, approximately quasi-isodynamic vacuum magnetic configurations. These configurations span a range of field period numbers and other geometric control parameters, including the magnetic axis shape and plasma elongation.  To evaluate each configuration, we use a broad set of measures, including effective ripple, sensitivity of the Shafranov shift to changes in plasma $\beta$, the prevalence of maximum-$\mathcal{J}$ trapped particles, and the Rosenbluth-Hinton residual, among others. This enables an exhaustive, thorough and quantitative characterization of the database.  Statistical analysis and modern machine learning techniques are then employed to find correlations, and identify key descriptors and heuristics to help understand tendencies that govern the behaviour of numerical optimization.  The database provides baseline configurations for further studies, and to serve as tailored initial conditions for optimization.  With this work we initiate a long term program to complete a systematic exploration of quasi-isodynamic stellarator design space.
\end{abstract}

\section{Introduction}
The promise of the stellarator \citep{spitzer1958stellarator,boozer1998stellarator,helander2014theory} as a means to achieving controlled magnetic confinement fusion has spurred activity in recent years \citep[see for instance][]{Parra_Diaz_2024, warmer2024overview}, with a particular focus on optimization, {\it i.e.} finding the appropriate shape of the magnetic fields to support sufficient confinement of the plasma \citep{mynick2006}, and other desirable properties. Many shapes can be discarded at the outset due to their lack of sufficient particle confinement, and it is therefore appropriate to specialise to particular subclasses of configurations that confine particles by design. In this paper we are concerned with one such family, namely, quasi-isodynamic (QI) stellarators \citep{Cary1997,gori1997quasi,Helander_2009,Nührenberg_2010}.

QI stellarators are defined as magnetic fields with nested flux surfaces with the following two properties. First, they belong to the larger class of \textit{omnigeneous stellarators} \citep{bernardin1986,Cary1997,hall1975,helander2014theory}; \textit{i.e.}, they, by construction, confine all collisionless charged particle orbits \citep{northrop1961guiding,littlejohn1983,wessonTok,blank2004guiding}, through a carefully tailored magnetic field magnitude $|\mathbf{B}|$ \citep{boozer1983transport,nuhren1988,Cary1997,parra2015less}. Second, the contours of $|\mathbf{B}|$ over flux surfaces are closed the poloidal way around the torus. The latter confers QI configurations some advantageous properties over other omnigenous configurations (especially the celebrated \textit{quasisymmetric} ones \citep{boozer1983transport,nuhren1988,rodriguez2020necessary,burby2020some}). In particular, the reduced level of bootstrap current \citep{Helander_2009} and narrower particle orbits are key distinctions, making QI stellarators an attractive stellarator class.

Following the above definition, the search for QI configurations is generally conducted via numerical optimisation \citep{gori1997quasi,mikhailov2002,goodman2023constructing,sanchez2023quasi,dudt2024magnetic}. A cost function that vanishes when the definition is met is required to guide navigation over the space of stellarator equilibria. This approach has proven successful in its many different iterations, providing in some cases entire collections of configurations of varying quality \citep{cadena2025constellaration}. However, this process is far from perfect, as optimization is highly-nonlinear procedure performed over a vast parameter space consisting of many non-physical hyperparameters \citep{laia2025data}. Thus, navigation of the space of possible QI configurations is hampered by the lack of a solid underlying structure to help guide and interpret its mysterious behaviour.  In addition, QI optimisation, more than other classes, is highly dependent on initial conditions \citep{goodman2023constructing}, and it is thus important to derive these in a systematic way. 

The near-axis expansion model \citep{mercier1964equilibrium,Solovev1970,lortz1976equilibrium,garrenboozer1991a} has proved to be an invaluable tool to tackle many of these limitations. This framework describes stellarator equilibria through their asymptotic expansion in the distance from the magnetic axis, providing a simplified, yet powerful perspective on the space of configurations. The model  has seen significant development for quasisymmetric stellarators \citep{landreman2020magnetic,landreman2021a,jorge2020naeturb,rodriguez2022phases,rodriguez2023constructing}, contributing to the understanding of their configuration space and enabling the generation of large databases of configurations. Recent examples of these can be found in  \cite{landreman2022mapping, giuliani2024direct, giuliani2025comprehensive}, which make direct use of the near-axis expansion. An equivalent development of the model has however lagged for QI stellarators. Only most recently, through a series of papers \citep{plunk2019direct, jorge2022c, camacho-mata-2022,Camacho2023helicity,rodriguez2023higher,rodriguez2024near,plunk2025-geometric,rodriguez2025near}, the QI side of the model has matured to the point where a first thorough attempt at the exploration of the space of QI configurations may be attempted {  on} par with the QS one. 

In this paper we present a first-of-its-kind database of more than 800,000 approximately QI vacuum near-axis configurations. The database is generated and structured such that each configuration is uniquely defined by a handful of scalar choices.  The resulting space of solutions, though vast, is therefore possible to thoroughly explore. The near-axis framework enables each configuration to be swiftly diagnosed within the framework taking many different physical properties into account \citep{rodriguez2025near}, including elements of MHD stability, neoclassical transport or coil complexity. The paper explores some of these physical properties in detail, characterising those configurations that excel in these aspects, and drawing physical and practical insight from it. Statistical and modern machine learning tools are also introduced to aid with the analysis of such vast data. Besides the interest of the individual stellarator fields themselves, and underlying physical and practical insights gained from the analysis, this paper also establishes the infrastructure for future more sophisticated, extensive and detailed analysis. 

\section{Construction of database}

We follow recently developed methodologies \citep{rodriguez2023higher, plunk2025-geometric} to construct approximately QI stellarator fields to second order in the near axis expansion.  In this section, basic details of this solution are described, and we refer the reader for the relevant details to mainly \cite{rodriguez2024near} and \cite{plunk2025-geometric}.

The near-axis expansion proceeds from zeroth to first to { then} second order, at each stage accumulating more detail in the input parameter space and the output solution space. At zeroth order in the expansion, we must specify the shape of the magnetic axis and the magnetic field strength along it. Following \cite{plunk2025-geometric}, we define the shape of the magnetic axis, ${\bf x}_0$, by prescribing curvature, $\kappa$, and torsion, $\tau$, as functions of the arc length $\ell\in[0,2\pi)$, and solving the Frenet-Serret (FS) equations \citep{frenet1852courbes, animov2001differential},
\begin{equation}
    \frac{d\mathbf{x}_0}{d\ell}=\t,\quad
    \frac{d\t}{d\ell}=\kappa\n,\quad
    \frac{d\n}{d\ell}=-\kappa\t + \tau \b,\quad
    \frac{d\b}{d\ell}=-\tau\n,  \label{eqn:FS_eqns}       
\end{equation}
where $\t$ is the tangent, and $(\n,\b)$ normal and binormal unit vectors. This approach to defining the curve enables direct control on the curvature and torsion, central ingredients in the near-axis framework. Thanks to this, we can easily impose two flattening points (of zero curvature) within each field period to comply with the requirements of a single-well QI field \citep{plunk2019direct}. {  To have a smooth framing of the curve we must allow curvature to be negative to be a smooth function, resulting upon integration of Eqs.~(\ref{eqn:FS_eqns}) in the signed Frenet-Serret frame \citep{plunk2019direct, Camacho2023helicity,rodriguez2024near}.} To ensure that a curve defined this way is properly closed, the curvature and torsion must be left some degrees of freedom to tweak, as detailed in \cite{plunk2025-geometric}. To accommodate these necessities, we parametrise curvature and torsion as,

\begin{align}
    \kappa =&  2\kappa_1\sin^u\left(\frac{N\ell}{2}\right)\cos^v\left(\frac{N\ell}{2}\right)\left[1+\sum_{n=2}^{N_\kappa}n\kappa_{n}\cos[(n-1)N\ell]\right],\label{eq:kappa-N1}\\
    \tau = & \tau_0+\sum_{n=1}^{N_\tau} \tau_{cn}\cos(n N \ell),\label{eq:tau-N1}
\end{align}
where $N\in\mathbb{N}^+$ is the number of field periods, and we choose $u=2$ and $v=3$. The parameters $\{\tau_0,\tau_{cn},\kappa_n\}$ can be thought of as the coefficients of a Fourier series, thus encompasing a larger space of functions for larger $N_\kappa$ and $N_\tau$. For the purpose of the construction in this paper, we choose $N_\tau=N_\kappa=2$; that is, the set of coefficients is $\{\kappa_1,\kappa_2,\tau_0,\tau_{c1},\tau_{c2}\}$. Of these coefficients, for $N\geq2$ configurations, $\kappa_1$ and $\tau_0$ are self-consistently found to ensure closure of the curve \citep[Sec.~2.2]{plunk2025-geometric}. The $N=1$ case is special, and it requires 4 free parameters, leaving only $\tau_0$ as input \citep[Sec.~2.1]{plunk2025-geometric}.

The curvature has flattening points at maxima and minima of order $u=2$ and $v=3$ respectively (the order of the first non-zero term in the expansion of $\kappa$ in $\ell$ at such points). The order of the zero at the bottom, $v$, is chosen to be odd to comply with QI \citep{plunk2019direct,rodriguez2023higher}. The choice of $u$ is made consistent with a $1/2$ helicity \citep{Camacho2023helicity}\citep[App.~A]{rodriguez2024near} field (see Appendix~\ref{app:helicity}). 

The magnetic field strength is taken to have the form
\begin{equation}
    B_0 = 1 + \Delta \left[(1 + \lambda_B)\cos(N \ell) + \left(\frac{1}{4} - 2 \lambda_B\right) \cos(2 N \ell) - \lambda_B \cos(3 N \ell)\right].
\end{equation}
Such choice has $B_0'' = 0$ at field minima, suitable to promote MHD stability \citep{rodriguez2024maximum,plunk2024back}. The choice of $v=3$ for the curvature is directly linked to this ``flattened'' minimum, needed to avoid introducing QI-breaking defects in $|\mathbf{B}|$ \citep{rodriguez2023higher}. In addition, the model for $B_0$ includes  additional shaping parameters in the form of a mirror ratio, $\Delta = (B_0(0)-B_0(\pi/N))/2$, and $\lambda_B$, which controls the width of the well. 

At first order in the construction we have freedom to specify the elliptical shape of the cross sections in the Frenet-Serret (FS) frame. We construct the ellipticity (\textit{i.e.}, ratio of major to minor axes of the ellipses) as a function of $\varphi${ , the toroidal Boozer angle,} in terms of an auxiliary function $\rho=\rho(\varphi)$ \citep{plunk2025-geometric}:

\begin{align}
    E(\varphi) &= \frac{1}{2}\left(\rho + \sqrt{\rho^2-4}\right).\label{eq:E-eqn}
\end{align}
The even function $\rho$ is given by
\begin{equation}
\rho = \rho_0 + \rho_1 \cos(N \varphi) + \rho_2 \cos(2 N \varphi),\label{eq:elongation-input-form}
\end{equation}
where $\rho_0$, $\rho_1$ and $\rho_2$ are free constants to be chosen. By controlling $\rho$, large elongation of flux surfaces can be prevented, avoiding the natural bias of QI fields. The drawback is that, in actuality, not all choices of parameters $\{\rho_n\}$ result in a valid solution through first order in the near-axis expansion \citep[Sec.~3.1]{plunk2025-geometric}, not even when $\rho\geq2$ is satisfied. Thus, we must check the construction a posteriori to populate our database with valid first order configurations.

The second order that completes the description of any configuration in our near-axis framework requires additional shaping inputs. Instead of including additional parameters that describe shaping, we follow the work in \cite{rodriguez2025near} and consider, associated to each first order choice, a whole family of minimally shaped configurations with different values of vacuum magnetic well, $W$ \citep[Sec.~6]{rodriguez2025near}. We shall often represent each of these families by its marginal representative, $W=0$, unless otherwise stated.

Together, the free constants determining the functions $B_0$, $\kappa$, $\tau$ and $\rho$ above, alongside $N$, uniquely define then a marginally stable vacuum configuration. We will refer as \textit{input features} to the minimal, unique set of input scalars $\{\tau_{c1},\tau_{c2},\kappa_2,\rho_0,\rho_1,\rho_2,\Delta,\lambda_B\}$. The case of $N=1$ is special (due to the role of $\tau_0$), and we shall generally treat this case separately. 

With the prescription detailed above, we explicitly construct a database of more than 800,000 vacuum, approximately QI stellarators, with half helicity and flattening class $(2,3)$. To populate such database we scan the input features over a rectangular domain in regular increments.\footnote{The choice of equidistant sampling points has the undesired effect of introducing granularity to the input parameter space, as we observe in the analysis that follows, which can difficult its statistical analysis.} The parameter domains and sampling is detailed in Table~\ref{tab:input_parameters} of Appendix~\ref{app:input_space}. These sampling domains must take into account the natural boundaries associated to each of the features. For instance, the solutions to the near-axis equations fails to exist when $\rho$ is too low \citep{plunk2025-geometric}, which sets a lower bound on the minimal achievable elongation.  Such limits are also present on the construction of the axis curves, which to comply with the desired helicity, cease to exist for large values of torsion. This sets an upper bound on parameters such as $\tau_{c1}$ (see Figure~3 in \cite{plunk2025-geometric}). Pushing the parameters to these extremes in the scan ensures that we are able to explore the full range of possible geometries captured within our model. Not every configuration constructed from such a scan can however be kept and archived in the database, only those that satisfy (a) closure of the axis curve, and (b) periodicity and realness of the first order solution of the $\sigma$-equation \citep[Eq.~(2.5)]{rodriguez2024near}\citep[Eq.~(3.1)]{plunk2025-geometric}, make the cut. Once the collection of configurations has been generated, their physical properties are assessed within the near-axis framework following \cite{rodriguez2025near,landreman2020magnetic, landreman2021a,rodriguez2024maximum}. 


\section{General statistical and physical features of the QI database}
The resulting database of near-axis QI configurations constructed following the prescription detailed in the previous section is made up of more than 800,000 different fields. There is no hope in studying the set on a one-by-one basis; a more practical, statistical approach to analysing the dataset is required. In this paper, we analyse the database by examining the behaviour of select physical properties across it. The dataset exhibits substantial variability in these properties, with many configurations showing unpractical values in many of these regards. Identifying and characterizing the features associated with the best-performing configurations provides valuable insights and shall be the focus of the analysis. To this end, we employ both descriptive statistical analyses and machine learning techniques, a detailed account of which is provided in Appendix~\ref{app:statistics_phaseI}. We choose to focus here, for the most part separately, on elements of the field shaping complexity, MHD stability, maximum-$\mathcal{J}$ness, neoclassical transport and finite $\beta$ sensitivity. A more comprehensive exploration of ``trade-offs'' between these properties, {\em i.e.} involving multi-variate considerations, is reserved for future work.


\subsection{Magnetic gradient scale, $L_{\nabla \mathbf{B}}$}
Each configuration in the database represents a different vacuum stellarator field. To physically sustain any of those magnetic fields, an appropriate arrangement of electric currents around the torus (electromagnetic coils) must be devised. In a reactor, it is desirable to place such coils sufficiently far from the fusing plasma, so that breeding blankets and neutron shielding may be fitted in between. However, there is an inherent cost to positioning coils far. The fine detail of the fields generated by coils, especially those with strong spatial variation, decay away from their generating currents. Thus, there is an inherent trade-off between coil distance and complexity, and it effectively becomes impossible to faithfully generate a given stellarator field at a sufficiently large distance. The $L_{\nabla\mathbf{B}}$ scalar measure is an attempt at describing this field-associated underlying scale.

We define the magnetic gradient scale, $L_{\nabla \mathbf{B}}$, as,
\begin{equation}
    L_{\nabla\mathbf{B}}=\mathrm{min}\left[\frac{2B}{||\nabla\mathbf{B}||^2}\right], \label{eqn:L_grad_B_full}
\end{equation}
following the original work of \cite[Eq.~(3.1)]{landreman2021a}, and later explorations by \cite{Kappel_2024}. The minimum is meant to be computed over some selected flux surface, so that $L_{\nabla\mathbf{B}}$ may be interpreted as a proxy for how close coils need to be placed to said surface. To make sense of its magnitude, we shall normalise $L_{\nabla \mathbf{B}}$ to the effective major radii of the configurations, defined as the length of the axis over $2\pi$. Thus, 'better' configurations will present larger values of $L_{\nabla\mathbf{B}}$, which we may interpret as an inverse aspect ratio of the coil set.

Within the near-axis framework, we shall compute $L_{\nabla\mathbf{B}}$ using the asymptotic form of Eq.~(\ref{eqn:L_grad_B_full}), originally derived in \cite[Eq.~(3.11)]{landreman2021a} for a general magnetic field. To leading order, the gradient $||\nabla\mathbf{B}||$ is determined by the axis and elliptical cross-section shapes (\textit{i.e.}, zeroth and first orders in the expansion), whose toroidal inhomogeneity makes it a function of the toroidal angle $\varphi$. 

Other field scales could also be defined (most notably $L_{\nabla\nabla\mathbf{B}}$ \citep{landreman2021a} which includes elements of second order and thus radial extent) and used as proxies. And ultimately, a highest fidelity description would necessitate an explicit coil construction (see for instance \cite{giuliani2022single}). Without going into the complexities of these measures, and noting that the database can be a good testbed ground for them, we opt for the simplest proxy in $L_{\nabla\mathbf{B}}$, backing its relevance on the evidence of \citep{Kappel_2024}. 

\par

\subsubsection{General statistical description}
\begin{figure}
    \centering
    \includegraphics[width=\textwidth]{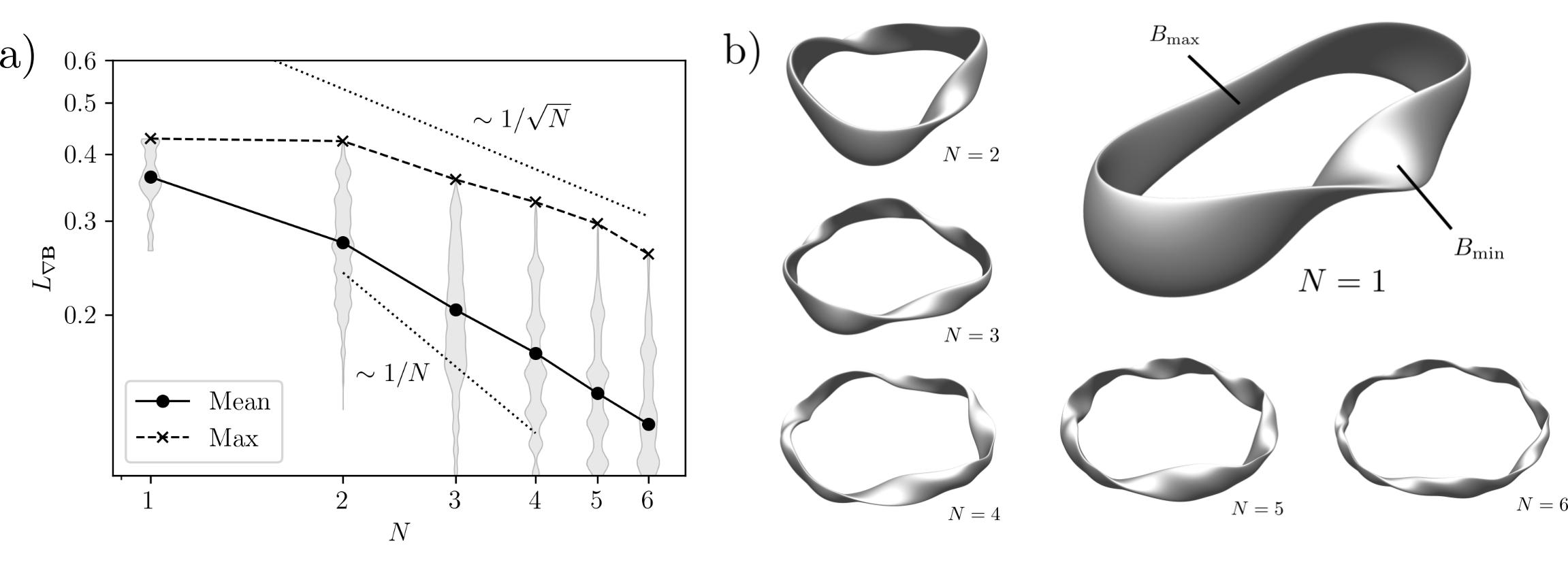}
    \caption{\textbf{$L_{\nabla\mathbf{B}}$ behaviour across different number of field periods.} a) Dependence of the maximum and mean value of $L_{\nabla\mathbf{B}}$ as a function of the number of field periods $N$ in the database. Reference scalings are given as dotted lines and the distribution of the configurations as violin plots. b) Rendering of the 3D finite aspect ratio flux surface of $L_{\nabla\mathbf{B}}$ maximising fields for each number of field periods in the database. As a result of the increased surface shaping with $N$, the finite-volume representation of the configurations are shown for increasingly higher aspect ratios. The locations of $B_\mathrm{min}$ and $B_\mathrm{max}$ are indicated for the $N=1$ configuration, which has the largest value of $L_{\nabla\mathbf{B}}$ of all (see Table~\ref{tab:summary_db}).}
    \label{fig:LgradB_field_period} 
\end{figure}

The scalar $L_{\nabla\mathbf{B}}$ exhibits an immediately evident trend across the database: stellarators with a lower number of field periods, $N$, favour longer field scales (see Figure~\ref{fig:LgradB_field_period}). Stellarators with $N=1,~2$ can exhibit values as low as $L_{\nabla \mathbf{B}}\sim0.4$, that is, coils with as low an aspect ratio as $A\sim2$. To put these values in context, we may compare them to the survey in \cite[Figure~7]{Kappel_2024}. The comparison to $L_{\nabla\mathbf{B}}$ values of other configurations must however be made with some words of caution. The near-axis $L_{\nabla\mathbf{B}}$ should be interpreted as a distance from the axis of the stellarator to coils, and not from the outermost flux surface of a finite aspect ratio configuration; that is, some sort of upper bound.\footnote{We remind the reader that when these near-axis fields are pursued beyond their asymptotic context as fully consistent global equilibria (say by following \cite{panici2025extending}), the value of $L_{\nabla\mathbf{B}}$ will change away from the axis. The precise way in which this occurs lies beyond this work. We consider that underlying fundamental scale instead.} With this in mind, and applying the necessary spatial scaling, the lowest $L_{\nabla\mathbf{B}}$ values in the database correspond to $L_{\nabla\mathbf{B}}^\star\sim7$, matching the best performing configurations in that work. It is no coincidence that those best performing configurations do also correspond to $N=1$ QI stellarators. 

This improved behaviour at lower $N$ translates also to an increased number of configurations found above any reasonable realistic threshold. We shall take as a reference value $L_{\nabla\mathbf{B}}=0.25$, which would amount to a distance of approximately 1.5~m from the plasma to the coils (in the ballpark of the needed space for various components in a fusion power plant \citep{Beidler_2001}) for an aspect ratio 10, $R\sim 10~\text{m}$ configuration. Setting a threshold for the analysis aligns with the notion of requiring a certain absolute physical plasma-coil distance to fit whichever elements are necessary (blanket, shielding, etc.). In this case, 96\% of the $N=1$ population exceeds the threshold, 56\% for $N=2$, and merely a per cent for $N=6$ (see the distributions in Figure~\ref{fig:LgradB_field_period}). This implies the need of larger aspect ratios for the practical realisation of larger $N$ stellarators. 

The shrinking of the $L_{\nabla\mathbf{B}}$ scale with field period number is expected from a purely geometric perspective. All other things being equal, a larger number of field periods requires any existing variation of the field to be packed into an increasingly narrower sector of the stellarator. Following this simple argument, we expect a linear scaling $L_{\nabla\mathbf{B}}\propto 1/N$. The dependence seen in the database is however weaker (especially at lower $N$, see Figure~\ref{fig:LgradB_field_period}), and although $N=1$ remains the best performing, excellent candidates may also be found for larger number of field periods. We show the configurations with the {  largest} value of $L_{\nabla\mathbf{B}}$, for each field period number, in Figure~\ref{fig:LgradB_field_period}b.

\subsubsection{Geometric dependencies of $L_{\nabla\mathbf{B}}$}
It is interesting to ask what these large $L_{\nabla\mathbf{B}}$ configurations have in common, and in particular, which of their defining features (if any) consistently yield the observed behaviour. At a qualitative level, the configurations in Figure~\ref{fig:LgradB_field_period}b share a small off-the-plane excursion (relative to the $x$-$y$ plane), yet an unevenly distributed torsion, particularly noticeable across the minimum of the magnetic field.

\begin{figure}
    \centering
    \includegraphics[width=\linewidth]{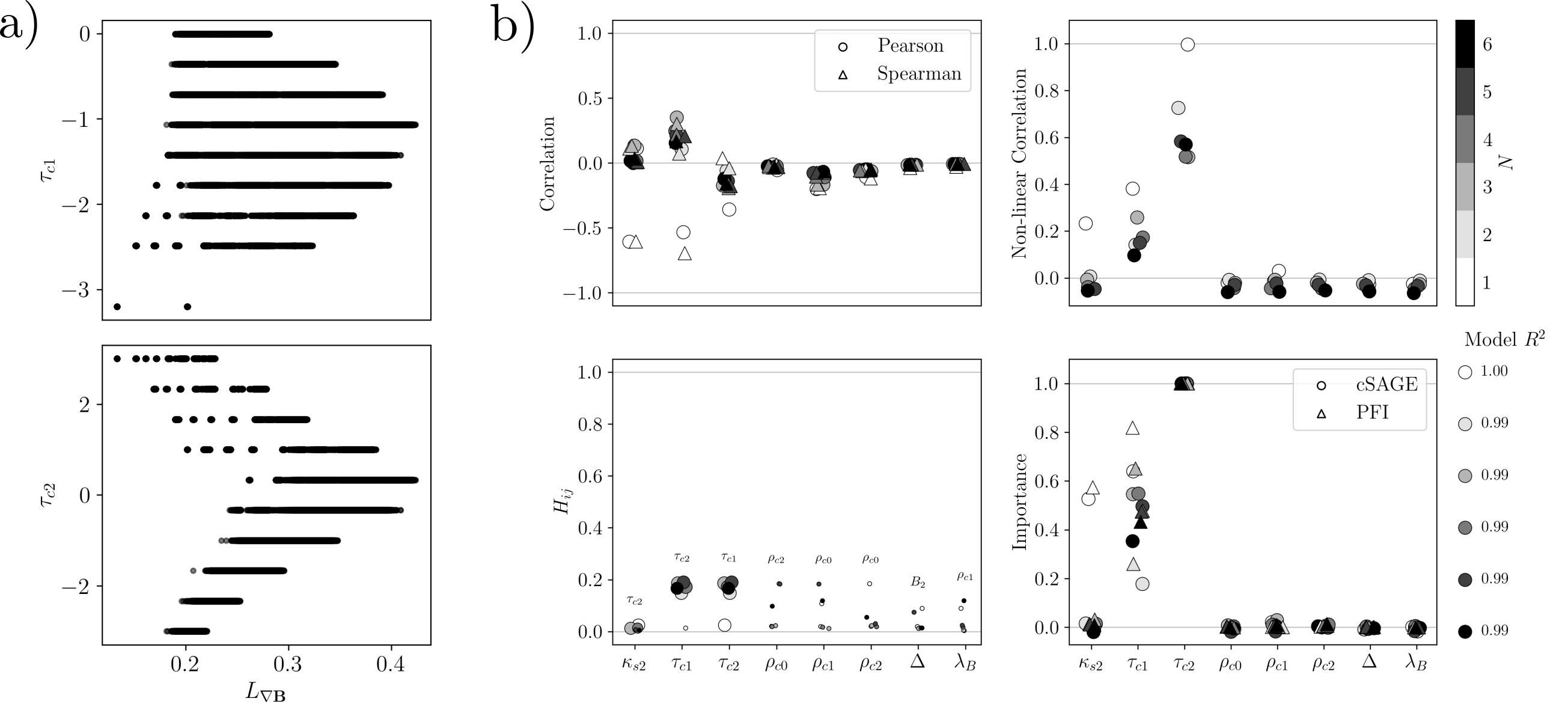}
    \caption{\textbf{Statistical summary of feature dependence for $L_{\nabla\mathbf{B}}$.} This figure is representative of the input feature dependence of derived quantities employed in the analysis and discussion of the database, here exemplified by $L_{\nabla\mathbf{B}}$. (a) Scatter plots illustrating any univariate dependence of $L_{\nabla\mathbf{B}}$ on the features $\tau_{c1}$ and $\tau_{c2}$, for the $N=2$ subset. (b) Summary of key statistical measures describing the dependence of $L_{\nabla\mathbf{B}}$ on input features. (Top left) Linear correlation between the input features and $L_{\nabla\mathbf{B}}$. The different symbols indicate the Pearson and Spearman coefficients, while the colour distinguishes different number of field periods. (Top right) Non-linear correlation coefficient, where the colour distinguishes different number of field periods. (Bottom right) Relative feature importance with symbols representing the PFI and cSAGE measures, and colour different number of field periods. The values listed to the right next to the scatter indicate the coefficient of determination of a multivariate SVM model fitted to $L_{\nabla\mathbf{B}}$ as a function of the input features. (Bottom left) Friedman H-statistic where the size of the scatter indicates the significance of the measure compared to a null reference distribution, and the colour represents different numbers of field periods. The labels indicate which other feature they are most closely linked to.}
    \label{fig:LgradB_stats}
\end{figure}

A more quantitative analysis of such observations requires us to be more systematic with the approach to the problem. In that direction we separate the features associated to each configuration in the database into: \textit{input features} (\textit{i.e.}, the minimal set of parameters uniquely describing configurations), and \textit{derived features} (\textit{i.e.}, derived properties describing the configuration, such as $L_{\nabla\mathbf{B}}$). The former constitutes the fundamental \textit{feature vector} (following the standard machine learning nomenclature), which by virtue of the near-axis construction represents not just labels, but precise physical description of the configurations: namely, their fundamental shape and field features. We will base our quantitative analysis on various statistical measures (detailed in Appendix~\ref{app:statistics_phaseI}) focused on the role of these input features, which by virtue of the near-axis construction constitute a low dimensional space.

Let us start the analysis of the influence of the input features on $L_{\nabla\mathbf{B}}$ by looking at linear correlation (see Figure~\ref{fig:LgradB_stats}b). The lack of a strong correlation (barring the special $N=1$ case) reflects the lack of a monotonic relation with any of the input features (see Figure~\ref{fig:LgradB_stats}a), but not necessarily a complete disconnection. We therefore need to supplement the analysis with an alternative that relaxes the monotonicity assumption. We construct a non-linear correlation measure (see Appendix~\ref{app:statistics_phaseI} for details) based on the goodness of fit of a univariate non-linear model. This measure, Figure~\ref{fig:LgradB_stats}b, unveils a significant non-monotonic dependence, highlighting details of the axis shape, with pre-eminence of $\tau_{c2}$ ($R^2\gtrsim 0.5$ across all different $N$). 

Despite the significance of the univariate description of $L_{\nabla\mathbf{B}}$, any such isolated model fails to capture the whole complexity of the database. We need to consider the dependence on multiple features simultaneously. When we consider all input features simultaneously, even a simple Support Vector Regression (SVR) \citep{scholkopf2002learning}\citep[Ch.~12.3.6.]{hastie2009elements} model for $L_{\nabla\mathbf{B}}$ can account for more than the 95\% of the database variability ($R^2\gtrsim 0.95$) without falling into over-fitting (see details in Appendix~\ref{app:statistics_phaseI}). Clearly, a multivariate description is more predictive, but does not unveil how much each feature contributes to predictions. To assess such relative importance we use Permutation Feature Importance (PFI) and cSAGE (see full details in Appendix~\ref{app:statistics_phaseI}). Both measures, Figure~\ref{fig:LgradB_stats}b, corroborate the dominant role of torsion, and $\tau_{c2}$ in particular, providing them high scores. 

The $\tau_{c1}$ parameter does also appear to be important, but its role appears to swing wildly between different $N$. This variability can in part be explained by the interaction of $\tau_{c1}$ with $\tau_{c2}$ in the $L_{\nabla\mathbf{B}}$ prediction. To describe this feature dependence we compute the Friedman H-statistic (see Appendix~\ref{app:statistics_phaseI}), which provides a measure of how much of the prediction really requires simultaneous knowledge of both features. In the case of $\tau_{c1}$, Figure~\ref{fig:LgradB_stats}b shows that approximately $\sim20\%$ of the behaviour requires the explicit involvement of both torsion features.

\subsubsection{Understanding the role of torsion}
Although it is unsurprising that torsion plays a role in setting $L_{\nabla\mathbf{B}}$, the relative unimportance of curvature may come as a surprise, as the gradients in $||\mathbf{B}||$ are controlled by curvature, and thus we naturally expect it to contribute to setting the length scales relevant to the field. 

An explanation of this observation necessitates a more precise consideration of the contributions to $L_{\nabla\mathbf{B}}$ at a theoretical level. In that direction, consider the leading asymptotic form of $L_{\nabla\mathbf{B}}$ \cite[Eq.~(3.11)]{landreman2021a}, written in the ideal QI limit\footnote{By ideal QI limit we mean the limit of the near-axis theory to first order in which the \textit{buffer regions} \citep{plunk2019direct} are ignored (especially simplifying the $\sigma$-equation). Although in general such buffers need to exist, this artificial consideration is nevertheless an insightful simplification that is particularly valid close in the neighbourhood of $B_\mathrm{min}$. Near $B_\mathrm{max}$ the validity breaks down, but behaviour in this region does not determine the minimum of $L_{\nabla\mathbf{B}}$ in practice.} as
\begin{multline}
    \nabla\mathbf{B}\approx\frac{B_0}{\ell'}\left(\frac{\bar{e}'}{2\bar{e}}-\frac{B_0'}{2B_0}\right)\hat{\pmb \kappa}\hat{\pmb \kappa} - \frac{B_0}{\ell'}\left(\frac{\bar{e}'}{2\bar{e}}+\frac{B_0'}{2B_0}\right)\hat{\pmb \tau}\hat{\pmb \tau}+\\
    +B_0\tau(\hat{\pmb \kappa}\hat{\pmb \tau}-\hat{\pmb \tau}\hat{\pmb \kappa})+B_0\kappa(\hat{\pmb t}\hat{\pmb \kappa}+\hat{\pmb \kappa}\hat{\pmb t})+\frac{B_0'}{\ell'}\hat{\pmb t}\hat{\pmb t}. \label{eqn:LgradB_QI}
\end{multline}
There are four physical sources of length scales in the field: the variation of the field strength along the axis ($B_0'$), the variation of elongation ($\bar{e}'$)\footnote{It is convenient here to use the function $\bar{e}$ as defined in \cite[Eq.~(3.6)]{plunk2025-geometric},     $\rho = \bar{e} + (1 + \sigma^2)/\bar{e}$, which is directly realted to the elongation of the elliptic cross-section, but does also involve $\sigma$, related to the rotation of the cross-sections \citep[Sec.~3.2.1]{rodriguez2023mhd}.}, the curvature ($\kappa$) and the torsion ($\tau$). The main curvature term is directly related to the bending ($\hat{\pmb \kappa}$) of field lines ($\hat{\pmb t}$), while the main torsion term clearly corresponds to twisting of the frame ($\hat{\pmb \tau}\hat{\pmb \kappa}-\hat{\pmb \kappa}\hat{\pmb \tau}$). In practice, in the context of the database, we may focus on these two contributions to explain the behaviour of $L_{\nabla\mathbf{B}}$. Both $B_0'$ and $\bar{e}'$ (which depends directly on $\tau$ and $\rho$) can be largely treated as secondary so long as their variations are tamed, which the direct control on $B_0$ and elongation in the construction enables. 

\begin{table}
\centering
\begin{tabular}{c c|c c|c c|c c|c c|c c}
\multicolumn{2}{c|}{$N=1$} & 
\multicolumn{2}{c|}{$N=2$} & 
\multicolumn{2}{c|}{$N=3$} & 
\multicolumn{2}{c|}{$N=4$} & 
\multicolumn{2}{c|}{$N=5$} & 
\multicolumn{2}{c}{$N=6$} \\
\hline
$\hat{\tau}$ & 0.995 & $\check{\tau}$ & 0.813 & $\check{\tau}$ & 0.759 & $\check{\tau}$ & 0.804 & $\sigma_\tau$ & 0.822 & $\sigma_\tau$ & 0.869  \\
$\tau_{\min}$ & 0.996 & $\tau_{\min}$ & 0.950 & $\mathcal{C}$ & 0.928 & $\hat{\tau}$ & 0.941 & $\mathcal{C}$ & 0.942 & $\mathcal{C}$  & 0.943  \\
$\tau_{\max}$ & 0.996 & $\mathcal{C}$ & 0.974 & $\tau_{\max}$ & 0.980 & $\mathcal{C}$ & 0.971 & $\check{\tau}$ & 0.965 & $\tau_\mathrm{max}$ & 0.961  \\
\end{tabular}
\caption{\textbf{Dominant extended features for $L_{\nabla\mathbf{B}}$.} The table summarises the top selected features by FSFS on the extended features detailed in Appendix~\ref{app:statistics_phaseI}. The numerical value represents the coefficient of determination $R^2$ of the $L_{\nabla\mathbf{B}}$ model using the variable on that row alongside those above. The meaning of the various parameters in the table are presented in Appendix~\ref{app:statistics_phaseI}: $\hat{\tau},\check{\tau}$ is the torsion at $B_\mathrm{max}$ and $B_\mathrm{min}$, $\tau_\mathrm{min},\tau_\mathrm{max}$ the minimum and maximum torsion respectively, $\sigma_\tau$ the standard deviation of torsion, and $\mathcal{C}=\sqrt{\kappa_\mathrm{max}^2+\tau_\mathrm{\kappa_\mathrm{max}}^2}$ with $\kappa_\mathrm{max}$ is the maximum curvature and $\tau_\mathrm{\kappa_\mathrm{max}}$ the value of torsion there. }
\label{tab:input_importance}
\end{table}

Following Eq.~(\ref{eqn:LgradB_QI}), $\kappa$ and $\tau$ contribute on equal footing to $||\nabla\mathbf{B}||$. The difference in their roles can be traced to the fact that the omnigenous (or rather the pseudosymmetric \citep{skovoroda2005}) requirement on the magnetic field constrains curvature in particular, forcing it to vanish at the maxima and minima of the magnetic field \citep{plunk2019direct}. The field gradient at these straight sections is thus uniquely set by torsion, whose value at the bottom we denote by $\check{\tau}$. Away from the flattening points, curvature grows, and the scale of the field is set by a combination of combining curvature and torsion that peaks somewhere half-way to the maximum. 

Making a direct connection between these theoretical observations and the role played by the input features is highly non-trivial. A prime example of this is the relation of the input features to $\check{\tau}$ (or any other value of torsion around the axis), which requires knowledge of $\tau_0$ after solving the axis closure problem, besides $\tau_{c1}$ and $\tau_{c2}$. It is then natural to set up a statistical analysis of the database with an extended feature vector that includes $\tau_0$ and $\kappa_1$. We may go beyond, and complement these with a plethora of other meaningful feature combinations (see a full account of these in Appendix~\ref{app:statistics_phaseI}), including, for instance, $\check{\tau}$ or the position of the maximum of curvature $\ell_\kappa$. A similar dependence question on $L_{\nabla\mathbf{B}}$ could be now posed on this extended set of features. The problem is however significantly different from the original, as the features considers are no longer strongly independent. Adapting the analysis of the problem to this reality then, we find it fit to adopt Forward Sequential Feature Selection (FSFS) \citep{whitney1971direct}. The importance of features in $L_{\nabla\mathbf{B}}$ is defined by sequentially choosing (and ranking) features so as to maximise the goodness of a model for $L_{\nabla\mathbf{B}}$ fitted to a sequentially increasing number of features (see the details in Appendix~\ref{app:statistics_phaseI}). The results are summarised in Table~\ref{tab:input_importance}.

\begin{figure}
    \centering
    \includegraphics[width=0.9\linewidth]{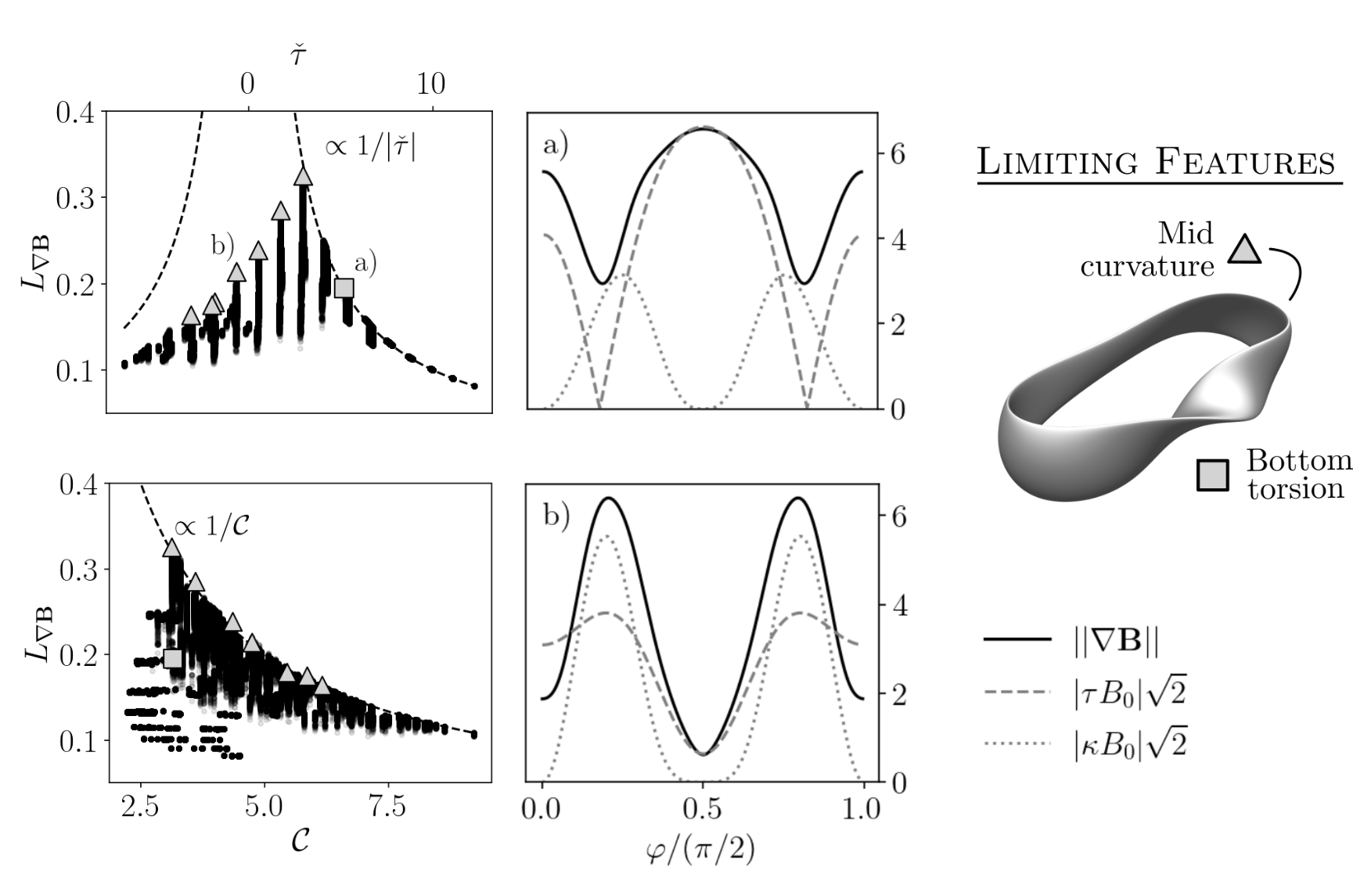}
    \caption{\textbf{Analysis of torsion dependence of $L_{\nabla\mathbf{B}}$.} The left plots show $L_{\nabla\mathbf{B}}$ in the $N=4$ subset (lightgray scatter) as a function of the value of torsion at the field minimum, $\check{\tau}$, and the combination of torsion and curvature at the point where curvature is maximum, $\mathcal{C}$. The black broken lines represent the two different predicted upper bounds for $L_{\nabla\mathbf{B}}$, illustrating the two limiting contributions indicated in the diagram (right). The behaviour of $||\nabla\mathbf{B}||$ in these two limits is exemplified for two example configurations a) and b) (right plots). Curvature and torsion are also shown for comparison. }
    \label{fig:L_grad_B_limit}
\end{figure}

The most relevant features appear directly related to torsion, in agreement with the reduced input feature analysis. Remarkably, the statistical analysis points at the torsion at the field minimum to be of particular importance (describing $\sim80\%$ of the data variance), connecting directly to our preceding theoretical discussion. Figure~\ref{fig:L_grad_B_limit} illustrates $L_{\nabla\mathbf{B}}$ (for the $N=4$ case) as a function of $\check{\tau}$, clearly showing that the value of the torsion at the field minimum sets an upper bound $L_{\nabla\mathbf{B}}\sim1/\check{\tau}$. Having an odd order of curvature at the field minimum forces a non-zero off-plane excursion of the axis (see Appendix~A of \cite{rodriguez2024near} or Appendix~A in \cite{plunk2025-geometric}), and thus generally an indelible finite torsion in the neighbourhood of $B_\mathrm{min}$, unlike near the maximum. The contribution to $||\nabla\mathbf{B}||$ is minimised there by keeping torsion spatially separated from curvature, and thus this finite torsion through the minimum is the least detrimental form of torsion that sets a minimal scale for $L_{\nabla\mathbf{B}}$.
This limiting torsion (\textit{i.e.} local twist) around $B_\mathrm{min}$ aligns with the configurations of Figure~\ref{fig:LgradB_field_period}, especially clear for $N=1$. 

The $N=1$ case is rather especial, in part because of its reduced range of input features (a single free parameter $\tau_0$ is scanned instead of $\{\tau_{c1},\tau_{c2},\kappa_1\}$). However, $\check{\tau}$ remains to capture $\sim99.5\%$ of the data variance, even if it does not appear in Table~\ref{tab:input_importance}. This highlights an important caveat of FSFS: it may hide features of similar importance just because a similar amount of information is given by an already included feature. This is partly what occurs at larger $N$ where the variation of the torsion along the axis, $\sigma_\tau$, overtakes $\check{\tau}$, even though the later remains highly important ($R^2>0.8$). This shift might nevertheless indicate an increased struggle to keeping the value of torsion tamed elsewhere, driven by $N$. 

This perspective, emphasizing torsion at $B_\mathrm{min}$, is however not complete, as it would suggest $\check{\tau}\rightarrow0$ to maximise $L_{\nabla\mathbf{B}}$, but Figure~\ref{fig:L_grad_B_limit} shows that this fails below some critical $\check{\tau}$ value. In this limit the configurations are no longer limited by the field gradients at $B_\mathrm{min}$, but instead a combination of curvature and torsion somewhere between $B_\mathrm{min}$ and $B_\mathrm{max}$ (see $||\nabla\mathbf{B}||$ for configuration b) in Figure~\ref{fig:L_grad_B_limit}). Such combination mentioned in the theoretical discussion does also appear in the FSFS analysis in the form of $\mathcal{C}=\sqrt{\kappa_\mathrm{max}^2+\tau_\mathrm{\kappa_\mathrm{max}}^2}$. This combination serves indeed as an upper bound $L_{\nabla\mathbf{B}}\sim1/\mathcal{C}$, see Figure~\ref{fig:L_grad_B_limit}. Despite this clear role, the FSFS analysis only gives $\mathcal{C}$ a secondary role. $\check{\tau}$ appears to capture its behaviour to a good degree; after all, torsion and curvature are tightly linked.  

The best configurations will therefore attempt to minimise both these features as much as possible. First, they will try to minimise torsion, being left with some remanent through $B_\mathrm{min}$. Second, they will limit the peak curvature to the largest extent possible, and spatially separate it from the torsion contribution.  It is then expected (and observed) for curves of maximal $L_{\nabla\mathbf{B}}$ to be rather flat, but retain a small off-the-plane excursion.  Extreme elongation variations or mirror ratio should also be avoided, but these issues are a lesser problem in practice.

\subsection{Critical aspect ratio of stabilised field, $A_c^\mathrm{mhd}$} \label{sec:Acmhd_anal}
One of the key developments of \cite{rodriguez2025near} was the possibility of constructing MHD stable near-axis fields with minimum amount of shaping and without the need of numerical optimisation. {  Here by "shaping" we refer to the variations in shaping of flux-surfaces to second order, which grow away from the axis and align (with the important toroidal dependence) with the classical notion of the shaping of cross-sections such as triangularity.} Given any first order construction, the second order shaping can always be directly chosen to achieve a desired level of stability. By MHD stability we mean the restricted notion of a vacuum magnetic well \citep{greene1997}, $W$ \citep[Eq.~(6.2)]{rodriguez2025near}, the leading low-beta limit of Mercier's criterion. Thus every first-order element in the database generates a whole family of minimally shaped second order near-axis fields, parametrised by the magnetic well value. We shall take as representative of such family the marginal $W=0$ point, {\em i.e.} this value is assumed in discussions below unless otherwise stated. 

The existence of such families do however pose an important question: how can we distinguish between different configurations with regard MHD stability? How can we say that a given configuration is more prone to stability than another when all can be made so? To address such questions we introduce a notion of shaping in the form of \cite[Definition~6.1c]{rodriguez2025near},
\begin{equation}
    A_c^\mathrm{mhd}=R/\mathrm{min}\left[r~|~\exists~\theta,\varphi: \mathcal{J}(r,\theta,\varphi)=0\right],
\end{equation}
for the marginally stable construction, where $\mathcal{J}$ is the Boozer coordinate Jacobian. $A_c^\mathrm{mhd}$ is the largest aspect ratio at which the nested flux surface construction of the near-axis model first ``breaks''. Beyond a measure of near-axis expansion validity, $A_c^\mathrm{mhd}$ can also be interpreted as a measure of intrinsic MHD stability: if $A_c^\mathrm{mhd}$ is small, it indicates that a small amount of flux-surface shaping\footnote{Note that a truly first order field with bare elliptical cross-sections has $A_c^\mathrm{mhd}\rightarrow0$ \citep{landreman2021a}. {  Thus, when talking about shaping, elongation, which in our approach is largely controlled, does not directly impact $A_c$.} Unlike in the $L_{\nabla\mathbf{B}}$ discussion, then, $A_c^\mathrm{mhd}$ is an intrinsically second order quantity.} suffices to achieve a magnetic well. 

\subsubsection{General statistical description}
\begin{figure}
    \centering
    \includegraphics[width=\linewidth]{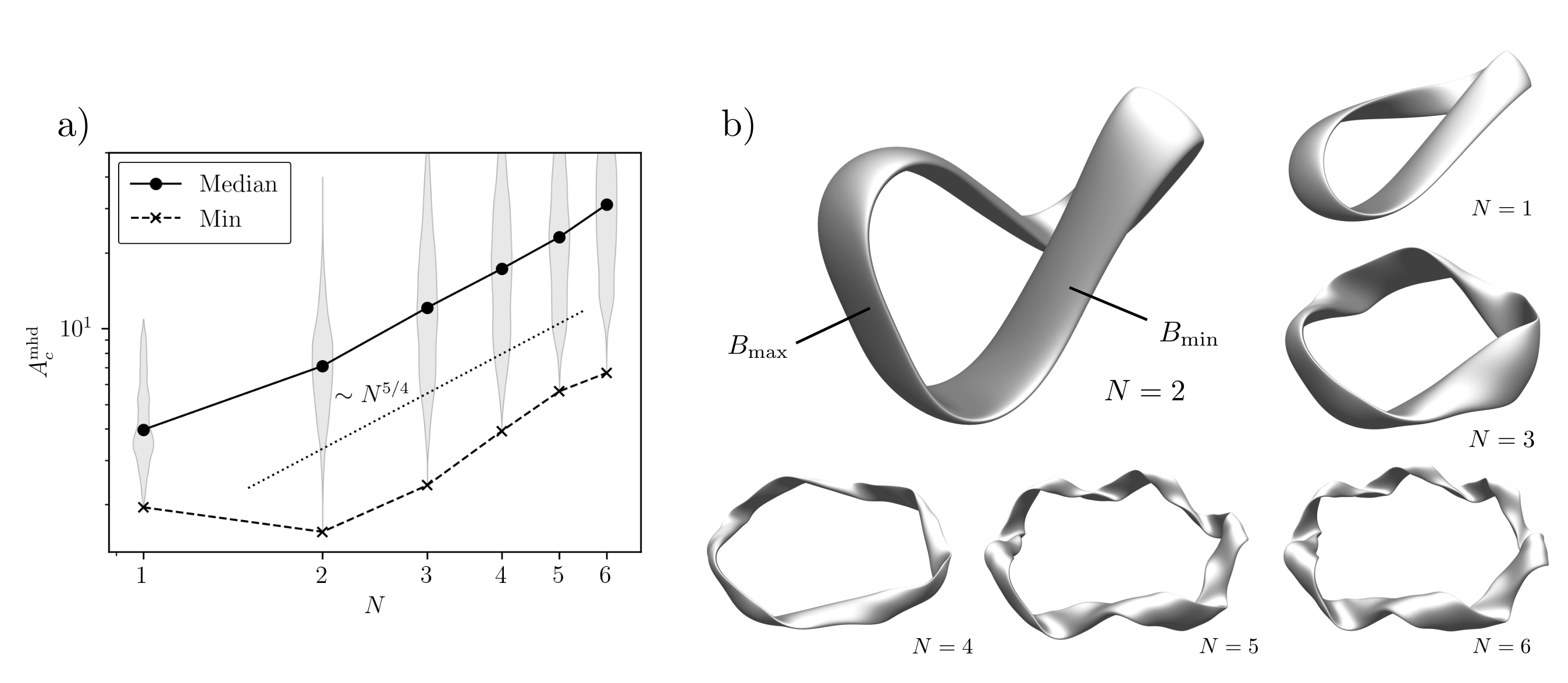}
    \caption{\textbf{$A_c^\mathrm{mhd}$ behaviour across different number of field periods.} a) Dependence of the minimum and median value of $A_c^\mathrm{mhd}$ as a function of the number of field periods $N$ in the database. Reference scaling is given as dotted line. b) Rendition of the 3D finite aspect ratio flux surface of $A_c^\mathrm{mhd}$ maximising fields for each number of field periods in the database. The locations of $B_\mathrm{min}$ and $B_\mathrm{max}$ are indicated for the $N=2$ configuration, which has the lowest value of $A_c^\mathrm{mhd}$.}
    \label{fig:Acmhd_field_period}
\end{figure}

As we did with $L_{\nabla\mathbf{B}}$, consider the variation with $N$ of $A_c^\mathrm{mhd}$ across the database, Figure~\ref{fig:Acmhd_field_period}a. Configurations with larger $N$ are increasingly shaped, owing to the need to pack the field in an increasingly narrow toroidal domain. A na\"{i}ve estimate of the scaling of the aspect ratio with $N$ based on the involvement of toroidal derivatives in the near-axis construction\footnote{The critical aspect ratio can be though as the aspect ratio for which the second order shaping contribution is strong enough to strongly affect the leading elliptical shaping of flux surfaces. Given the involvement (formally) of two $\varphi$-derivatives in going from first to second order in the near-axis construction we expect the magnitude of the second order corrections to scale roughly like $S_{2^\mathrm{nd}}/S_{1^\mathrm{st}}\sim N^2$. Balancing then $r S_{1^\mathrm{st}}\sim r^2S_{2^\mathrm{nd}}$, $r\sim1/N^2$, yielding $A\sim N^2$.} gives a pessimistic $A \sim N^2$, while the database shows a scaling closer to, but slightly above, linear ($A_c^\mathrm{mhd}\sim N^{5/4}$). A linear dependence on $N$ can be { pictured} as the natural change in aspect ratio for a stellarator put together stitching together $N$ identical sectors toroidally. This stark discrepancy is not necessarily due to the imposition of MHD stability, but rather the minimal choice of second order shaping. This appears clear when a similar analysis is performed through the database choosing the 2nd order completion of the configurations without targeting the magnetic well. In that case, $A_c\sim N^{3/4}$, roughly a factor of $\sim 2N^{1/2}$ smaller than the stability constrained solutions.

In practice, the deterioration in shaping is important, but not forbidding: configurations with reasonable aspect ratios can be found at larger $N$ (see Figure~\ref{fig:Acmhd_field_period}). We take as a measure of `reasonable' aspect ratio $A_c^\mathrm{mhd}\leq10$, which is similar to that of W7-X and typical QI stellarator power plant designs \citep{Beidler_2001, LION_2025, Hegna_2025}. Given the $N$ dependence of $A_c^\mathrm{mhd}$, such an absolute reference becomes increasingly restrictive with $N$, \textit{e.g.}, corresponding to a 1\% of the database population for $N=6$.  We note, however, that there is no fundamental reason that large aspect ratios should be excluded from consideration absolutely.

We illustrate configurations at different $N$ showing those with the lowest $A_c^\mathrm{mhd}$ (for each $N$) in Figure~\ref{fig:Acmhd_field_period}b. The absolute lowest critical aspect ratio in the whole database is $A_c^\mathrm{mhd}\approx1.6$, and is achieved by the \textit{figure-8} configuration at $N=2$. Unlike the $L_{\nabla\mathbf{B}}$ maximising configurations in Figure~\ref{fig:LgradB_field_period}b, configurations can present a strong off-the-plane excursion, epitomised by the figure-8 \citep{plunk2024back}. Such a trait does however not appear to be a necessity, a point that we shall consider in detail in the following sections. Interestingly, when the lowest aspect ratios are calculated for configurations constructed at a stronger magnetic well strength, take $W=-3\%$, the shaping does overall increase and the absolute lowest $A_c^\mathrm{mhd}$ no longer resides on the $N=2$ figure-8, but instead shifts to $N=3$. 

Before moving on to a discussion on the different ways of achieving low $A_c^\mathrm{mhd}$, we must make an aside on interpreting exceedingly low values of $A_c^\mathrm{mhd}$. Any attempt to construct such configurations will quickly encounter significant limitations. Besides the deterioration of the near-axis model reliability (likely the main concern), unintended flux surface overlap could also occur. This is true even if $A>A_c^\mathrm{mhd}$ \citep[Sec.~4]{landreman2021a}), as is apparent near the crossing of the figure-8. In that particular case, the true bound on the aspect ratio becomes $A\approx2.3$.\footnote{The $A_c^\mathrm{mhd}$ measure can only predict local breakdown in flux surfaces (\textit{i.e.,} coordinate singularities). To assess non-local breakdowns such as the two sections in the figure-8 crossing touching, we need to measure the minimum distance between different points on a flux surface constructed at a finite $r$, and see when this vanishes. This is a rather numerically expensive calculation (even after using improved algorithms such as Bounding Volume Hierarchy \citep{kay1986ray,ericson2004real}), as it requires computing intersections between many point-pairs. The calculation has been nevertheless performed on the entire database, but we shall not consider it further as it generally leads to no dramatic change.}


\subsubsection{Dependencies of $A_c^\mathrm{mhd}$} \label{sec:Acmhd_dep}
\begin{figure}
    \centering
    \includegraphics[width=\linewidth]{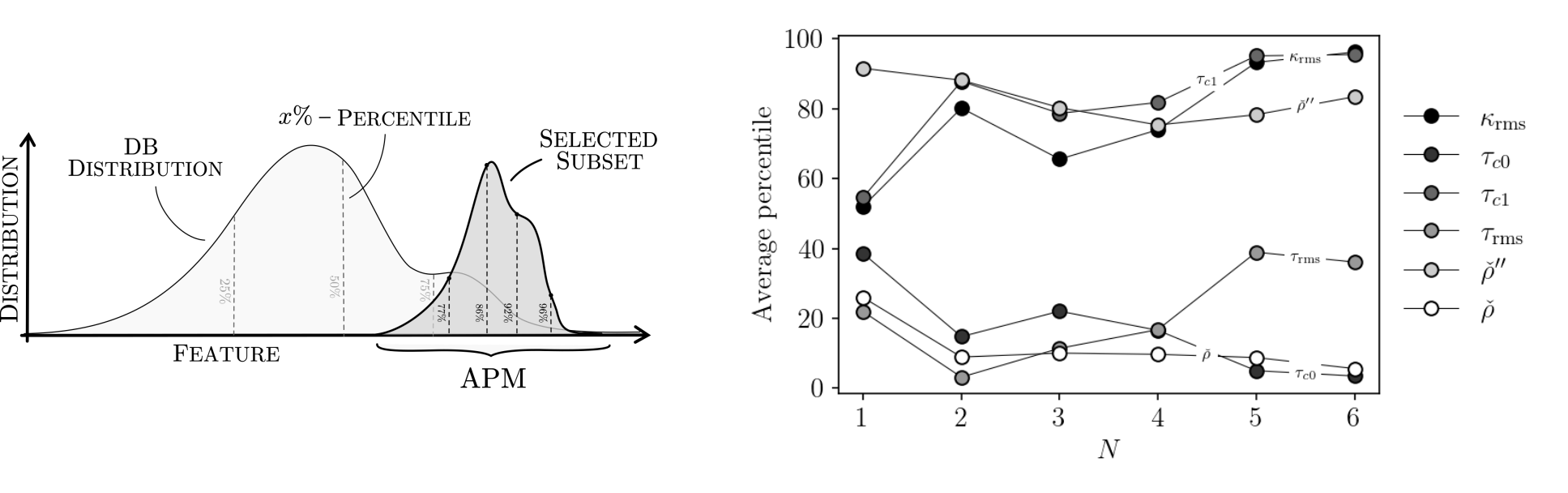}
    \caption{\textbf{Average percentile measure (APM) for lowest $A_c^\mathrm{mhd}$.}  (Left) Schematic diagram for the average percentile measure (APM) calculation. (Right) APM of the best $A_c^\mathrm{mhd}$ 150 configuration subset (roughly the lowest 1000th quantile for $N\geq2$) respect to the total population as a function of the number of field periods.}
    \label{fig:distribution_Acmhd}
\end{figure}

Let us now delve into the dependencies of $A_c^\mathrm{mhd}$ on input features (see Figure~\ref{fig:Acmhd_stats}). The correlation analysis indicates that torsion is the most important feature, dominated by $\tau_{c1}$. A significant linear correlation with $\tau_{1c}$ ($R^2\gtrsim0.6$) indicates that the critical aspect ratio of configurations tends to decrease as $\tau_{c1}$ (related to the inclination of the configurations \citep[Fig.~6]{plunk2024back}), increases (see Figure~\ref{fig:Acmhd_stats}a), pushing the lowest $A_c^\mathrm{mhd}$ configurations towards the sampling boundary of $\tau_{c1}$.  To assess this stark difference with the behaviour for $L_{\nabla\mathbf{B}}$, we introduce an \textit{average percentile measure} (APM). The measure calculates (for each $N$) the database population percentile (in the feature of interest, say $\tau_{c1}$) to which each configuration in a small group of interest (say, the lowest 1000th quantile, 150 configurations, in $A_c^\mathrm{mhd}$) belongs, and averages it (see diagram in Figure~\ref{fig:distribution_Acmhd}). Applied to $\tau_{c1}$ (see Figure~\ref{fig:distribution_Acmhd}) it shows a clear skewness (above 80\%), a sign that the sampling approach to the database may be missing configurations nearby and beyond this point. Although there is some truth to that (a point which a future extension of the database shall consider), the bound on the sampling space is not arbitrary, as it reflects to an intrinsic limit encountered in the construction of family of magnetic axes with the desired half helicity \citep[Fig.~6]{plunk2025-geometric}.

\begin{table}
    \centering
\begin{tabular}{c c|c c|c c|c c|c c|c c}
\multicolumn{2}{c|}{$N=1$} &
\multicolumn{2}{c|}{$N=2$} &
\multicolumn{2}{c|}{$N=3$} &
\multicolumn{2}{c|}{$N=4$} &
\multicolumn{2}{c|}{$N=5$} &
\multicolumn{2}{c}{$N=6$} \\
\hline
$\tau_{0}$ & 0.935 & $\tau_{\max}$ & 0.514 & $\tau_{\max}$ & 0.659 & $\tau_{\max}$ & 0.524 & $\check{\tau}$ & 0.558 & $\check{\tau}$ & 0.501  \\
$\hat{\rho}''$ & 0.963 & $\int\kappa\,d\ell$ & 0.634 & $\check{\rho}$ & 0.782 & $\check{\rho}$ & 0.679 & $\hat{\tau}$ & 0.737 & $\hat{\tau}$ & 0.717  \\
$\check{\rho}$ & 0.970 & $\check{\rho}$ & 0.701 & $\tau_{\kappa_\mathrm{max}}$ & 0.854 & $\int\kappa\,d\ell$ & 0.798 & $\check{\rho}$ & 0.866 & $\check{\rho}$ & 0.862  \\
$\Delta$ & 0.977 & $\hat{\tau}$ & 0.748 & $\hat{\tau}$ & 0.898 & $\hat{\tau}$ & 0.839 & $\check{\kappa}^{(3)}$ & 0.881 & $\hat{\rho}$ & 0.894  \\
\end{tabular}

\caption{\textbf{Most important combined input features for $A_c^\mathrm{mhd}$.} The table summarises the top selected features by FSFS on the combined input features detailed in Appendix~\ref{app:statistics_phaseI}. The numerical value represents the coefficient of determination $R^2$ of the $A_c^\mathrm{mhd}$ model using the variable on that row alongside those above. The meaning of the various parameters in the table are presented in Appendix~\ref{app:statistics_phaseI}: $\hat{\tau},\check{\tau}$ is the torsion at $B_\mathrm{max}$ and $B_\mathrm{min}$, $\tau_0$ the integrated torsion, $\tau_\mathrm{rms}$ the root-mean-square torsion, $\check{\rho}$ and $\hat{\rho}$ the value of $\rho$ at $B_\mathrm{min}$ and $B_\mathrm{max}$ respectively, $\hat{\rho}''$ the variation of elongation at $B_\mathrm{max}$, $\int\kappa\mathrm{d}\ell$ the integrated curvature over half a period, and $\tau_{\kappa_\mathrm{max}}$ the torsion at the point of maximum curvature, $\check{\kappa}^{(3)}$ the third derivative of curvature at $B_\mathrm{min}$ and $\Delta$ the mirror ratio. }
    \label{tab:Acmhd_feature_importance}
\end{table}

Besides the involvement of torsion, flux surface elongation also plays a non-trivial part with $A_c^\mathrm{mhd}$. We can see this in both the non-trivial importance of $\rho_1$ as well as the behaviour of $\rho$-related features in Figure~\ref{fig:distribution_Acmhd}. Unlike with $L_{\nabla\mathbf{B}}$, field shaping beyond the axis geometry has direct important practical implications. This is also evident on the FSFS analysis summarised in Table~\ref{tab:Acmhd_feature_importance}, where a description of $A_c^\mathrm{mhd}$ based only on torsion is clearly lacking. Elements of elongation are thus key, as are some consideration of curvature (especially at large $N$).

\subsubsection{The physics and dependencies of $A_c^\mathrm{mhd}$}
To be able to frame the statistical observations made above, we must introduce some additional theoretical notions involving MHD stability and shaping of configurations. 

Let us start with the notion of MHD stability, or more precisely, that of a vacuum magnetic well. When a field has a magnetic well it means that, on average, the gradient of the magnetic field magnitude points outwards, leaving the core a magnetic minimum. Formally, $W\sim\mathrm{d}(1/B^2)/\mathrm{d}\psi$, which in the context of a near-axis field is \citep[Eq.~(3.6)]{landreman2020magnetic}\citep[Eq.~(6.1-2)]{rodriguez2025near}
\begin{equation}
    W\propto\left[\int_0^{2\pi}\frac{\mathrm{d}\varphi}{B_0^2}\right]^{-1}\int_0^{2\pi}\frac{\mathrm{d}\varphi}{B_0^4}\left[3\left(B_{1s}^2+B_{1c}^2\right)-4B_0B_{20}\right], \label{eqn:mag_well_nae}
\end{equation}
where quantities have their usual near-axis meaning. The first term in the integrand involves the variation of the magnetic field strength due to the curvature of the axis, $B_1$, and its contribution to the well is always adverse (\textit{i.e.}, the term is always positive, pushing towards $W>0$).\footnote{The reason behind such term follows from the expansion of $1/B^2$, and noting that the average of $B_1$ vanishes (due to its $\theta$ dependence).} The poloidal average of the radial $|\mathbf{B}|$ gradient, $B_{20}$, can in turn contribute positively to stability if $B_{20}>0$. Thus, if we take these two contributions as independent of each other, minimising curvature and making $B_{20}$ large and positive would be seen to promote stability.

However, the problem is not so simple, as $B_{20}$ is the consequence of a complex balance between geometry and the field. Although it can be strongly altered by choice of flux surface shaping at second order in the expansion (\textit{i.e.}, triangularity), it also inherits part of its behaviour from lower orders. This is particularly apparent at $B_\mathrm{min}$ (where unlike at $B_\mathrm{max}$ deviations from QI are by construction minimal), which following the analysis in \cite{rodriguez2024maximum} and \citep[Eq.~(15)]{plunk2024back}
\begin{equation}
    B_{20,\mathrm{b}} = \left. B_{20}\right|_{\varphi = \pi/N} = \frac{\rho}{4}\left(\frac{d\varphi}{dl}\right)^2 \left [-\frac{B_0^{\prime\prime}}{2B_0} + \frac{1}{2} \frac{\rho^{\prime\prime}}{\rho} -\left(\tau \frac{\mathrm{d}\ell}{\mathrm{d}\varphi}\right)^2 \right],\label{eq:magnetic-well-contributions-QI}
\end{equation}
where primes denote toroidal derivatives. Although such measure is by definition local, it can serve as guidance, as it exhibit sufficient correlation with $A_c^\mathrm{mhd}$ across the database and exposes clear physical contributions. The $B_0''$ term represents the detrimental inwards bending of field lines that trace the shrinking cross-sections (at constant flux) away from $B_\mathrm{min}$, \textit{i.e.}, breathing. By virtue of belonging to the $(2,3)$-class, this local contribution vanishes.  
The same cannot be said of the $\rho''$ term, which favours configurations whose elongation grows away from the minimum, \textit{i.e.}, stretching. This is reflected in the database, as both the average percentile measure of $\check{\rho}''$ and $\check{\rho}$ in Figure~\ref{fig:distribution_Acmhd} and the FSFS analysis, Table~\ref{tab:Acmhd_feature_importance}, show.\footnote{A connection could also be made directly to $\rho_1$, which we saw to be the most important input feature of $\rho$. Elongation behaviour at $B_\mathrm{min}$ can be described by $\check{\rho}''\propto\rho_1+4\rho_2$ and $\check{\rho}=\rho_0+\rho_1+\rho_2$, which clearly involves $\rho_1$. However, it also involves $\rho_2$, and thus cannot explain the prominence of $\rho_1$ over $\rho_2$. Only upon consideration of the elongation at $B_\mathrm{max}$ we can distinguish between these two. $\rho_1$ contributes in opposite ways at the minimum and maximum, while $\rho_2$ (or $\rho_0$) does not. Following Eq.~(\ref{eq:magnetic-well-contributions-QI}), it makes thus sense that $\rho_1$ changes $B_{20}$ more significantly than $\rho_2$ does, given the opposed natures of $B_\mathrm{min}$ and $B_\mathrm{max}$.} Finally, we have the always detrimental contribution of torsion, representing the inwards pointing curvature of any helical field line. This universality in its contribution presents minimising torsion everywhere as a solid principle to ameliorate stability. This is reflected in the involvement of $\tau_0$ and $\tau_\mathrm{rms}$ in the FSFS analysis, as well as APM in Figure~\ref{fig:distribution_Acmhd}.

The APM analysis is particularly insightful here, showing a consistent low integrated torsion $\tau_{0}$ throughout the low $A_c^\mathrm{mhd}$ configurations across the database (below the 20\% mark for $N\geq2$). This aligns with the configurations living near the $\tau_{c1}$ domain boundary, as noted above. In this limit the average torsion of the constructed curves sharply drops, as they are able to release their `tension' through large inclination (see Figure~3 in \cite{plunk2025-geometric}). Despite these low values, it must be noted that $\tau_0\approx0$ curves are seldom seen in the database, likely impeded by the sharpness of the transition and the granularity of the sampling. Curves with small total twist sharply increase their curvature to remain closed, which leads to curvature signs in Table~\ref{tab:Acmhd_feature_importance} and Figure~\ref{fig:distribution_Acmhd}, and could also play a regularising role through Eq.~(\ref{eqn:mag_well_nae}). Although this integrated torsion picture appears to nicely tie the database and theory, we must note that such curves do not have a pointwise vanishing torsion (with the exception of the figure-8). In that sense, then, is the torsion in the field really being minimised? Is minimising $\tau_\mathrm{rms}$ not closer to the theoretical argument? Apparently not: minimising integrated torsion appears a priority over $\tau_\mathrm{rms}$, especially at large $N$.

We must discuss shaping to explain this observation, and in particular the sensitivity of the ellipse rotation to the mean torsion. As it follows from the so-called $\sigma$-equation \citep[Eq.~(B3)]{plunk2025-geometric}\citep[Eq.~(A26)]{landreman2019}, in the vanishing integrated torsion scenario a solution to the equation with a small $\sigma$ is possible, minimising ``shaping proliferation''. Therefore $\tau_0$ and $\tau_\mathrm{rms}$ are not the same, and their difference in value grows with $N$ for $N\geq2$, explaining the trends in Figure~\ref{fig:distribution_Acmhd}, and hinting an explanation of the high-$N$ burden that stability imposes on shaping requirements ($A_c^\mathrm{mhd}/A_c\sim N^{1/2}$).

In summary, a few strategies for reducing $A_c^\mathrm{mhd}$ are apparent: (i) reduce torsion as much as possible, (ii) limit the amount of shaping in the construction by minimising $\tau_0$ (and if possible stick to lower $N$), (iii) avoid extreme curvature values, and (iv) leverage elongation stretching. 

\subsubsection{Different forms of field}
\begin{figure}
    \centering
    \includegraphics[width=0.9\linewidth]{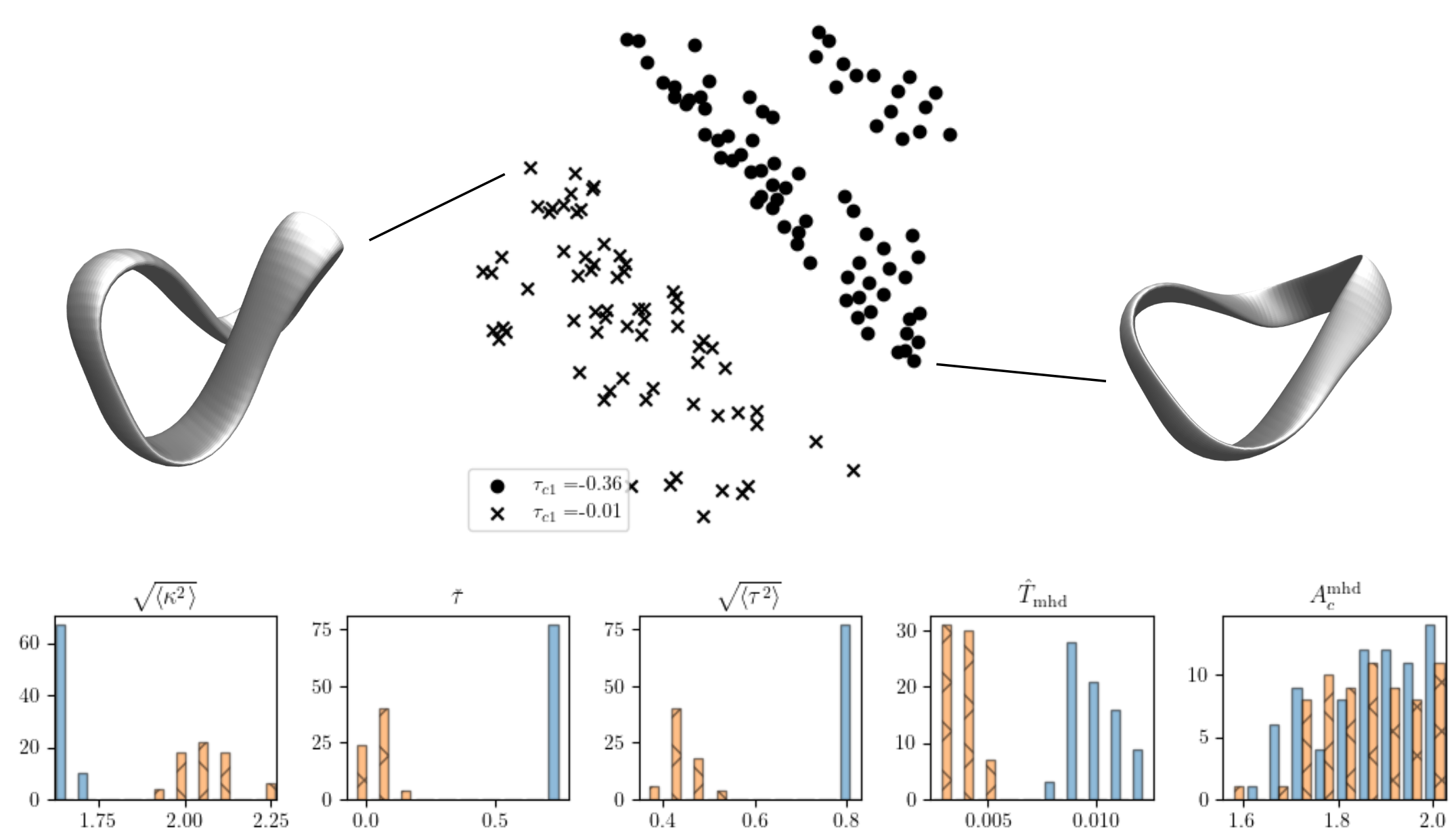}
    \caption{\textbf{Clustering of least shaped $N=2$ configurations.} The figure shows the clustering of the 1 thousandth percentile (a total of 145 configurations) of least shaped $N=2$ configurations represented as a cloud of points in 2D. MDS was used to perform the dimensionality reduction. Examples of the two resulting clusters, matching different $\tau_{c1}$ values, are illustrated. The figure-8 like configurations are represented by crosses - orange, and flat ones circles - blue. The distribution of some of their properties are shown in the histograms below. The discreteness of the database precludes a definitive consideration about inherent nature of these clusters.}
    \label{fig:clust_N_2}
\end{figure}

The broad criteria for achieving low $A_c^\mathrm{mhd}$ is fitting to understand the statistical behaviour of the database. Such principles are however not rigid, especially given the complexity of the measures involved. This opens the door to a larger variety of configurations compared to the $L_{\nabla\mathbf{B}}$ scenario, as was appreciated in Figure~\ref{fig:Acmhd_field_period}. 

To investigate this, let us look for different configurations within the group of the smallest $A_c^\mathrm{mhd}$, $N=2$ configurations (the same 1000th quantile as for the APM).\footnote{The conclusions of the analysis can be reached using larger fractions of the population as well. Restricting to smaller numbers does however make the discussion clearer.} To distinguish `significantly' different configurations, we proceed systematically by applying clustering algorithms (see details on the methodology in Appendix~\ref{app:clustering}) and representing the input-feature space in an approximately distance-preserving lower dimensional projection using Multidimensional Scaling (MDS) \citep{borg2005modern}, see Figure~\ref{fig:clust_N_2}. We observe configurations to cluster into two groups, each tagged by a different value of $\tau_{c1}$. The group with $\tau_{c1}\approx0$ corresponds to the markedly inclined \textit{figure-8} family; the other contains flatter configurations. This separation is likely artificial, a result of the granularity of the database. It is nevertheless clear that there is a degree of flexibility in achieving low $A_c^\mathrm{mhd}$ with crowns and flat configurations as limits. The figure-8 family tightly conforms to the guiding principles presented above: smallest torsion and minimal shaping (see the bottom of Figure~\ref{fig:clust_N_2}). The more open configurations are able to maintain a similar $A_c^\mathrm{mhd}$ despite requiring a moderate increase in shaping (as captured by $\hat{T}_\mathrm{mhd}$)\footnote{The measure $\hat{T}_\mathrm{mhd}$ is an absolute measure of second order shaping \citep[Def.~6b]{rodriguez2024near}, clearly indicating the need for a larger amount of shaping, even though it remains sufficiently small so as not to drive $A_c^\mathrm{mhd}$.} due to having larger torsion, although in exchange for a smaller curvature. 

\begin{figure}
    \centering 
    \hspace*{-0.1\textwidth} 
    \begin{minipage}[c]{0.75\textwidth}
        \centering
        \includegraphics[width=\linewidth]{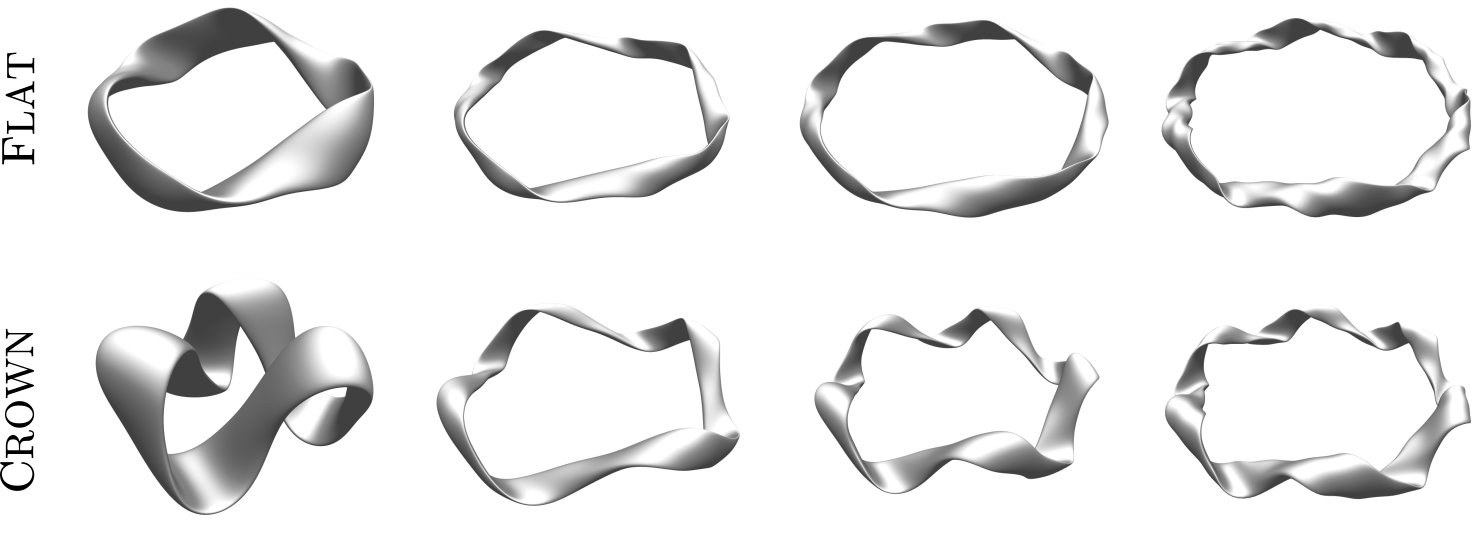}
    \end{minipage}
    \hspace{0.01\textwidth} 
    \begin{minipage}[c]{0.1\textwidth}
        \centering
        \begin{tabular}{ccc}
        N & Flat & Crown \\\hline
        2 & 1.6 & 1.6 \\
        3 & 2.4 & 3.1 \\
        4 & 3.9 & 4.1 \\
        5 & 6.5 & 5.9 \\
        6 & 8.1 & 6.7 \\
    \end{tabular}
    \end{minipage}%
     \caption{\textbf{Examples of least shaped marginally stable configurations of the "flat" and "crown" classes.} The figures are a 3D rendition of the representative least shaped, marginally stable configurations continuing the two $N=2$ families in Figure~\ref{fig:clust_N_2}. The table shows the $A_c^\mathrm{mhd}$ value for each of these (including the $N=2$ as reference).}
    \label{fig:table_image}
\end{figure}

A similar analysis and conclusion applies for larger $N$, which we represent in Figure~\ref{fig:table_image}. We have separated configurations between \textit{flat} configurations and \textit{crowns}. The latter can be thought of as extensions of the figure 8 to higher $N$, showing a degree of inclination, as the $N=3$ example nicely illustrates. Note that only milder versions of such crowns are found at higher $N$, perhaps inhibited by the increase in the curvature, or some other theoretical limit in the construction of the axis curves, although a finer resolution in the database would presumably help assess this question. Despite this trend, both ways of achieving low $A_c^\mathrm{mhd}$ (small and modest mean torsion) remain accessible with increasing $N$.

\begin{figure}
    \centering
    \includegraphics[width=\linewidth]{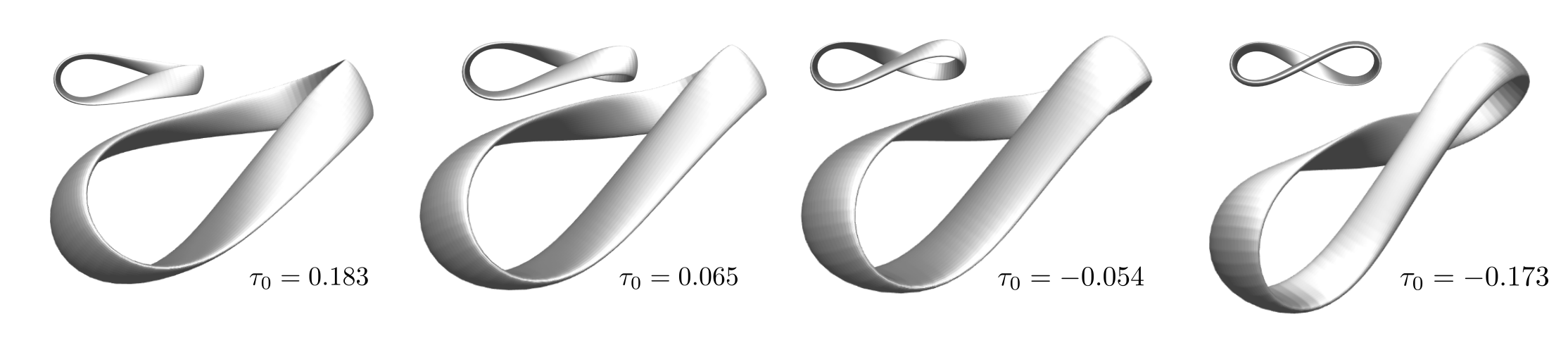}
    \caption{\textbf{Family of $N=1$ configurations.} Examples of configurations belonging to the low $A_c^\mathrm{mhd}$, $N=1$ family of configurations. A continuum of stellarators parametrised by the integrated torsion appears to exist, their corresponding critical aspect ratios are, in order, $A_c^\mathrm{mhd}=2.1,~2.0,~2.3,~2.3$.}
    \label{fig:N_1_Acmhd}
\end{figure}

The keen reader will have noticed that we have deliberately left the $N=1$ configurations out of this discussion. These configurations clearly belong to a wide continuum of configurations, as explicitly shown in Figure~\ref{fig:N_1_Acmhd}. This family is naturally parametrised by their integrated torsion, which is actually the sampling parameter used to construct them as part of the database. The latter circumvents some of the limitations for the other $N$ inherited from the discreteness and sensitivity on $\tau_{1c}$, unveiling a continuous family more clearly. The lowest $A_c^\mathrm{mhd}$ is still found under conditions of near-vanishing integrated torsion (see Figure~\ref{fig:N_1_Acmhd}), as expected from the above discussion. 

\subsection{Maximum-$\mathcal{J}$ behaviour, $f_\mathcal{J}$}
The radial gradient of the magnetic field strength is key not only for MHD stability, but also the magnetic drift precession of charged particles over flux surfaces. We expect to find the most MHD stable configurations, which push for a positive gradient, prone to achieving maximum-$\mathcal{J}$ \citep{rosenbluth1968} configurations \cite[p.~24]{helander2014theory}. A field attains maximum-$\mathcal{J}$ when all its trapped particles rotate against the diamagnetic drift direction. This precludes certain trapped-particle-resonance-driven instabilities \citep{rosenbluth1968,proll2012resilience,helander2013collisionless}, and is regarded as a desirable property of stellarator fields. 
\par
To explore this property across the configurations in the database, we define $f_\mathcal{J}$ to be the fraction of trapped particles that precess in the direction of maximum-$\mathcal{J}$ in a slender volume of radius $r$ about the magnetic axis. Note we are only interested in their direction rather than magnitude. Adapting \cite{rodriguez2024maximum},
\begin{equation}
    f_\mathcal{J}=\frac{1}{\mathcal{N}}\int_0^{r_\mathrm{ref}}r  \mathrm{d}r \int_0^{2\pi}\mathrm{d}\alpha\int_{1/B_\mathrm{max}}^{1/B_\mathrm{min}}\Theta[q\omega_\alpha(\lambda, \alpha, r)]\hat{\tau}_b(\lambda, \alpha, r)\mathrm{d}\lambda, \label{eqn:frac_max_J}
\end{equation}
where $\omega_\alpha$ is the bounce averaged poloidal drift,
\begin{equation}
    \omega_\alpha = \frac{mv^2}{q} \oint  \left(\frac{\partial B}{\partial \psi}\right)_{\alpha, \varphi} 
    F(\lambda,B) \; d\varphi  
      \bigg\slash \oint \frac{d \varphi}{B \sqrt{1-\lambda B}},\label{eq:precess_wa_exact}
\end{equation}
in a shearless vacuum. Here $F(\lambda,B) = (1-\lambda B/2)/B^2 \sqrt{1-\lambda B}$, and $\lambda$ is the usual velocity space label. The normalised bounce time $\hat{\tau}_b$ measures the relative population of trapped particles, assuming a basic Maxwellian distribution. $\mathcal{N}$ is defined by taking the same integral as in Eq.~(\ref{eqn:frac_max_J}) but without the Heaviside function in the integrand. We identify a maximum- (minimum) $\mathcal{J}$ configuration with a field for which $f_\mathcal{J}=1$ ($=0$). 
\par
Evaluating the contributions to Eqn.~\ref{eq:precess_wa_exact} within the near-axis leads to a number of terms at different orders that can be traced to omnigenous and non-omnigenous behaviour; a detailed discussion is found in Appendix \ref{app:fJ}.  We find that a single second order omnigenous contribution appears to dominate in practice, so that we may approximately write

\begin{equation}
    \omega_{\alpha} \approx \omega_{\alpha,\mathrm{vac}} = \frac{2mv^2}{q \bar{B}}\oint F(\lambda,B_0)\left[B_{20}-\frac{1}{4}\left(\frac{B_0^2d^2}{B_0'}\right)'\right]\mathrm{d}\varphi\Bigg/\oint\frac{\mathrm{d}\varphi}{B_0\sqrt{1-\lambda B_0}}. \label{eqn:nae_w_alpha}
\end{equation}

\subsubsection{Two approaches to large fractions}

One natural approach to achieving large $f_\mathcal{J}$ would be indeed to seek a dominant and positive $B_{20}$ value. The field equilibrium can be pushed in this direction leveraging large second order shaping.\footnote{Doing so will also generally make $B_{2c}$ large driving the non-omnigeneous term so that $\omega_{\alpha,\mathrm{vac}}\sim |\omega_{\alpha,0}^\mathrm{non-QI}|$; see Appendix \ref{app:fJ}.}  The resulting fields are however hardly realistic, as extremely high aspect ratios would be generally required for their realisation. It is also true that the shaping in the database has not been chosen with this particular strategy in mind, but rather to achieve a desired value of $W$. Unlike $f_\mathcal{J}$, the magnetic well measure only cares about average behaviour and not pointwise values.  Thus, even when strongly shaped, we cannot expect to achieve a fully maximum-$\mathcal{J}$ field. 

\begin{figure}
    \centering
    \includegraphics[width=\linewidth]{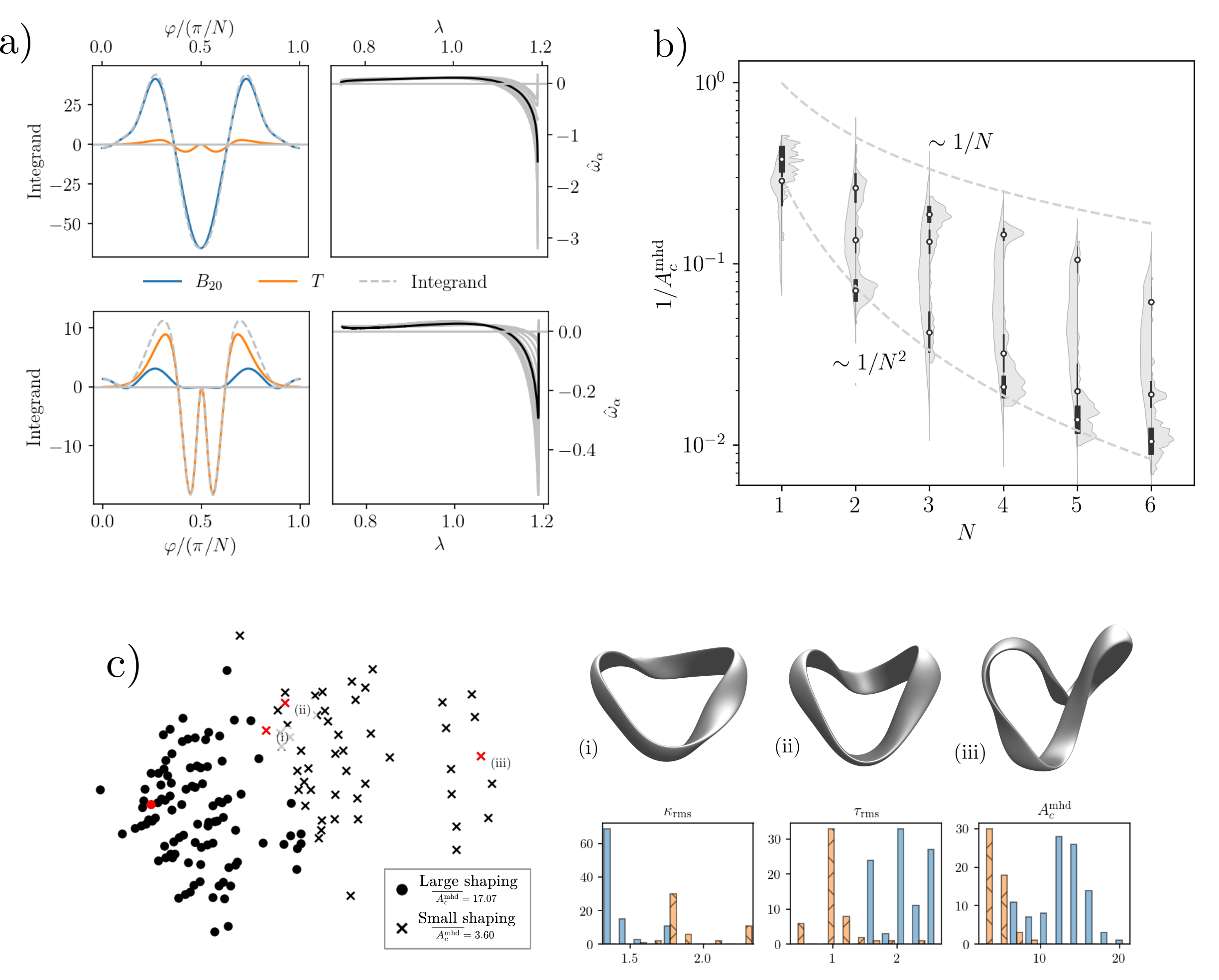}
    \caption{\textbf{High and low shaping approaches to $f_\mathcal{J}$.} a) Near-axis detail of the high (top) and low (bottom) shaping approaches to $f_\mathcal{J}$. Configurations are $N=2$ corresponding to the red scatter from the shaped cluster and (ii) in (c). The left plots show the contributions to the integrand of $\omega_{\alpha,\mathrm{vac}}$ in Eq.~(\ref{eqn:nae_w_alpha}), showing the $B_{20}$ and $T$-dominated limits. The plots on the right show the particle precession at $r=0.1$ as a function of $\lambda$, with the gray lines representing the precession for different $\alpha$ field lines.  b) Behaviour of distribution of $A_c^\mathrm{mhd}$ across different $N$, both of the whole database (left half of the violin plots) and the top $f_\mathcal{J}$ 1000th percentile subset (right half). The presence of separate clusters is apparent especially for $N=2-4$, and is highlighted by representing the mean and standard deviation of fitted Gaussian Mixture Models through errorbars \citep[Ch.~3.2]{wierzchon2018modern}. The line width represents the weight of the Gaussian. Reference scalings are given, following each of the cluster evolutions. c) Illustration of the $N=2$, top $f_\mathcal{J}$ 1000th percentile in a MDS representation (left). The clustering distinguishes the high and low shaping scenarios, and three different configurations in the latter are shown labelled (i), (ii) and (iii). The distribution of some features distinguishing the two clusters are shown as histograms (orange hashed corresponding to the moderately shaped fields).}
    \label{fig:two_approaches_frac_mj}
\end{figure}

Even if second order shaping was chosen more fittingly, we know there is an intrinsic part of $B_{20}$, near flattening points \cite{rodriguez2024maximum,rodriguez2024near}, that cannot be altered by it. In practice, the lack of correlation between $B_{20,b}$ and $f_\mathcal{J}$ implies that this limitation has no impact.  This irrelevance derives from the flatness of the field about $B_\mathrm{min}$, which dilutes the contribution of $B_\mathrm{min}$ with its shaped surroundings. The two examples of Figure~\ref{fig:two_approaches_frac_mj}a illustrate this, showing a very sudden transition in the particle precession $\omega_\alpha$ for extreme values of $\lambda$ corresponding to trapped particles, $\lambda=1/B_\mathrm{min}$. Most of the population, corresponding to smaller values of $\lambda$, is therefore not affected.

The second contribution inside the square bracket in Eq.~(\ref{eqn:nae_w_alpha}) is
\begin{equation}
    T=\left(B_0^2d^2/B_0'\right)'/4.
\end{equation}
The origin of this term is understood as follows.  In a QI field, the leading (first order) poloidal magnetic drift does not contribute to precession due to its parity, forced by omnigeneity. However, if one takes into account the change in the parallel velocity of the particles due to the magnetic field variation in the neighbourhood of the axis, \textit{i.e.}, $B_1$ contribution, a non-zero, second order net poloidal precession due to first order variations is obtained. The result is $T$.\footnote{The physical origin of this term can also be elucidated in more precise terms by revisiting the derivation of Eq.~(\ref{eqn:nae_w_alpha}) in \cite{rodriguez2024maximum}.} Stellarator symmetry forces this term to be even in $\varphi$, necessarily vanishing at both $B_\mathrm{min}$ and $B_\mathrm{max}$, \footnote{For a $B_0'\sim(\varphi-\varphi_0)^{u-1}$ and $d\sim(\varphi-\varphi_0)^v$, we have $\left(B_0^2d^2/B_0'\right)'/4\sim(\varphi-\varphi_0)^{2v-u}$. Thus, for $2v-u>0$ (necessary to avoid QI-breaking puddles) \citep{rodriguez2023higher}, $T$ vanishes at $\varphi=\varphi_0$.} and satisfying $T\geq0$ ($T\leq0$) in their neighbourhood respectively. The sign of $T$ changes roughly at the point of maximum $d$ (also the point of maximum curvature when we control elongation), making a majority of trapped particles contribute positively to $f_\mathcal{J}$ (as bounce times are biased towards barely trapped particles). Although this approach cannot attain $f_\mathcal{J}=1$ on its own, maximal profit is achieved if the bottom of the well is as wide as possible and the curvature maximum as close as possible to it. That way, the $T\geq 0$ contribution is restricted to a smaller fraction of deeply trapped particles, and the positive contribution maximised. For the first time in the context of this paper, curvature does not necessarily act to the detriment of the desired behaviour.


To investigate how these two extreme approaches materialise in the database, we investigate the subset of configurations with the largest values of $f_\mathcal{J}$ for each field period. As shown explicitly in Figure~\ref{fig:two_approaches_frac_mj}b, the distributions of these configurations exhibit clear traces of both the large and small shaping approaches, especially for $N=2,~3$. The $N$-scaling of the critical aspect ratio is indicative of this essential difference, with a scaling $A_c^\mathrm{mhd}\sim N^2$ matching strong shaping and $A_c^\mathrm{mhd}\sim N$ the lower one. At larger $N$ it appears that the distribution of configurations with high $f_\mathcal{J}$ is biased towards the larger shaping configurations, as the natural magnitude of second order quantities grow in this limit. 

The presence of these two groups can also be observed by application of clustering algorithms (see Appendix~\ref{app:clustering}), which naturally split the population according to this criterion (see Figure~~\ref{fig:two_approaches_frac_mj}c). Some representative shapes of the low shaping (\textit{i.e.}, $T$-dominated) cluster are given, recovering shapes seen in the analysis of $A_c^\mathrm{mhd}$: configurations with a limited torsion that limit amount of shaping and promote $B_{20}>0$. In the $T$-dominated regime the value of $B_{20}$ can still play a relevant role, in particular near points where $T=0$. More generally, a compromise between $B_{20}$ and $T$ contributions may be needed, especially if close-to-unity values of $f_\mathcal{J}$ are sought; the contribution from $B_{20}$ must amend the mishaps near $B_\mathrm{min}$ through $T$. 

\begin{figure}
    \centering
    \includegraphics[width=\linewidth]{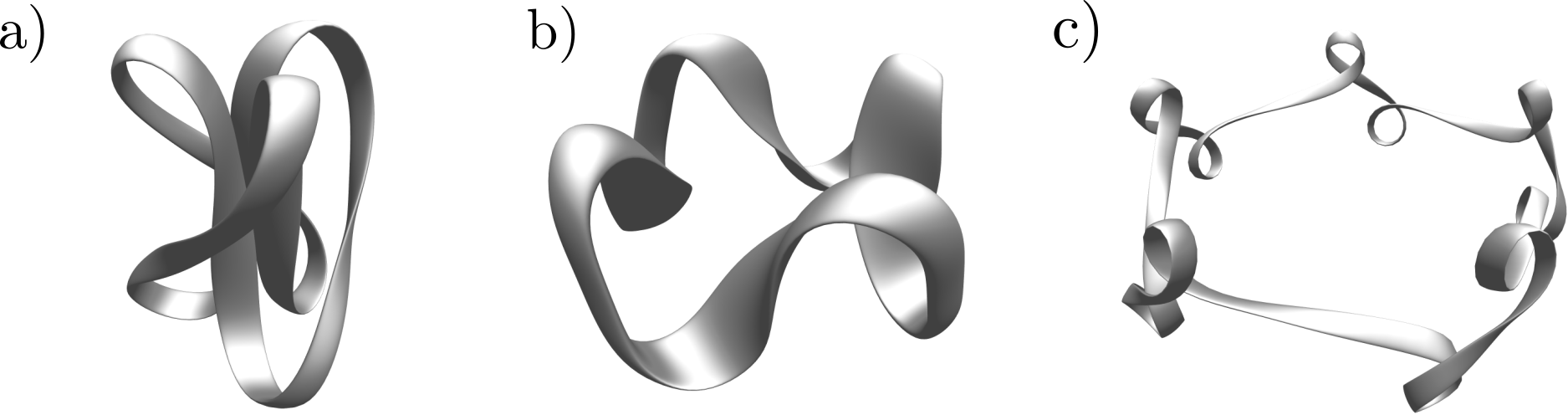}
    \caption{\textbf{Large curvature fields in the large $f_\mathcal{J}$ subsets.} The figure presents three examples of large curvature fields for a) $N=3$, b) $N=4$ and c) $N=5$, as part of the low shaping clusters. These show a knotted configuration, a crown and a curled alternative. Both a) and c) are more a curiosity than practical alternatives, but emphasise the breadth of the database, as well as the large curvature solutions. All these configurations remain half helicity stellarators, following the generalised definition in Appendix~\ref{app:helicity}}
    \label{fig:high_curv_solutions}
\end{figure}

Unlike in the $A_c^\mathrm{mhd}$ analysis, here curvature is not an impediment, allowing highly exotic configurations to come to the forefront, with especially large inclinations (and even knotted forms), such as those illustrated in Figure~\ref{fig:high_curv_solutions}

\subsubsection{Statistical description of $f_\mathcal{J}$}
Armed with the perspective above, we now proceed to the statistical analysis of $f_\mathcal{J}$ across the database (see Table~\ref{tab:summary_db}). 
The database does not present a clear dependence of $f_\mathcal{J}$ on $N$, similar best values $f_\mathcal{J}\gtrsim 80$\% being achievable for any $N$. This seeming immutability is in part due to $f_\mathcal{J}$ not depending on the magnitude of the precession of particles, but rather their directions. That is, even if precession may scale with $N$, the fraction does not necessarily do so. Having said that, at larger $N$ we expect the role of $B_{20}$ to become more prominent in the fight with $T$ in Eq.~(\ref{eqn:nae_w_alpha}), given its natural $\sim N^2$ scaling. We see this to be indeed the case in the two-distribution analysis of Figure~\ref{fig:two_approaches_frac_mj}b.

\begin{table}
    \centering
\begin{tabular}{c c|c c|c c|c c|c c|c c}
\multicolumn{2}{c|}{$N=1$} &
\multicolumn{2}{c|}{$N=2$} &
\multicolumn{2}{c|}{$N=3$} &
\multicolumn{2}{c|}{$N=4$} &
\multicolumn{2}{c|}{$N=5$} &
\multicolumn{2}{c}{$N=6$} \\
\hline
$w_\mathrm{\hat{B}}$ & 0.708 & $\Delta$ & 0.429 & $\ell_{B_0}$ & 0.259 & $\tau_{0}$ & 0.257 & $\tau_{0}$ & 0.252 & $\tau_{0}$ & 0.266  \\
$\hat{\tau}$ & 0.821 & $\check{\tau}$ & 0.534 & $\hat{\kappa}''$ & 0.538 & $\ell_{B_0}$ & 0.506 & $w_\mathrm{\check{B}}$ & 0.448 & $\ell_{B_0}$ & 0.443  \\
$\check{\rho}$ & 0.893 & $w_\mathrm{\hat{B}}$ & 0.695 & $\check{\tau}$ & 0.678 & $\ell_{\kappa_\mathrm{max}}$ & 0.617 & $\hat{\rho}''$ & 0.631 & $\hat{\rho}''$ & 0.621  \\
$B_0'$ & 0.932 & $\ell_{\kappa_\mathrm{max}}$ & 0.871 & $\hat{\rho}''$ & 0.738 & $\hat{\rho}''$ & 0.720 & $\check{\kappa}^{(3)}$ & 0.726 & $\ell_{\kappa_\mathrm{max}}$ & 0.714  \\
\end{tabular}
\caption{\textbf{Most important combined input features for $f_\mathcal{J}$.} The table summarises the top selected features by FSFS on the combined input features detailed in Appendix~\ref{app:statistics_phaseI}. The numerical value represents the coefficient of determination $R^2$ of the $f_\mathcal{J}$ model using the variable on that row alongside those above. The meaning of the various parameters in the table are presented in Appendix~\ref{app:statistics_phaseI}: $\ell_{B_0}$ is the position of the $B_\mathrm{min}-B_\mathrm{max}$ transition, $\Delta$ the mirror ratio, $w_{\check{B}},w_{\hat{B}}$ the toroidal extent of $B_\mathrm{min}$ and $B_\mathrm{max}$, $B_0'$ the sharpness of the $B_\mathrm{min}-B_\mathrm{max}$ transition, $\ell_{\kappa_\mathrm{max}}$ the position of the curvature maximum, $\hat{\kappa}''$ the second derivative of the curvature at $B_\mathrm{max}$, $\check{\kappa}^{(3)}$ the third derivative of the curvature at $B_\mathrm{min}$, $\hat{\tau},\check{\tau}$ the torsion at $B_\mathrm{max}$ and $B_\mathrm{min}$, $\tau_0$ the integrated torsion, $\check{\rho}$ the value of $\rho$ at $B_\mathrm{min}$, and $\hat{\rho}''$ the variation of elongation at $B_\mathrm{max}$. }
    \label{tab:fracmj_feature_importance}
\end{table}

The correlation analysis (see Figure~\ref{fig:fmj_stats}) reflects the complexity of $f_\mathcal{J}$, indicating a distributed importance amongst elements of all curvature, torsion, magnetic field, and elongation. It is particularly remarkable how elements of the magnetic field shape, $\lambda_B$, and curvature, $\kappa_2$, have gained relevance here (see also Table~\ref{tab:fracmj_feature_importance}). The former controls the location (in the toroidal domain) at which the magnetic field transitions from $B_\mathrm{min}$ to $B_\mathrm{max}$, while $\kappa_2$ determines the position of peak curvature. Both these features have direct control on the trapped particle population and their drifts.

Besides curvature and $B_0$, FSFS does also indicate the necessary involvement of torsion (particularly strong at higher $N$) and elongation. Both of these contributions should not come as a surprise given the involvement of $B_{20}$; see analysis of $A_c^\mathrm{mhd}$.  Exploiting this, $f_\mathcal{J}$ can be improved (especially for large $N$) by minimising $\tau_\mathrm{rms}$, and pushing variation of elongation near $B_\mathrm{max}$ (and not $B_\mathrm{min}$). Although all these features are clearly key and necessary, they hardly amount to a 50-70\% description of the total variance in the data, pointing at the complexity of the measure and the existing competing approaches to $f_\mathcal{J}$.

\subsubsection{Improving maximum-$\mathcal{J}$ through additional shaping}
\begin{figure}
    \centering
    \includegraphics[width=\linewidth]{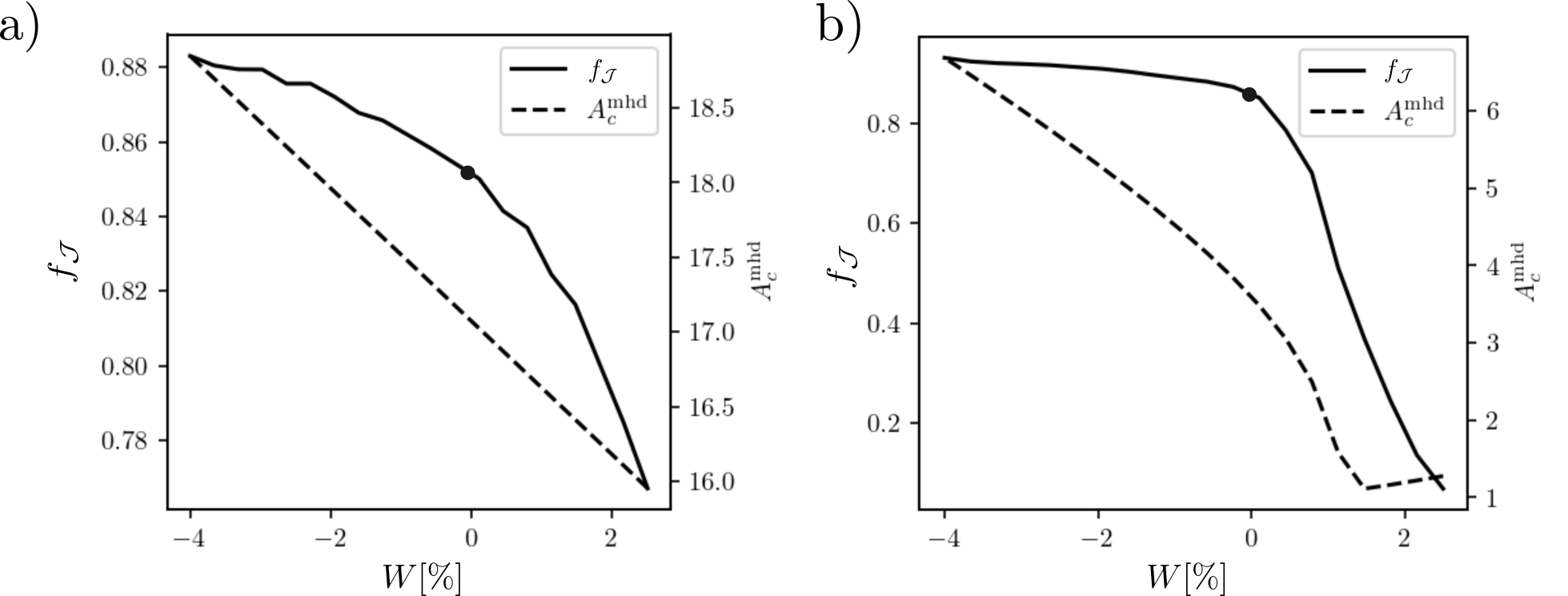}
    \caption{\textbf{Evolution of $f_\mathcal{J}$ with different amounts of MHD stabilising shaping.} The plots show the value of both the fraction of maximum $\mathcal{J}$ for different amounts of shaping tuned to achieve a magnetic well (negative values) desired, $W$. Plot a) corresponds to the highly shaped configuration in Figure~\ref{fig:two_approaches_frac_mj}c, while b) is configuration (ii). The plots also show the explicit evolution of the shaping through $A_c^\mathrm{mhd}$.}
    \label{fig:fmj_evol_mw}
\end{figure}

The database behaviour above considered near-axis constructions achieving marginal stability minimally \citep{rodriguez2025near}. We noted that such a choice was suboptimal for $f_\mathcal{J}$, and given the synergy between the magnetic well and precession, we expect shaping for larger $W$ values would improve $f_\mathcal{J}$. As a matter of fact, a practical design of a stellarator would {  require} a finite value of the magnetic well, and thus such constructions are highly relevant. We present two examples of the behaviour of $f_\mathcal{J}$ for different target-$W$ values in Figure~\ref{fig:fmj_evol_mw}.

The figures show a clear improvement in $f_\mathcal{J}$ with increasing magnetic well (more negative $W$). The slope is however highly variable, opening the door to future considerations regarding robustness of configurations. As $f_\mathcal{J}$ is improved, the amount of shaping needed to achieve this does as well. Other physically relevant features, such as the omnigenous behaviour of the field, will also change.

Before closing this section, we note that these considerations of precession have been taken in the context of a vacuum. It is well known that finite plasma $\beta$ increases precession of trapped particles in the direction of maximum $\mathcal{J}$. Thus, $f_\mathcal{J}$ will only increase.

\subsection{Neoclassical measure, $\epsilon_\mathrm{eff}$}
The database is made up of approximately quasi-isodynamic near-axis fields, property that imposes severe constraints on $|\mathbf{B}|$ \citep{bernardin1986, Cary1997,Nührenberg_2010,parra2015less}. Some of these constraints are exactly satisfied {\em by construction} using near-axis framework \citep{plunk2019direct,rodriguez2023higher}, for instance the existence of flattening points of the magnetic axis, but omnigeneity can only be achieved approximately \citep{plunk2019direct,rodriguez2024near}. To gauge how omnigenous a configuration is or is not, we need a way of measuring the deviation from the ideal behaviour.

We quantify the significance of departures from omnigeneity in terms of the neoclassical measure $\epsilon_\mathrm{eff}$, the effective ripple \citep{nemov1999}. This geometric scalar can be regarded as an average measure of the net radial drift of trapped particles, proportional (to the power of $3/2$) to the neoclassical heat transport in the $1/\nu$ regime. A perfectly omnigenous configuration will have $\epsilon_\mathrm{eff}=0$, but perfect omnigenity is not needed in reactor designs and larger values of $\epsilon_\mathrm{eff}$ are deemed acceptable especially if fast particles are confined.  From experience, such configurations tend to have $\epsilon_\mathrm{eff}\lesssim 1\%$ \citep[see for instance][]{aeries-cs}. Effective ripple is a useful measure quantifying deviations from omnigeneity but is not all-encompassing. Other alternative measures of this deviation can be more fitting for certain purposes, including efforts focused on the second adiabatic invariant $\mathcal{J}$ \cite{goodman2024quasi} or ideal fields \cite{goodman2023constructing,dudt2024magnetic}.

The near-axis form of $\epsilon_\mathrm{eff}$ was presented in detail in \cite[Sec.~4]{rodriguez2025near}, where the effective ripple at a reference radial position $r_\mathrm{ref}$ was estimated to be $\epsilon_\mathrm{eff}^\mathrm{edge}$,
\begin{equation}
    \epsilon_\mathrm{eff}^\mathrm{edge}=\left(\epsilon_\mathrm{eff}^{3/2,(0)}+r_\mathrm{ref}^2\epsilon_\mathrm{eff}^{3/2,(2)}\right)^{2/3}.
\end{equation}
The first term, $\epsilon_\mathrm{eff}^{3/2,(0)}$, Eq.~(4.8) in \cite{rodriguez2025near}, quantifies the breaking of omnigeneity at first order (by ``buffers''), but may be in practice ignored \citep{rodriguez2025near}. The second term, $\epsilon_\mathrm{eff}^{3/2,(2)}$, is the key contribution to $\epsilon_\mathrm{eff}^\mathrm{edge}$, measuring the degree of omnigeneity at second order. We remind the reader that through its asymptotic construction this measure has ignored contributions to the ripple from sources such as misalignment of maxima and minima of $|\mathbf{B}|$ or the appearance of additional ripple wells \citep{rodriguez2025near}. 

\subsubsection{Theoretical perspective on small $\epsilon_\mathrm{eff}^\mathrm{edge}$} \label{sec:theory_eps_eff}

By definition, the effective ripple is a weighted average of the net radial drift squared of trapped particles over the whole population. Configurations that minimise this drift will also tend to minimize $\epsilon_\mathrm{eff}$, although contributions also depend in detail on the number of different trapped particles, which depends on how the magnetic field $B_0$ varies with $\varphi$, its mirror ratio $\Delta$ {\it etc}. 

Looking at the formal contributors to $\epsilon_\mathrm{eff}^{3/2,(2)}$, \citep[Eq.~(4.11)]{rodriguez2025near}
\begin{equation}
    \epsilon_\mathrm{eff}^{3/2,(2)}=\frac{\pi}{8\sqrt{2}}\frac{(\bar{R}\bar{B})^2}{(\mathcal{G}^{(1)})^2}\int_{1/B_0^{\mathrm{max}}}^{1/B_0^{\mathrm{min}}}\lambda\frac{(h^{(2)})^2}{I^{(0)}}\mathrm{d}\lambda, \label{eqn:eps_eff_nae_2}
\end{equation}
where
\begin{subequations}
\begin{align}
    h^{(2)} &= -2\int_{\varphi_-}^{\varphi_+}\mathcal{H}(\lambda,B_0)\frac{\Delta B_{2c}^\mathrm{QI}}{\bar{B}}\,\mathrm{d}\varphi, \label{eqn:h2} \\
    \Delta B_{2c}^\mathrm{QI} &= B_{2c}^\mathrm{QI}-\frac{1}{4}\partial_\varphi\left(\frac{B_0^2d^2}{B_0'}\cos2\alpha_\mathrm{buf}\right), \label{eqn:qi_2nd_order}
\end{align}
\end{subequations}
and $B_{2c}^\mathrm{QI}=-(B_{2c}\cos2\bar{\iota}_0\varphi+B_{2s}\sin2\bar{\iota}_0\varphi)$, where $\bar{\iota}_0 = \iota_0 - N/2$. 

The function $B_{2c}(\varphi)$ is central to $\epsilon_\mathrm{eff}$, its pointwise value having to be of a very particular form, $\Delta B_{2c}^\mathrm{QI}=0$, so as to ensure $\epsilon_\mathrm{eff}^{3/2,(2)}=0$. Unlike many of the previously discussed measures, this is a local equality condition for all $\varphi$. We cannot therefore easily come up with a line of attack at achieving low effective ripple by considering extreme limits. Instead we must take into account every contribution to the $T$-terms and $B_{2c}^\mathrm{QI}$. Elements of curvature clearly arise in $T$, with shaping playing a direct role in $B_{2c}$. Generally, the choice of stabilising shaping does not minimise $\epsilon_\mathrm{eff}$ \citep{rodriguez2025near}, whose intrinsic behaviour we may also study as we did with $B_{20}$. The value of $B_{2c}$ at $B_\mathrm{min}$, $B_{2c,\mathrm{b}}$ is partially correlated (with $R^2\sim0.7$) to the ripple, and can be written, assuming perfect QI at first order (see 3.10 of \cite{rodriguez2024near}), as
\begin{equation}
    \left. \frac{B_{2c}}{B_0}\right|_{\varphi = \pi/N} =-\frac{1}{4(\ell')^2}\frac{\bar{B}}{B_0}\left[\bar{e}''\left(1+\frac{1}{\bar{e}^2}\right)+\bar{e}(\tau\ell')^2\left(3+\frac{1}{\bar{e}^2}\right)\right]. \label{eqn:B2c_b}
\end{equation}
This shows that elongation and torsion are both important; after all, they control the bending of field lines over flux surfaces, and thus the variation of $|\mathbf{B}|$ over them. From Eq.~(\ref{eqn:B2c_b}) and the QI condition, it is necessary that $\bar{e}''\leq0$ and $|\tau|\geq0$ for $B_{2c}$ to vanish, so if elongation grows away from the bottom of the well to aid with the magnetic well, then a finite amount of torsion is necessary. There is a clear tension between this and the behaviour of the magnetic well, as pointed in \cite{rodriguez2024maximum} and \cite{plunk2024back}.

\subsubsection{Statistics of $\epsilon_\mathrm{eff}^\mathrm{edge}$}
With the theoretical grounds laid, we consider the statistics of $\epsilon_\mathrm{eff}^\mathrm{edge}$. Aligning with the trends of previous features, the effective ripple of the bulk of the configurations becomes poorer the larger the number of field periods. In fact, the value of the population median is not far from the natural $\sim N^{8/3}$ scaling, making the bulk of the configurations have a forbiddingly large value of $\epsilon_\mathrm{eff}^\mathrm{edge}$.\footnote{By natural scaling we mean the result of the second order field being $B_2\sim N^2$, and thus $\epsilon_\mathrm{eff}\sim(N^2)^{2\times 2/3}\sim N^{8/3}$. From the median of the data it is hard to distinguish a precise scaling though, and nearby powers could be equally reasonable.} 
\begin{figure}
    \centering
    \includegraphics[width=\textwidth]{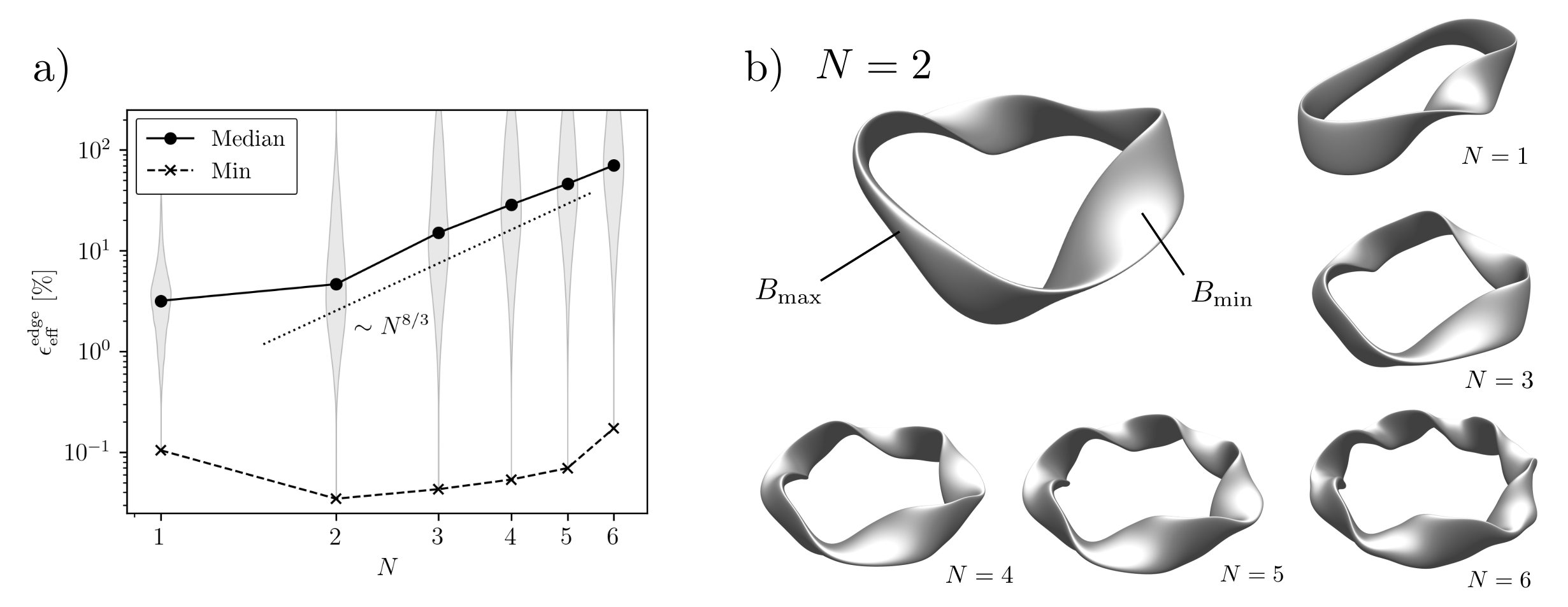}
    \caption{\textbf{$\epsilon_\mathrm{eff}^\mathrm{edge}$ behaviour across different number of field periods.} a) Dependence of the minimum and median value of $\epsilon_\mathrm{eff}^\mathrm{edge}$ as a function of the number of field periods $N$ in the database. Reference scaling is given as dotted line and distributions represented as vertical violin plots. b) Rendition of the 3D finite aspect ratio flux surface of $\epsilon_\mathrm{eff}^\mathrm{edge}$ minimising fields for each number of field periods in the database. The locations of $B_\mathrm{min}$ and $B_\mathrm{max}$ are indicated for the $N=2$ configuration, which has the lowest value of $\epsilon_\mathrm{eff}^\mathrm{edge}$.}
    \label{fig:eps_eff_field_period}
\end{figure}

Unlike the bulk of the population, though, the best configurations (lowest values of $\epsilon_\mathrm{eff}^\mathrm{edge}$) do not exhibit such a strong deterioration with $N$. One is able to find configurations with values of ripple well below a percent for all $N$, with the best candidate corresponding to $N=2$. Depictions of the best performing configurations are shown in Figure~\ref{fig:eps_eff_field_period}b. This stark difference between the bulk of the population and the extremes can be attributed to the necessity to carefully balance the different geometric elements of the field to achieve an omnigenous form of $|\mathbf{B}|$, as we have discussed. If such balance is not struck, then the mismatch will grow by an amount directly related to the magnitude of any of the ingredients involved, in this case the $N^2$ of the second order field. Because of the need to strike such a balance, and the restricted forms used for the different inputs of the problem ($\kappa$ $\tau$, $\rho$, {\it etc}), it is likely that configurations exist beyond the database which achieve lower $\epsilon_\mathrm{eff}$. Optimisation within a larger input-feature near-axis space is left for future work.

\begin{figure}
    \begin{minipage}[c]{0.4\textwidth}
        \centering
\begin{tabular}{c c|c c|c c}
\multicolumn{2}{c|}{$N=1$} &
\multicolumn{2}{c|}{$N=2$} &
\multicolumn{2}{c}{$N=3$} 
\\
\hline
$\check{\tau}$ & 0.443 & $\check{\tau}$ & 0.436 & $\check{\tau}$ & 0.522  \\
$\Delta$ & 0.620 & $\Delta$ & 0.583 & $\Delta$ & 0.621  \\
$\check{\rho}$ & 0.692 & $\check{\rho}$ & 0.650 & $\int\kappa\,d\ell$ & 0.709 \\
$w_\mathrm{\check{B}}$ & 0.778 & $\int\kappa\,d\ell$ & 0.717 & $\check{\rho}$ & 0.793 \\\hline\hline
\multicolumn{2}{c|}{$N=4$} &
\multicolumn{2}{c|}{$N=5$} &
\multicolumn{2}{c}{$N=6$} \\
\hline
$\int\kappa\,d\ell$ & 0.478 & $\int\kappa\,d\ell$ & 0.497 & $\int\kappa\,d\ell$ & 0.489  \\
$\check{\tau}$ & 0.634 & $\check{\tau}$ & 0.654 & $\check{\tau}$ & 0.618  \\
$\check{\rho}$ & 0.752 & $\Delta$ & 0.737 & $\Delta$ & 0.709  \\
$\Delta$ & 0.836 & $\check{\rho}$ & 0.816 & $\check{\rho}$ & 0.797  \\

\end{tabular}
    
    \end{minipage}%
    \hspace{0.03\textwidth} 
    \begin{minipage}[c]{0.6\textwidth}
            \centering \vspace*{0.7cm}
    \includegraphics[width=0.9\linewidth]{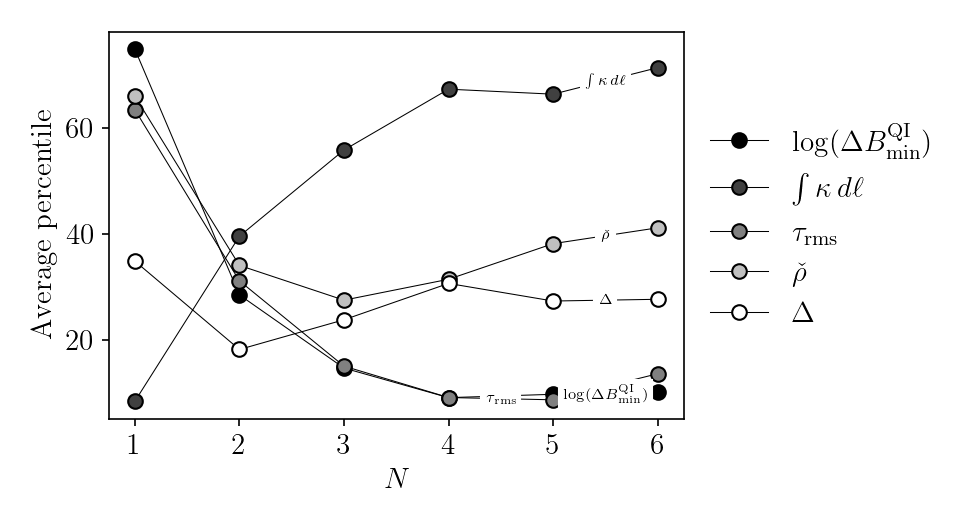}

     \end{minipage}
    \caption{\textbf{FSFS summary for $\epsilon_\mathrm{eff}^\mathrm{edge}$ and average percentile of lowest $\epsilon_\mathrm{eff}^\mathrm{edge}$ subspace.}
    The left table summarises the top selected features by FSFS on the combined input features detailed in Appendix~\ref{app:statistics_phaseI}. The numerical value represents the coefficient of determination $R^2$ of the $\epsilon_\mathrm{eff}^\mathrm{edge}$ model using the variable on that row alongside those above. We considered $\epsilon_\mathrm{eff}^\mathrm{edge}$ in log-scale. The figure on the right presents the average percentile (of the whole population across different field periods) to which the subset of lowest 1000th percentile of $\epsilon_\mathrm{eff}^\mathrm{edge}$ belong. The meaning of the various parameters in the table are presented in Appendix~\ref{app:statistics_phaseI}: $\hat{\tau},\check{\tau}$ is the torsion at $B_\mathrm{max}$ and $B_\mathrm{min}$, $\check{\rho}$ the value of $\rho$ at $B_\mathrm{min}$, and $\rho_{\kappa_\mathrm{max}}$ the value of $\rho$ at the point of maximum curvature, $\int\kappa\mathrm{d}\ell$ the integrated curvature over half a period.}
    \label{fig:eps_eff_distribution}
\end{figure}

The correlation analysis provides evidence of the complexity of the measure and the underlying careful balance. Although torsion of the axis appears to be most important, and in particular $\tau_{c1}$ (see Figure~\ref{fig:epseff_stats}), the relatively low value of the non-linear correlation shows that there is no clear simple dominating trend. The importance analysis shows that other aspects such as elongation and magnetic shaping do also play an important role. We proceed with FSFS to further investigate this dependence, summarising the results in Figure~\ref{fig:eps_eff_distribution}.
The dependence on torsion is the most apparent, with a particular preference for its value at the bottom of the magnetic well, where its role was explicitly described by Eq.~(\ref{eqn:B2c_b}). Unsurprisingly, elongation also appears, although in no clear direction. The APM analysis of Figure~\ref{fig:eps_eff_distribution}) reflects the requirement of a finite amount of torsion argued theoretically (see low $N$), but an emphasis to avoid large values at higher $N$ to avoid overwhelming shaping.

As part of the FSFS analysis, we also obtain two other expected ingredients, as are the integrated curvature and the mirror ratio. We already touched on the physical grounds of both these elements, which we can connect to the radial drifts and trapped population size respectively. Thus, it is reassuring to see them come out of the analysis. Despite this, we cannot say that configurations that achieve low ripple minimise their integrated curvature; see Figure~\ref{fig:eps_eff_distribution}. Although the integrated curvature is kept small (especially at lower $N$), there is a clear shift toward prioritizing low torsion as $N$ increases, reflecting the growing necessity to control shaping. Finally, the discreteness in the mirror ratio precludes any detailed conclusion regarding its role. 

\subsubsection{"Good" configurations}
The analysis above remains somewhat unsatisfactory, as no simple recipe to achieving low effective ripple has really been found. This is in part because of the many elements that go into the complex balance in $\epsilon_\mathrm{eff}^\mathrm{edge}$. 

We could proceed like previously and study the behaviour of the lowest $\epsilon_\mathrm{eff}$ subset. However, we are in a position to pose a more interesting problem now. Instead of considering $\epsilon_\mathrm{eff}$ in isolation, we consider it alongside $L_{\nabla \mathbf{B}}$ and $A_c^\mathrm{mhd}$, defining \textit{good} configurations to be those that simultaneously satisfy
\vskip 10pt
\begin{enumerate}
    \item  $L_{\nabla\mathbf{B}}>0.25$
    \item  $A_c^\mathrm{mhd}<10$
    \item  $\epsilon_\mathrm{eff}^\mathrm{edge}<0.01$.
\end{enumerate}
\vskip 10pt
A configuration meeting these criteria achieves a marginal magnetic well with minimal shaping and sufficiently small effective ripple (see previous sections for the informed choice of such thresholds). By studying the subclass of such configurations, we can see how constraining each of these requirements is in practice. An exhaustive study of physics trade-offs across the database is not presented in this piece of work. 

\begin{figure}
    \centering
    \includegraphics[width=\linewidth]{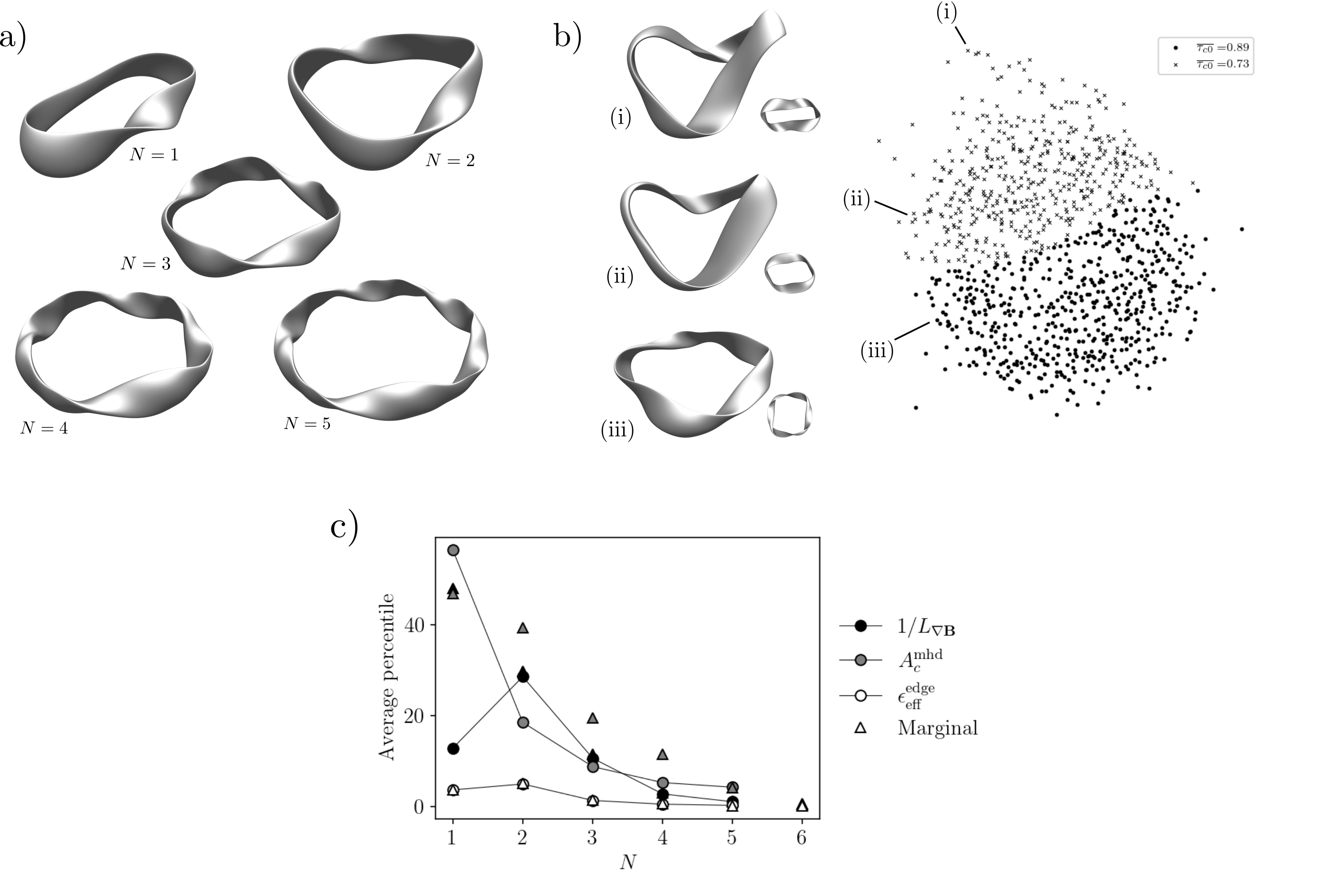}
    \caption{\textbf{`Good' configurations in the database.} (a) 3D rendition of the largest $L_{\nabla\mathbf{B}}$ configurations in the `good' configuration subset per field period. (b) Example of variety of good $N=2$ configurations in a lower dimensional MDS representation showing larger and lower torsion clusters. 3 examples are illustrated numbered (i)-(iii), bearing features of the configurations excelling in the various physics features discussed before. (c) Average percentile measure (APM) across different $N$ for `good' configurations. Note that there is no `good' configurations for $N=6$. We also indicate as reference the marginal APM values, meaning the APM of the subset satisfying each of the thresholds separately, to illustrate the influence of each threshold on the others. }
    \label{fig:eps_eff_good}
\end{figure}


`Good' configurations with the largest $L_{\nabla\mathbf{B}}$ for each $N$ are shown Figure~\ref{fig:eps_eff_good}. These are prime candidates as initial conditions of further stellarator optimisation pursues. But that is not our main focus here; we are interested in learning how restrictive each of the requirements is, {\it i.e.} how strongly they shape the overall population. We thus apply our average-percentile-measure (APM) to the good-configuration subsets (see Figure~\ref{fig:eps_eff_good}c). The shaping and ripple measures behave quite differently through $N$. The variation of the $\epsilon_\mathrm{eff}^\mathrm{edge}$ APM with $N$ appears weakest, but it proves to be the single most limiting feature: the measure stays below $\sim5\%$ for all $N$. This is particularly noticeable at the lowest $N$, where the APM of both $L_{\nabla\mathbf{B}}$ and $A_c^\mathrm{mhd}$ are significantly larger (even though the $\epsilon_\mathrm{eff}$ constraint significantly hampers $A_c^\mathrm{mhd}$ as shown in Figure~\ref{fig:eps_eff_good}). This is consistent with the loss of standing of the figure-8-like configurations over the less inclined configurations at low $N$ (even though there still remains an ample variety of configurations, as shown in Figure~\ref{fig:eps_eff_good}b). The implications of shaping and coil compatibility requirements ($L_{\nabla\mathbf{B}}$ and $A_c^\mathrm{mhd}$) significantly worsen at larger $N$, becoming similarly constraining as $\epsilon_\mathrm{eff}^\mathrm{edge}$.  Ultimately, no $N=6$ configuration exists within the database that satisfies all requirements to be considered ``good''. It is worth noting that additional low-ripple configurations might be found by extending the input-model flexibility, which could alleviate some of the burden imposed by the requirement on $\epsilon_\mathrm{eff}$.

\subsection{Other measures of interest}

We now touch on several properties that differ from those so far discussed in that they favour larger number of field periods, as is necessary to understand why intermediate values of $N \sim 3-5$ are favoured in QI stellarator reactor design proposals \citep{Beidler_2001, LION_2025, Hegna_2025}. The two properties we consider are: (i) sensitivity of equilibrium to finite $\beta$, and (ii) finite orbit width related effects.

The resilience of an equilibrium field to plasma $\beta$ is an important consideration for designing a reliable confinement field. Failing to make fields robust to changes in plasma pressure could undermine the key external control necessary to support critical features such as the divertor. The basic change in the field due to pressure comes in the form of a relative shift of flux surfaces, the Shafranov shift \citep{shafranov1963,wessonTok} (as well as an associated deformation). Although such shift may appear benign, it may when sufficiently large lead to flux surfaces touching, {\it i.e.} the breakdown of flux surface geometry \citep{loizu2017equilibrium}. 

Measures of this displacement were introduced and defined within the near-axis framework in \citep{rodriguez2025near} following previous work \citep{landreman2021a,rodriguez2023mhd}. The fractional displacement of the surface (keeping the axis unchanged) with a change in plasma beta was encapsulated in the measure $\hat{\mathcal{S}}_\mathrm{max}$ \citep[Def.~7a]{rodriguez2025near}.\footnote{ A critical $\beta$ representing the value of plasma $\beta$ leading to the touching of flux surfaces, $\beta_\Delta$, could also be defined, but we prefer $\hat{\mathcal{S}}$ for the analysis due to its simplicity.} The value of $\hat{S}_\mathrm{max}$ captures the fight of the plasma pressure against the magnetic field, in the tokamak limit against the poloidal magnetic field. This leads in that scenario to the sensitivity scaling inversely with the rotational transform $\sim1/\iota^2$ \citep[Eq.~(7.6)]{rodriguez2025near}. 

\begin{figure}
    \centering
    \includegraphics[width=\linewidth]{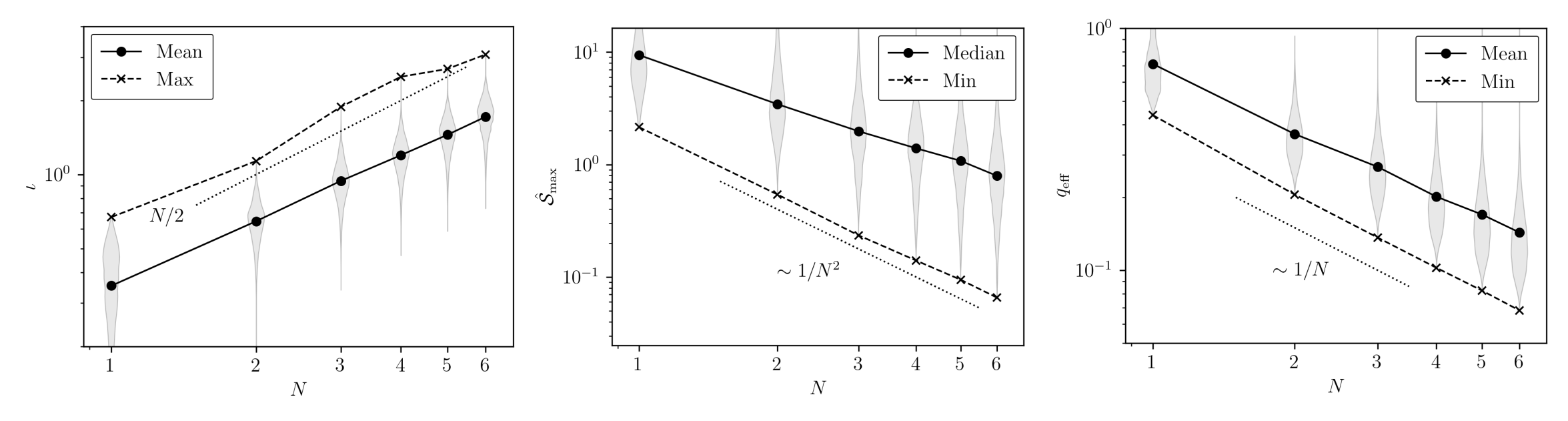}
    \caption{\textbf{Behaviour of high-$N$ prone properties across number of field periods.} The plots show the behaviour of the mean and maximum (or minimum) values of the database as a function of the number of field periods $N$ of (a) rotational transform, $\iota$, (b) relative shift of surfaces with plasma beta, $\hat{\mathcal{S}}_\mathrm{max}$, and (c) effective $q$ for finite orbit width physics, $q_\mathrm{eff}$. The dotted line show reference scalings, in the case of (a) the function $\iota=N/2$. }
    \label{fig:N_scale_beta_qeff}
\end{figure}

Following this simple tokamak argument we may explain the quadratic $N$-dependence of $\hat{\mathcal{S}}_\mathrm{max}$ through the database (see Figure~\ref{fig:N_scale_beta_qeff}), as the rotational transform grows linearly with $N$. The latter is the result of the $\iota$ generated per field period being to a large extent limited by helicity. This means that $\iota=N/2$ serves as an approximate bound for rotational transform, as indicated in Figure~\ref{fig:N_scale_beta_qeff}a.

The dependence on $\iota$ does however not fully explain $\hat{\mathcal{S}}_\mathrm{max}$ (the correlation between $\iota$ and $\hat{\mathcal{S}}_\mathrm{max}$ is $R^2\approx 0.9-0.6$, decreasing with increasing $N$). In a stellarator the larger the shaping, the harder one expects it to be for the plasma pressure to push through, and thus the smaller the sensitivity. This effect should increase with $N$, overshadowing somewhat the effects of $\iota$. A sign of this occurring is the large integrated torsion of configurations with smallest $\hat{\mathcal{S}}_\mathrm{max}$ (in the sense that the one thousandth percentile of $\hat{\mathcal{S}}_\mathrm{max}$ belongs to the average integrated torsion $\sim80\%$ percentile of the total population). 

Let us now turn to the other property we mentioned in the opening of the section; namely, finite orbit width related physical phenomena. There are many pieces of physics behaviour which emanate from the orbit width of charged particles, prominent examples being the bootstrap current and the linear response of the plasma to an externally applied electric field (\textit{i.e.}, the so called Rosenbluth-Hinton residual \citep{rosenbluth1998poloidal,xiao2006short,monreal2016residual}). In both cases, larger orbits lead to `worse' behaviour: potentially larger bootstrap currents and also smaller residual (often linked to weaker zonal flow support).

The orbit width magnitude results from a balance of two key ingredients. One, the magnitude of the radial drift that leads to the departure of charged particles off flux surfaces. And second, the distance over which the particles travel. If we consider trapped particles, the distance travelled is directly linked to the connection length, meaning the distance along the field line from the minimum of $|\mathbf{B}|$ to the maximum. Quasi-isodynamic configurations are essentially different from axisymmetric or quasisymmetric fields in this regard, as they are capable of shortening the connection length by restricting this to a field period. In the other scenarios, one must follow the field line a number of toroidal turns to the point that, through the finite rotational transform of the field, the field line wraps poloidally, bringing the rotational transform in a prime form into the problem. In the QI scenario the rotational transform plays a subsidiary role. 

With this in mind, orbit widths will decrease with an increasing number of field periods. Returning to the two physics elements described above, this suggests that larger number of field periods will naturally decrease the magnitude of the bootstrap current, as well as increase the zonal flow residual. In an exact QI field the bootstrap current will exactly vanish, due to there being an equal number of particles drifting outwards and inwards. Deviations from exact omnigeneity will scale with the magnitude of the orbits, which being small, will generally lead to controlled finite values of bootstrap current. 

We focus then on the residual measure, motivated by the idea that a large value indicates a propensity to supporting larger zonal flows. We follow the recent work in \cite{rodriguez2024zonal} (also explored for quasisymmetric fields in \cite{zhu2025collisionless}), and write the residual flow (the fraction of the flow remaining at long times),
\begin{equation}
    \frac{\phi(t\rightarrow\infty)}{\phi(0)}=\frac{1}{1+1.6q_\mathrm{eff}^2/\sqrt{\Delta}}.
\end{equation}
This form of the expression is equivalent to that in \cite{rosenbluth1998poloidal} for the tokamak limit, for which $q_\mathrm{eff}=1/\iota$, with $\Delta$ being the mirror ratio. In the case of a QI stellarator, an effective safety factor $q_\mathrm{eff}$ needs to be defined, and can be done approximately using the flux expansion geometry and $\Delta\ll1$ as done in \cite[Eq.~(C26)]{rodriguez2024zonal}. Details may be found in that publication. In short, $q_\mathrm{eff}\sim \text{Connection length} \times \text{Flux expansion}$. As expected, this measure does also scale favourably with $N$ (see Figure~\ref{fig:N_scale_beta_qeff}c). We may take the linear scaling $\sim1/N$ as a result of the shortened connection length.\footnote{The flux expansion, $1/|\nabla\psi|$, remains largely unchanged with $N$, as do the curvature, elongation and $\sigma$.} The scaling is the same that one would expect in a tokamak, where $q_\mathrm{eff}\sim1/\iota\sim1/N$.

\section{Conclusions}

In this paper we present a first-of-its-kind database of approximately QI, stable stellarator configurations, employing the near-axis expansion to enable a systematic and controlled way to explore and diagnose this space. Construction of the database is enabled by recent developments in the underlying theoretical foundations, in particular the systematic solution of fields to a sufficiently high order in the asymptotic description, such that key equilibrium properties are captured to allow an assessment as candidates for stellarator optimization, {\it e.g.} MHD stability and neoclassical transport, amongst others. Every stellarator configuration found is diagnosed over a extensive range of physics within the same theoretical framework, allowing a fast and detailed analysis over the entire database.  This database, which includes over 800,000 configurations, represents a significant step in mapping out the space of QI stellarators, but leaves ample room for further extensions to refine and broaden the parameter space.

By restricting to a limited set of mostly geometric input parameters, we tried to capture a relevant part of the space of QI stellarators.  Beyond providing a potentially valuable set of baseline configurations for further studies, the database also opens a window onto trends and behaviour across such configurations, {\em i.e.} suggesting statistically-informed heuristics regarding design parameters. As a practical guide to readers, we now briefly summarise these heuristics.  We consider first the role of the most basic design parameter, the field period number $N$:
\begin{enumerate}[leftmargin=2em,itemindent=-1em]
    \item \quad Coil compatibility ($L_{\nabla \mathbf{B}}$), shaping and stability ($A_c^\mathrm{mhd}$), and low neoclassical transport ($\epsilon_\mathrm{eff}$) are easier to achieve at low field period number. {  When considered alongside each other, } the latter appears to be the most constraining at low $N$, while at large $N$ shaping becomes equally (if not more) constraining, requiring larger aspect ratios, { see Figure~\ref{fig:eps_eff_good}c}.
    \item \quad The Maximum-$J$ property can be approximately achieved in a vacuum at all $N$, although lower $N$ allow this without the need of excessive shaping.
    \item \quad Several properties, like resilience to plasma pressure (low Shafranov sensitivity $\hat{\mathcal{S}}$), rotational transform, and measure of orbit widths that controls zonal flows and plasma currents ($q_{\mathrm{eff}}$) all favour high field period number.
\end{enumerate}

Taken together, these observations align with experience from stellarator optimization, namely that there is some optimal field period number to achieve the desired balance of properties. The observed flexibility, however, leaves the possibility that attractive designs can still be found at unconventional field period numbers, a prime example being the figure-8 at $N=2$.

Because of the manner in which the database has been constructed \citep[see ][]{plunk2025-geometric}, the geometry of the magnetic axis is placed front and centre in the analysis. A deeper understanding of the relation between torsion and curvature in the special class of QI-closed curves requires further work, as this trade-off proves key in understanding behaviour throughout the relevant space. Nevertheless, we can already report on some interesting basic trends in the statistics:

\begin{enumerate}[leftmargin=2em,itemindent=-1em]
    \item \quad  Axis torsion is in many respects `bad' in that large values can increase the magnitude of shaping with varied consequences for achieving good designs. This is particularly true of the net integrated torsion, which leads to \textit{crown}-like configurations that minimise it. 
    \item \quad Small torsion is also beneficial for stable configurations ($A_c^\mathrm{mhd}$); as a consequence, configurations with very low aspect ratios can be found at small field period numbers, with the most compact examples in the entire database corresponding to $N=2$ figure-8-like examples with $A \sim 2.3$; For $N > 2$, small integrated torsion can only be achieved while also having finite torsion locally.
    \item \quad The analysis also confirms that curvature generally plays an adverse role by affecting $|\mathbf{B}|$ directly. We find that its peak value plays a critical role in coil compatibility ($L_{\nabla \mathbf{B}}$), playing a limiting role as strong as the torsion (which appears unavoidable at the field minimum). 
    \item \quad A notable exception to this role of curvature is maximum-$\mathcal{J}$, for which large curvature can be beneficial. This requires careful placement of peak curvature relative to a widely shaped magnetic trapping well.
    \item \quad In the analysis of neoclassical transport ($\epsilon_\mathrm{eff}$), the necessity of torsion especially at the point of weakest field is broadly supported in the database.  The precise term balance required to achieve a low value of $\epsilon_\mathrm{eff}$ makes this property particularly sensitive and hard to achieve, although precise balance does not appear to depend significantly on $N$. 
\end{enumerate}

Away from the magnetic axis, the ellipticity of magnetic surfaces also affects the physics of the configurations, albeit to a lesser extent:

\begin{enumerate}[leftmargin=2em,itemindent=-1em]
    \item \quad Elongation seems not to play a very strong role so long as it is not allowed to get too large, as we are able to enforce in the  construction by providing it as an input \citep{plunk2025-geometric}. 
    \item \quad MHD stability depends on elongation, as the magnetic well responds to stretching of surfaces near the field extrema \citep{plunk2024back}; the maximum-$J$ property also shows sensitivity to stretching at the field maximum, for similar reasons.
\end{enumerate}

We close the paper with comments about future directions.  Having established the framework and basic tools required to explore the space of QI configurations we could construct more extensive databases, with greater detail. For instance, we can extend the relatively simple models used for the magnetic axis curve, or on-axis magnetic field strength.  Indeed, it is straightforward to include more input parameters, making the near-axis description of equilibrium fields arbitrarily general.  However, the size of space and resulting database may quickly become unmanageable, and a practical way forward would be to perform optimisation, {\it e.g.} using gradient descent within the near-axis space, as was done by \cite{landreman2022mapping}.  Exploring other helicities or curve classes and configurations qualitatively outside the space explored in this paper, not necessarily requiring a larger space of parameters, is also of interest, and will be the subject of future work.
\par
Looking beyond the framework of the near-axis model, the database can be leveraged to provide large sets of starting points for further design optimisation efforts, employing realistic global equilibria tailored specifically to meet practical requirements for those studies. {  Experience has shown this to be a valuable tool to explore an optimisation landscape that tends to show many local minima. It remains to be seen the extent to which the heuristics and lessons from this analysis remain in such optimisation efforts, but we know at least should hold true close to the axis.}  
\par
The database can also serve as test bed of more many interesting questions such as details on realistic coil construction or transport questions including turbulence.  To spur such work, the database is made open and publicly accessible.

\section*{Data availability}
The database and scripts that support this study are openly available at the Zenodo repository with DOI/URL 10.5281/zenodo.18220133.

\section*{Acknowledgments}  This work has been carried out within the framework of the EUROfusion Consortium, funded by the European Union via the Euratom Research and Training Programme (Grant Agreement No 101052200 — EUROfusion).  Views and opinions expressed are however those of the author(s) only and do not necessarily reflect those of the European Union or the European Commission.  Neither the European Union nor the European Commission can be held responsible for them.

\section*{Declaration of interests}  The authors declare no competing interests.

\appendix

\section{Input parameter search space} \label{app:input_space}

\begin{table}
    \centering
        \begin{tabular}{c|cc|cc|cc|cc|}
    & \multicolumn{2}{c|}{$\kappa_{2}$} 
    & \multicolumn{2}{c|}{$\tau_{c1}$} 
    & \multicolumn{2}{c|}{$\tau_{c2}$} 
    & \multicolumn{2}{c|}{$\tau_0$} \\
    
    \hline
    $N$ & {\tiny min} & {\tiny max} & {\tiny min} & {\tiny max} 
       & {\tiny min} & {\tiny max} & {\tiny min} & {\tiny max} \\
    \hline
    
    1 & \cellcolor{gray} & \cellcolor{gray} & \cellcolor{gray} & \cellcolor{gray}
      & \cellcolor{gray} & \cellcolor{gray} & -0.41 & 1.37 \\

    2 & -0.3 & 0.4 & -3.2 & -0.005 & -3.0 & 3.0 
      & \cellcolor{gray} & \cellcolor{gray} \\

    3 & -0.45 & 0.45 & -5.72 & 0.89 & -4.0 & 4.0 
      & \cellcolor{gray} & \cellcolor{gray} \\

    4 & -0.4 & 0.45 & -8.25 & 1.8 & -5.3 & 5.3 
      & \cellcolor{gray} & \cellcolor{gray} \\

    5 & -0.4 & 0.45 & -10.78 & 2.71 & -6.6 & 6.6 
      & \cellcolor{gray} & \cellcolor{gray} \\

    6 & -0.4 & 0.45 & -13.32 & 3.64 & -7.9 & 7.9 
      & \cellcolor{gray} & \cellcolor{gray} \\

    \hline
    \hline
    Points & \multicolumn{2}{c|}{10} & \multicolumn{2}{c|}{10}
           & \multicolumn{2}{c|}{10} & \multicolumn{2}{c|}{16} \\\hline
\end{tabular}
\\\vspace{12pt}
\begin{tabular}{cc|cc|cc|cc|cc}
      \multicolumn{2}{c|}{$\rho_{0}$}
      & \multicolumn{2}{c|}{$\rho_{1}$}
      & \multicolumn{2}{c|}{$\rho_{2}$}
      & \multicolumn{2}{c|}{$\Delta$}
      & \multicolumn{2}{c}{$\lambda_B$} \\
\hline
    4.0 & 5.0
      & -2.0 & 2.0
      & -1.0 & 1.0
      & 0.15 & 0.25
      & -0.056 & 0.124 \\
\hline\hline
    \multicolumn{2}{c|}{3}
       & \multicolumn{2}{c|}{5}
       & \multicolumn{2}{c|}{5}
       & \multicolumn{2}{c|}{2}
       & \multicolumn{2}{c}{5} \\
\end{tabular}

    \caption{\textbf{Parameter search intervals for database construction.}  The tables summarise the ranges of the input features and number of points used for its discretisation in the construction of the database. The top table lists the axis curve parameters, showing the difference between $N=1$ and the other field periods. The second table gives ranges of $\rho_{0}$, $\rho_{1}$, $\rho_{2}$, $\Delta$, and $\lambda_B$ common to all $N$. Note the smaller dimensionality of the parameter space for $N = 1$, which is due to the larger number of constraints required to achieve curve closure \citep{plunk2025-geometric}.}
    
    \label{tab:input_parameters}
\end{table}

Table \ref{tab:input_parameters} summarizes the space of input parameters scanned during the construction of the database.  For each $N$, a uniform grid in the shown intervals is used and solutions attempted for each point.  The Frenet-Serret and $\sigma$ equation solvers are initialized to neighbouring solutions already successfully calculated. This warm start helps with the highly non-linear curve closure problem. If the solver succeeds altogether in finding a solution for the axis curve ${\bf x}_0$ and first order field $\sigma$ (\textit{i.e.}, satisfy closure criteria, has the right helicity, and is periodic to desired precision) then the solution is added to the database.  The final parameter space, corresponding to these valid solutions, is non-rectangular but "connected".

\section{Statistical analysis of feature dependence} \label{app:statistics_phaseI}
In this appendix we present a detailed account of the statistical tools we employ throughout this paper to describe the dependence of different properties on input features across the database. There is no single statistical measure in isolation that can capture such dependence unequivocally, and any insightful analysis requires a number of them. 

\subsection{Set-up of the problem}
Let us first, for clarity, define the fundamental problem we are trying to understand formally. We are considering a database consisting of data uniquely defined by an \textit{input feature} vector $\mathbf{x}$. The input vector is a label of a point in $\mathbb{R}^8$, each entry representing one of $\{\tau_{c1},\tau_{c2},\kappa_2,\rho_0,\rho_1,\rho_2,\Delta,\lambda_B\}$, which we can succinctly refer to by an integer index $\{x_i\}$. The labels of a particular instance in the database $j$ we write as $\mathbf{X}^j$. Each instance is characterised further by another vector of properties, $\mathbf{y}$, consisting of the different physics properties of interest (\textit{e.g.}, $L_{\nabla\mathbf{B}}$ or $\epsilon_\mathrm{eff}^\mathrm{edge}$). It is our goal to understand how $\mathbf{x}$ affects $\mathbf{y}$, or put differently, what the role played by each $x_i$ is in the function $f_i$ such that $y_i=f_i(\mathbf{x})$. 

\subsection{Correlation measures}
The most basic dependence between $y_i$ and the input features ${x_i}$ can be probed considering each $x_i$ in isolation, \textit{i.e.}, univariate dependence. We are interested in gauging how much about $y_i$ can be learnt given knowledge of $x_i$ (and ignoring all remaining $\{x_j\}$ for $j\neq i$), and we shall resort to variance based statistics for that purpose. 

We introduce a key basic auxiliary measure for each pair of features $(x_i,y_j)$, the coefficient of determination \citep[pg.~117]{dodge2008concise},
\begin{equation}
    R^2_{(i,j)}=\frac{\sum_k\left(\hat{f}_j(X_i^k)-\Bar{Y}_j\right)^2}{\sum_k\left(Y_j^k-\Bar{Y}_j\right)^2}.
\end{equation}
The overbar represents the average over the data and the function $\hat{f}_j$ is a guessed model relating $y_j$ from $x_i$. $R^2_{(i,j)}$ is then a scalar that measures what fraction of the total variance of $y_i$ in the database is described by the model $\hat{f}_j$, \textit{i.e.}, how much it explains. 

As given, this determination coefficient is equally a measure of the fidelity of the model $\hat{f}_j$ as it may be be of the underlying dependence. To skew the balance towards the latter, one must choose a model that tries to maximise $R^2_{(i,j)}$, in which scenario, the maximum possible such value is a direct measure of the underlying unconditional dependence; we call that the \textit{correlation} value. In the case of there existing an ideal underlying function $f_i$ such that $y_i=f_i(x_i)$, the answer would then be $R^2_{(i,j)}=1$, \textit{i.e.}, we can know $y_i$ completely. Even if such an ideal relation existed, in practice, we can only approximate. Different choices for such an approximation leads to different forms of correlation. 

\begin{figure}
    \centering
    \includegraphics[width=0.8\linewidth]{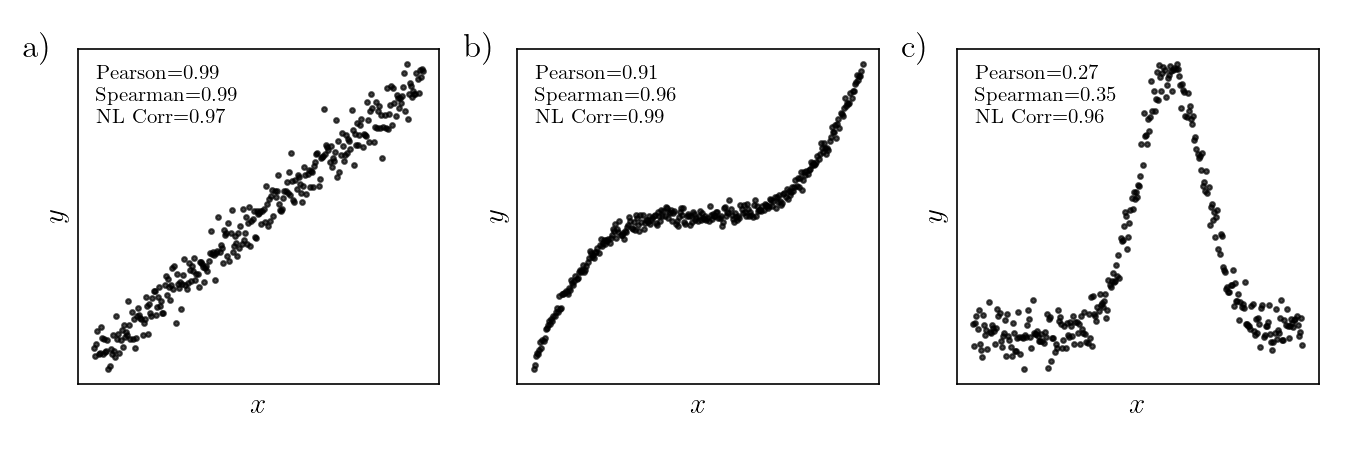}
    \caption{\textbf{Illustration of different correlation measures.} Examples of different data distributions illustrating the limitations of the different correlation measures (Pearson, Spearman and non-linear). The artificial data is generated using (a) a linear function, (b) a cubic function and (c) a Gaussian, with respective noise.}
    \label{fig:corr_limitations}
\end{figure}

The simplest form is \textit{Pearson correlation} \citep[Pgs.~117-118]{dodge2008concise}, which considers maximising $R^2_{(i,j)}$ only allowing for $\hat{f}_i$ to be a linear function (\textit{i.e.}, with two degrees of freedom). It is often defined by taking a square root and giving it the sign of the linear model slope. With such simple choice of model, a low Person correlation does not necessarily indicate lack of correlation. In an attempt to incorporate a larger group of functions to the approximation of $\hat{f}_j$, the \textit{Spearman coefficient} \citep[Pgs.~502-505]{dodge2008concise} considers the Pearson correlation of not the values but the ranks of the input and output data. This way, it can capture complicated monotonic relations that could mislead the Pearson measure (see Figure~\ref{fig:corr_limitations}). These two measures do in practice present similar traits, and thus will be presented alongside the other.
\par

\begin{figure}
    \centering
    \includegraphics[width=\linewidth]{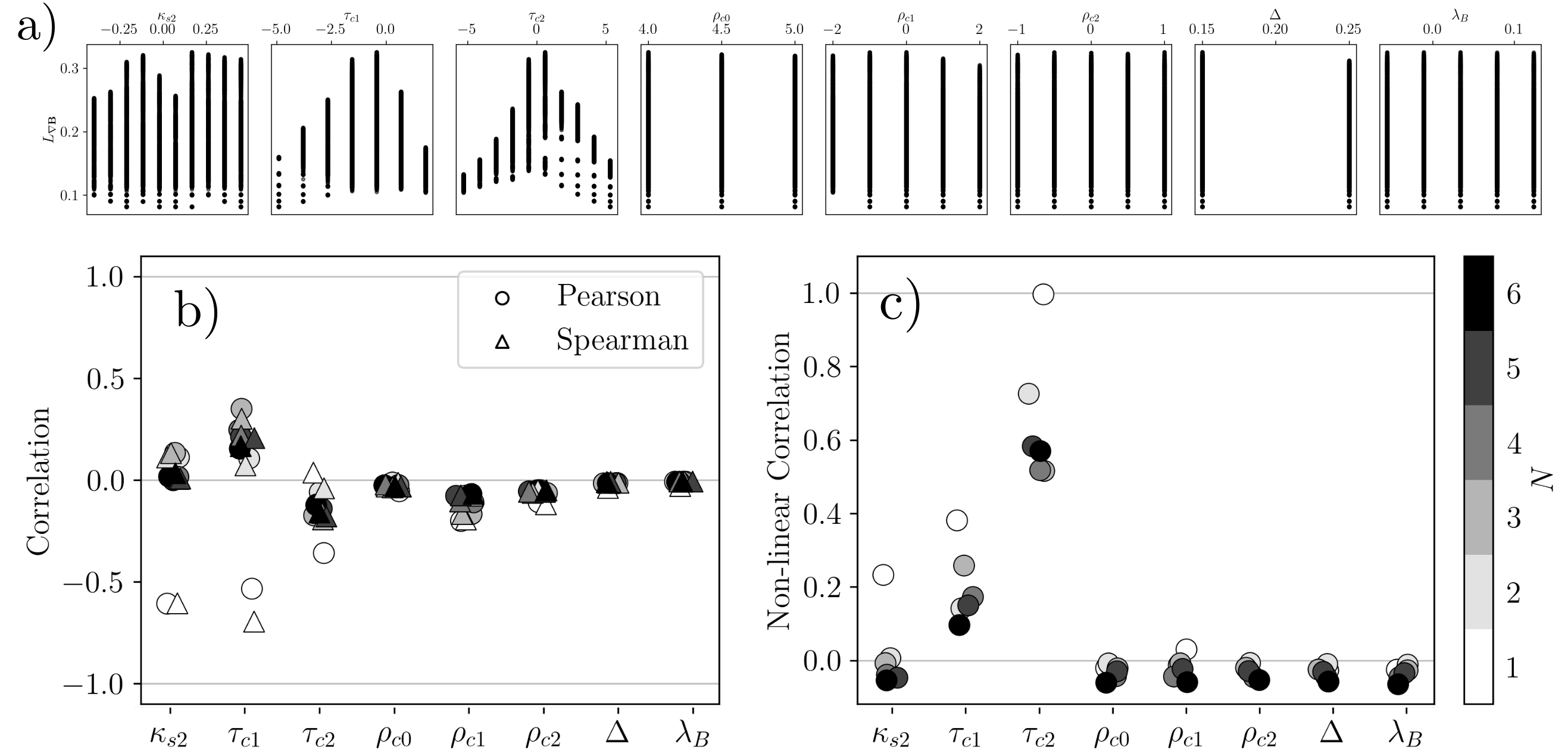}
    \caption{\textbf{Correlation analysis of $L_{\nabla\mathbf{B}}$ with input features.} a) $L_{\nabla\mathbf{B}}$ distribution as a function of the input features for the $N=4$ subset of the database. b) Pearson (circles) and Spearman (triangles) correlation for different input features and different number of field periods (see colour bar). c) Non-linear correlation for different input features and different number of field periods (see colour bar).}
    \label{fig:L_grad_B_correlation_example}
\end{figure}

Both these measures, most notably, fail to recognise peaked dependencies (see Fig.~\ref{fig:corr_limitations}) which do in practice often occur. To incorporate such dependencies, we extend the class of function models $\hat{f}_j$ to a more general class of smooth non-linear functions in Support Vector Regression (SVR) \citep{scholkopf2002learning}\citep[Ch.~12.3.6.]{hastie2009elements} models. This class of functions is broad enough to capture complex underlying dependence, but avoids pathological behaviour that may incur in overfitting that other approaches such as Random Forests \citep[Ch.~15]{hastie2009elements} may easily fall into.\footnote{In the limit of having no repeated $X_i$ in the data, a fully unconstrained dependence consideration would always lead to $R^2_{(i,j)}=1$, even if the underlying choice of model was a completely random function joining the existing data points. To avoid such non-physical relations in a granular data such as that of the database (see Figure~\ref{fig:L_grad_B_correlation_example}a), SVRs are more suitable, even though more costly.} We define a \textit{non-linear correlation} as the $R^2_{(i,j)}$ score of the fitted non-linear radial-kernel SVR model,\footnote{As is standard in the machine learning context, we consider optimisation of the model to maximise the $R^2$-score on K-fold cross-validation, to avoid overfiting.} optimised over common hyperparameter values. 

Armed with these three measures, we may construct a preliminary picture of the dependence of any given property of interest, $y_i$, on the input features. We present an example of this for $y_i=L_{\nabla\mathbf{B}}$, summarised in Figure~\ref{fig:L_grad_B_correlation_example}, and discussed in length in Section~\ref{sec:Acmhd_anal}. Linear correlation is generally weak (below a $|R|<0.5$), without much difference between Pearson and Spearman, meaning that none of the input features can single-handedly explain any significant part of the behaviour of $L_{\nabla \mathbf{B}}$ through a simple monotonic relation. The $N=1$ scenario (white scatter) appears to be somewhat of an exception, as it shows a distinct sign of linear correlation with $\kappa_{s2}$ and $\tau_{c1}$. This is partially due to the reduced population in the $N=1$ subset, which reduces significantly the complexity of the subset.

The dependence on $\tau_{c2}$, Figure~\ref{fig:L_grad_B_correlation_example}a, illustrates the shortcoming of the linear correlation coefficient, incapable of describing appropriately the existing peaked dependence. The non-linear correlation measure in Figure~\ref{fig:L_grad_B_correlation_example}c does cleanly capture this dependence of $\tau_{c2}$, and to a lesser extent, $\tau_{c1}$. Other parameters appear to exhibit $R^2\approx0$, suggesting their relative unimportance. However, it is premature to conclude that these features are unimportant. All we can say is that as isolated parameters, they are not very useful in making predictions, at least not the way that torsion is. But even the torsion can only explain a fraction of the total data variance, roughly 50-70\%.

\subsection{Sequential feature selection}
Performing an analysis like the above, we may identify the feature that in isolation best predicts the measure of interest. However, one feature is hardly ever sufficient to capture the complexity of any feature of interest (see the moderate values of $R^2$ in Figure~\ref{fig:L_grad_B_correlation_example} as evidence). Thus, it is natural to ask what the next most important features are, and how much more information is obtained by including these in the prediction.

There are different ways of approaching this question that falls within the Machine Learning problem of Feature Selection \citep{liu2012feature, shalev2014understanding}. Here we consider what has come to be known as Forward Sequential Feature Selection \citep{whitney1971direct} (also called greedy feature selection in \cite{shalev2014understanding}, and one-best look-ahead in \cite{liu2012feature}). In FSFS one proceeds as follows. First, the single best feature that, in isolation, describes $y_i$ is ranked 1st, using say the non-linear correlation $R^2$ as score. Then, one considers all pairings of this rank 1 feature with the other features left, and a two feature model is fitted for each pairing. The best performing pair is then selected and the second feature of the pair is then given ranked 2nd. The procedure continues this way to provide a hierarchy of features. By logging the evolution of the coefficient of determination of the fitted models, the relevance of features can be tracked as part of the diagnostic.

Unlike the univariate correlation considerations, this approach takes interaction between features into account. That is also why it is significantly more computationally intensive, as it requires many model fits. To make the FSFS manageable with readily available resources, we consider once again fitting Support Vector Regression models with a radial kernel. Other models could also be considered, as could be the case of other algorithms. 

A key observation of FSFS is that, by construction, it avoids ranking two features one right after the other which provide essentially the same information about $y_i$. This allows us to explore the problem not only considering the bare form of the input parameters $\mathbf{x}$, but other more elaborate combinations that could be more insightful. In the case of our database, although the input features are complete and have a very specific physical meaning, they hide complex underlying interrelations to field features that are more immediately involved in $\mathbf{y}$. A clear example of this is the role of torsion and curvature parameters, which only in a highly non-linear way will lead to a consistent pair of functions $\kappa(\ell)$ and $\tau(\ell)$ describing a closed curve. If more meaningful combinations of the parameters could be concocted, one could then use FSFS to gain additional insight on the key dependencies of $y_i$. 

\begin{figure}
    \centering
    \includegraphics[width=0.5\linewidth]{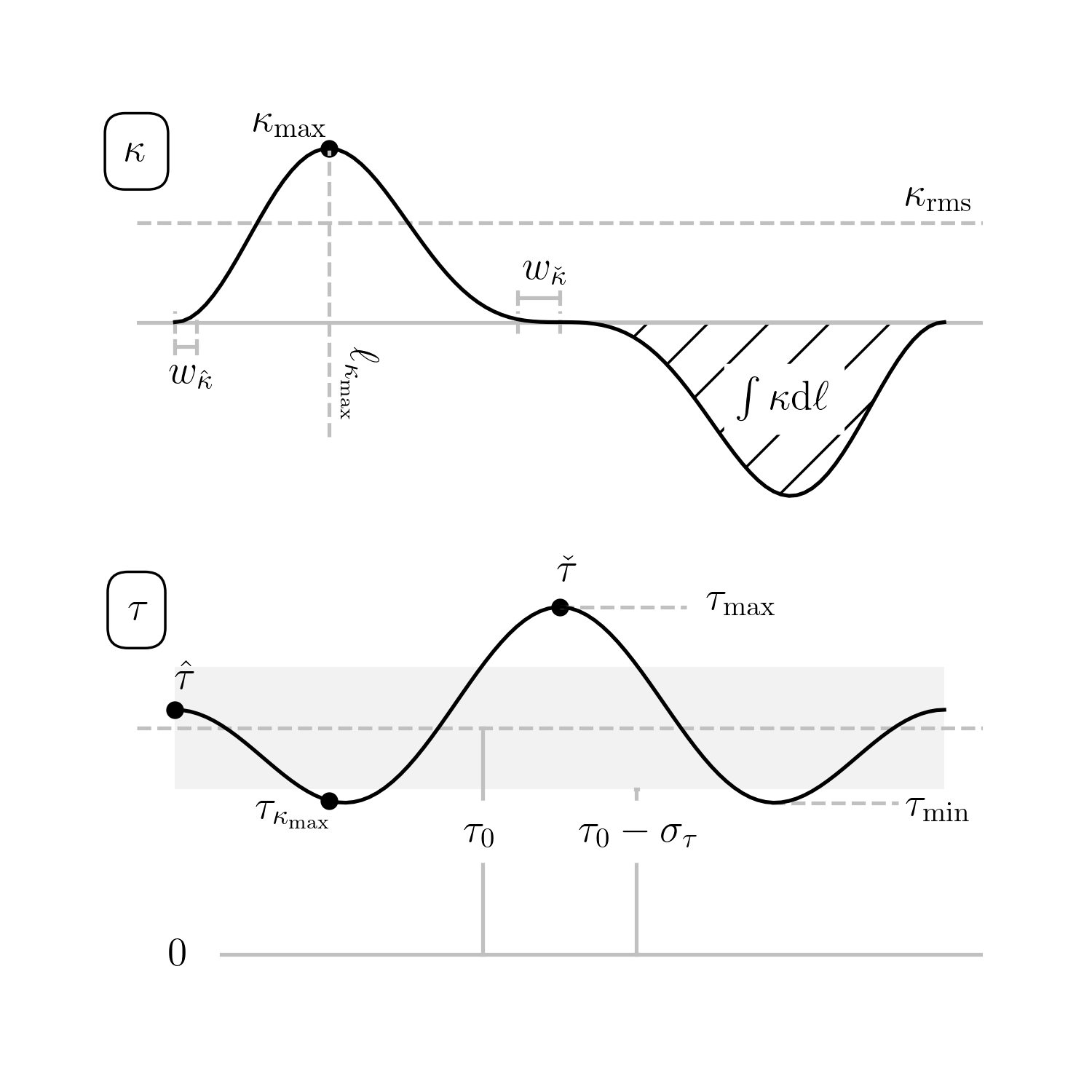}\includegraphics[width=0.5\linewidth]{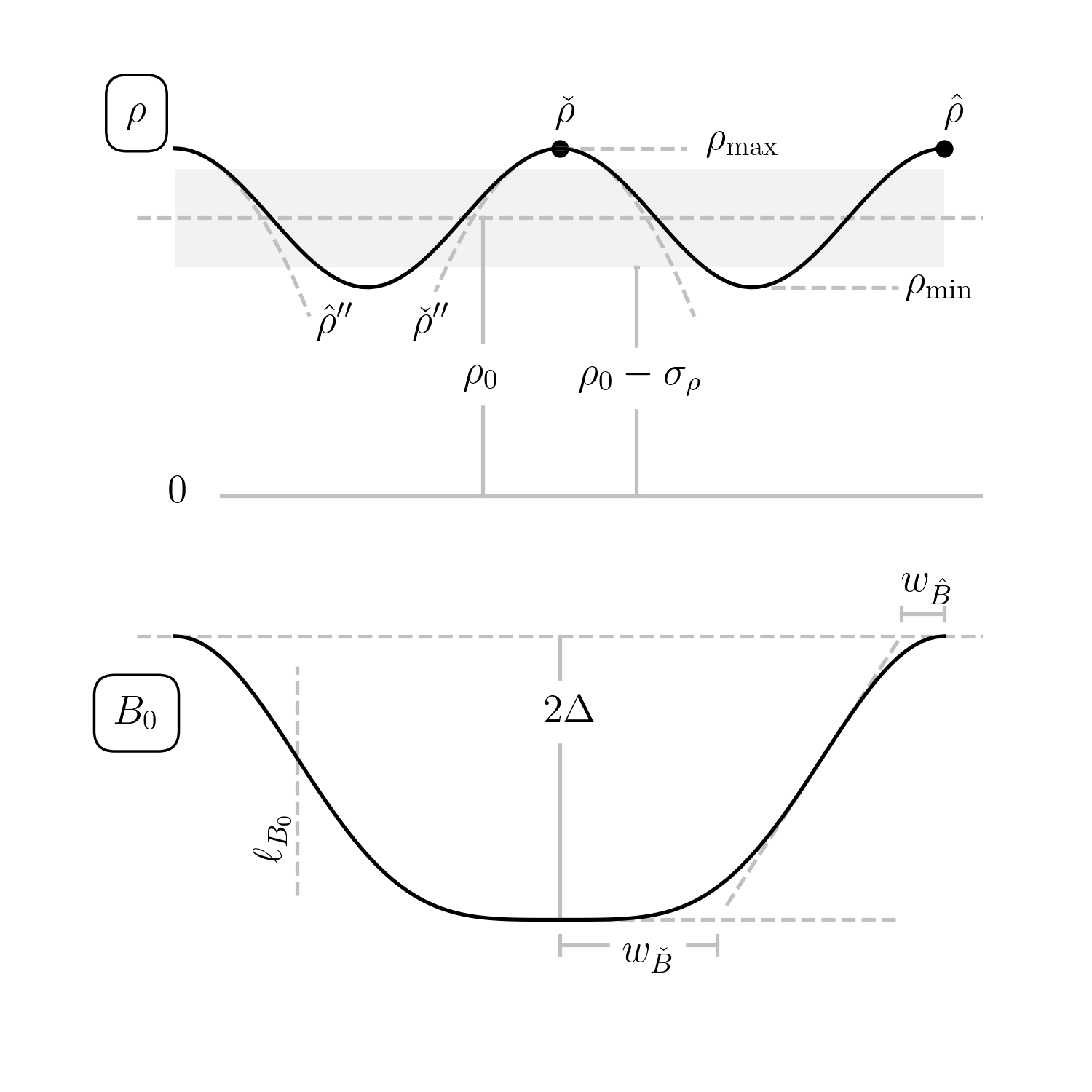}
    \caption{\textbf{Characteristic features of the basic input near-axis functions.} Diagram showing the definition of characteristic features of curvature, torsion elongation and magnetic field. These features have a clearer meaning compared to the true input parameters,  }
    \label{fig:input_feature_comb}
\end{figure}

Let us then use this opportunity to concoct derived quantities out of input features. At a basic level, we could distinguish torsion-related, curvature-related, elongation-related and magnetic strength related quantities that describe some of their key features. Let us illustrate this with torsion, defined in Eq.~(\ref{eq:tau-N1}), which begs us to include $\tau_0$ as part of our extended $\mathbf{x}$. With that inclusion, we may then describe directly the average torsion, the maximum and minimum torsion values, the torsion at the bottom and the top of the magnetic well, as well as other features illustrated in Figure~\ref{fig:input_feature_comb}. A similar consideration applies to curvature in Eq.~(\ref{eq:kappa-N1}), for which $\kappa_1$ needs to be promoted. In that case, we can define the root mean square curvature, the maximum curvature, its location in $\ell$ amongst others (see Figure~\ref{fig:input_feature_comb}). 

These derived quantities can be explicitly constructed in closed form. As a way of example, let us consider $\ell_{\kappa_\mathrm{max}}$, the location of the curvature maximum. Requiring $\partial_\ell\kappa=0$, it follows that
\begin{equation}
    \cos^2\left(\frac{N\ell_{\kappa_\mathrm{max}}}{2}\right)=\frac{1}{56\kappa_2}\left(-5+30\kappa_2+\sqrt{25+36\kappa_2+228\kappa_2^2}\right), \label{eqn:l_k_max}
\end{equation}
for $\kappa_2\neq0$, which defines a closed form function $\ell_{\kappa_\mathrm{max}}(\kappa_2)$ monotonically decreasing with $\kappa_2$. Note that this is defined in the domain $\kappa_2\in(-1/2,1/2)$, limits imposed by the requirements on the orders of the zeroes. The procedure is similar with other features and descriptive derived constructions (see Figure~\ref{fig:input_feature_comb} for details).


Despite the clear distinct meaning of each of these measures, under the premises of the models used, they are strongly dependent. Hence, we must always proceed with care when discussing the resulting hierarchy from FSFS. In particular, it may happen that, let's say, $\tau_\mathrm{max}$ is selected as the most important feature over $\tau_{0}$, even though the difference in $R^2$ is minute. The latter will then likely be excluded from the FSFS hierarchy, due to its significant information overlap with the first rank choice. Because such choice could come down to minor differences in $R^2$, the hierarchy obtained should be interpreted with this caveat in mind. 

\begin{figure}
    \centering
    \includegraphics[width=\linewidth]{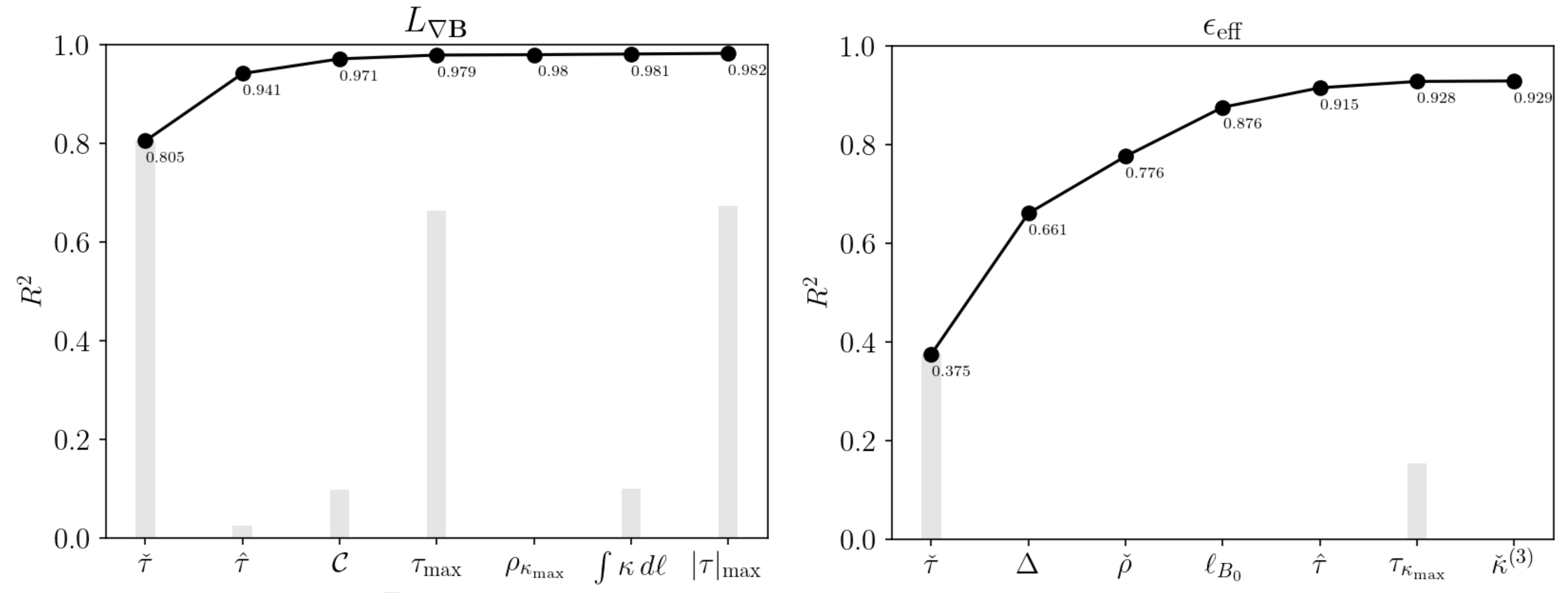}
    \caption{\textbf{Forward sequential feature selection examples.} The two plots show the evolution of the $R^2$ coefficient of determination of the random forest models fitted as each of the features is added to the set of inputs. The bars denotes the coefficient of determination of a univariate model including the feature. (Left) Example for $N=4$, $L_{\nabla\mathbf{B}}$ considerations. (Right) Example for $N=2$, $\epsilon_\mathrm{eff}$. These illustrate two different cases in which many features or few are needed to faithfully reproduce the behaviour in the database.}
    \label{fig:fsfs}
\end{figure}

Let us then put these considerations into practice, with the example case of the input dependence of $L_{\nabla \mathbf{B}}$ for $N=4$, and $\epsilon_{\mathrm{eff}}$ for $N=2$, summarised in Figure~\ref{fig:fsfs}. The hierarchy shows in quite a pristine way the dominant role of $\check{\tau}$ for $L_{\nabla\mathbf{B}}$, the torsion value at the bottom of the well. A univariate model already accounts for as much as $\sim80$\% of the variance, consistent with the correlation analysis. The following feature is also torsion related, the standard deviation $\sigma_\tau$, whose univariate description (see the bar plot) of $L_{\nabla\mathbf{B}}$ is similarly descriptive. A physical and theoretical explanation of this observation, and a reading into the significance of the observed differences is considered in the main text. The curvature makes its appearance only in a third place, with an added 1\% accuracy. Marginal improvements ensue after that. The situation in the $\epsilon_\mathrm{eff}$ case is rather different, Figure~\ref{fig:fsfs}b, showing at least four different features being needed to capture $\sim$90\% of the data variance. In the main text, we shall present FSFS data not in the form of plots, which difficults a condensed presentation for multiple $N$, but rather in the form of ranked tables. The first few relevant ranked features will be presented, next to the $R^2$ value of the model fitted by including the features at that rank and below (see, for example, Table~\ref{tab:input_importance}).

\subsection{Global importance measures}
The forward sequential feature selection is insightful, but it is by no means complete. It is based on a hierarchical construction of increasingly more complex models, but its decisions give a large weight to the simplest ones. This may not always be appropriate. 

We now turn to the notion of feature importance as an alternative diagnostic of the relevance of each feature in the problem. The notion of \textit{importance} is commonplace in Machine Learning (especially \textit{explainable} ML), and it is precisely devoted to understanding how a fitted model $\hat{f}(\mathbf{x})$ depends on the different features $\{x_i\}$. It is a direct assessment of the fitted model $\hat{f}$, which can take into consideration all features at the same time. No unique definition of importance exists \citep{ewald2024guide, kamath2021explainable}, but here we choose to focus on two global, model agnostic measures (we shall not consider other local measures such as PCP, LIME, SHAP, etc.).

The first measure is Permutation Feature Importance (PFI) \citep{breiman2001random,fisher2019all}\citep[Pg.~154]{molnar2020interpretable}. This scalar measures the change in the prediction of the model $\hat{f}$ when the inputs of a particular feature are randomly shuffled. If the result changes significantly, then one is led to think that the modified feature is important. Relative comparison of those changes define a normalised PFI value which then be considered to rank features. This relies on the fitted model being \textit{good}, as it is a description of the \textit{model} as much as the actual data. 

Although insight may be gained this way, PFI ignores possible feature dependence. This leads to a prototypical failure of PFI when a model has to be fit to two features with a cunning resemblance. More often than not, the model will choose to fit one of them preferentially, leading to PFI misleadingly indicating unimportance of the other feature. In order to incorporate to the analysis some measure that takes some of that dependence into account, we introduce conditional Shapley Additive Global importancE (cSAGE) \citep{covert2020understanding,ewald2024guide}. cSAGE measures the importance of each feature by assessing how the predictions of different sub-models constructed from the original full model marginalised over the conditional distributions of features fare \citep[Sec.~6]{ewald2024guide}. It then compares fairly all of these differences using Shapley values \citep{shapley1953value, winter2002shapley}, a way of distributing payoffs given the contribution of each feature. The relative cSAGE values serve then as a complement of PFI ones. In practice, evaluation of cSAGE is quite computationally costly and therefore we shall resort to Random Forest models (in particular the \texttt{ExtraTrees} algorithm in \texttt{scikit-learn} \citep{scikit-learn}) for the underlying function, which is significantly faster.

\begin{figure}
    \centering
    \includegraphics[width=0.6\linewidth]{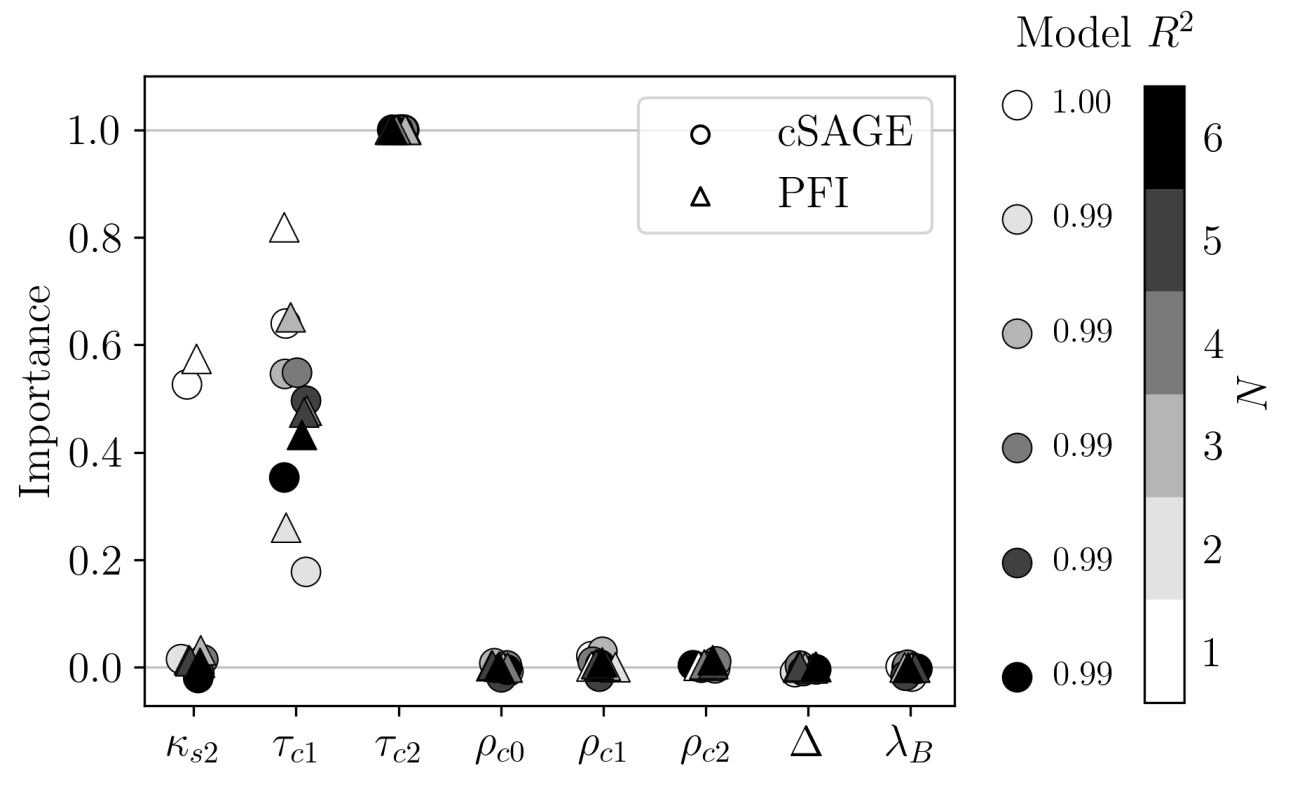}
    \caption{\textbf{Feature importance analysis for $L_{\nabla\mathbf{B}}$ dependence on input features.} PFI and cSAGE feature importance for the input features for predicting $L_{\nabla\mathbf{B}}$ of the $N=4$ subset of the database. The goodness of the model fit is shown on the right margin of the plot, in this case showing a good $R^2=0.97$ fit.}
    \label{fig:fi}
\end{figure}

We show an example of the importance calculation in Figure~\ref{fig:fi}, which includes a measure of goodness of the models fitted to each $N$ subset of the database (the standard cross-correlation average $R^2$ measure). It is important  for such model to be sufficiently accurate (as in the example, $R^2\gtrsim 0.97$) so that importance is meaningful. With that said, the importance measures show, both PFI and cSAGE, that the torsion feature is indeed key, as the non-linear correlation analysis pointed at. The larger spread in $\tau_{c2}$ shows the secondary importance of $\tau_{c1}$ with a potential dependence on the dominant $\tau_{c2}$.

\subsection{Mutual feature dependence}
In many cases, multiple features may present similar importances. Or in the case of FSFS the ranking choice may be taken only through very small margins. That is why we want some measure to assess how much information two given features share when it comes to determining a particular $y_j$. 

To that end we consider Friedman's H-statistic \citep[Eq.~(44)]{friedman2008predictive}. For every pair of features $x_i$ and $x_j$, $H_{ij}$ measures the fraction of the variance predicted by the bi-variate model not explained by the addition of the univariate models. These models $\hat{F}$ are constructed by marginalising over the unaccounted features the original model $\hat{f}$ fitted to all input features over. To be more quantitative,
\begin{equation}
    H_{ij}^2=\frac{\sum_{\mathrm{data}}\left[\hat{F}_{ij}(x_i,x_j)-\hat{F}_{i}(x_i)-\hat{F}_{j}(x_j) \right]^2}{\sum_\mathrm{data}\hat{F}_{ij}(x_i,x_j)^2}
\end{equation}
A value $H_{ij}=0$ implies no interaction between features $x_i$ and $x_j$, with higher values indicating stronger interdependence. To make sense of the magnitude of this statistic, we follow \cite[Sec.~8.3]{friedman2008predictive} and construct a reference null distribution over which we may compute $H_{ij}$. If $H_{ij}$ for the true data is larger than the corresponding value for the reference distribution, then it can be taken to be significant. The magnitude of $H_{ij}$ can then be interpreted respect to one as a measure of interdependence. The H-statistic is not without its limitations, and can lead to misleading conclusions in case of very strongly correlated features, but we should use it to enrich the discussion nonetheless. 

\begin{figure}
    \centering
    \includegraphics[width=0.6\linewidth]{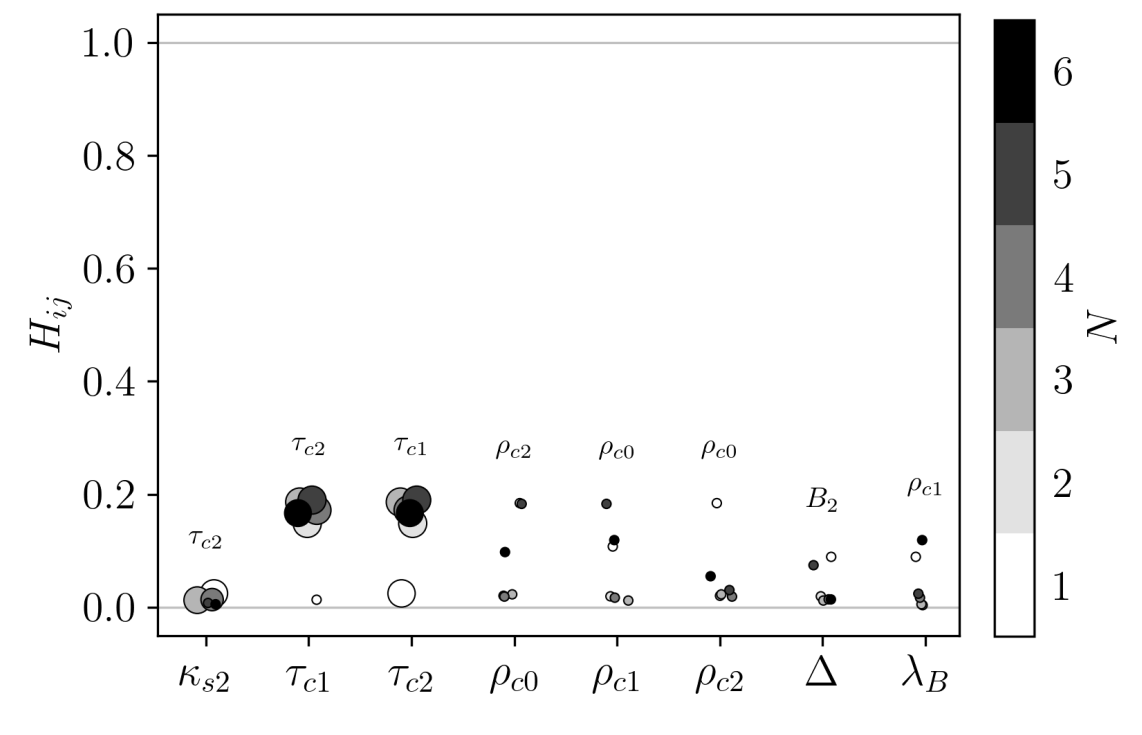}
    \caption{\textbf{Friedman H-statistic for $L_{\nabla\mathbf{B}}$ dependence on input features.} Summary of the feature overlap quantified by the Friedman H-statistic as a function of input feature for different number of field periods. The value of the statistic shown is the maximum value of $H_{i,j}$ for fixed $i$ (the feature labelling the abscissa), with the maximising pair of the largest value (amongst all $N$) explicitly indicated. The size of the scatter comes to indicate the significance of $H_{i,j}$; its area is defined to be proportional to $1-H_{i,j}^\mathrm{null}/H_{i,j}$ with a lower bound. }
    \label{fig:Friedman}
\end{figure}

As a way of example, let us consider once again $L_{\nabla\mathbf{B}}$, for which the H-statistics analysis is summarised in Figure~\ref{fig:Friedman}. The plot indicates a significant cross-talk between the torsion components, giving some standing to the observation made when analysing importance. The $N=5$ case does appear to show some mutual dependence between elongation features as well. One must nevertheless frame this observation in the context of the importance of these features being small, and the behaviour being an outlier respect to other $N$.

\section{Characterising different classes of configurations} \label{app:clustering}
In a dataset with a large number of elements it is daunting, or dare say impossible, to go one-by-one and check how many of those are truly distinct. This is however an important task if we are to have a thorough understanding of the database, and in particular, make sense of the optimal configuration options available. The difficulty of the task has two facets to it: what makes two configurations similar and, the large number of elements (hundreds in the most benign of cases). 
\par
The most basic of notions of distance, at least formally, may be defined in input feature hyperspace. Each element in the set is uniquely defined by an input vector $\mathbf{x}$ of $N_\mathrm{in}$ components, and thus we may define $d(i,j)=||\mathbf{X}^i-\mathbf{X}^j||$ as Euclidean distance in this $\mathbb{R}^{N_\mathrm{in}}$ space. This is however quite artificial, as some inputs refer to degrees of freedom of the magnetic axis, while others refer to the field strength more directly or the shape of cross-sections. Even though this is true, if it is the case that $d(i,j)\gg d(i,k)$ we could say that $j$ is more similar to $i$ than $k$ is. To make sense of this distance though, given our mixing of different dimensions, we will at least have to normalise each dimension in some way. We adopt the extended machine learning practice of normalising these in such a way so as to have zero-averaged, unit spread populations in each dimension. This is a fair a priori, if we are to weight all components in the problem equally, assuming no fundamental difference in their domains. 
\par
In practice, we also find convenient to use not the bare input feature space of size $N_\mathrm{in}$, but rather its extended form including $\tau_0$ and $\kappa_1$. This extended form more fully captures the shape of the field in its real space embedding, and thus is convenient to include. This extension of the space is not without some potential peril, in particular because these added parameters are continuous in a sense that the sampled inputs are not. This could mistakenly lead to additional linkage in space, an element to keep in mind throughout the analysis.
\par
With our hyperspace set-up, and the standard Euclidean definition of distance, what constitutes a short distance? When can we deem two configurations similar and when distinct? Without a notion of a meaningful absolute distance, we must make sense of it relative to the distribution of distances with the remaining elements of the set. That way, one may determine whether two configurations are close or not. Naturally, this calls for a need to \textit{cluster} the set into families of configurations sharing certain features. The entirety of the dataset can then be described through this smaller number of families, each with its own distribution. Even if the clustering was not particularly accurate, it is an important digestive tool to process the overwhelming immensity of data. The question is then, how do we cluster the data?
\par
To cluster our set $\mathcal{S}$ of configurations into $M$ clusters $\{\mathcal{S}_i\}$ for $i=1,\dots, M$, we resort to the extensive literature in machine learning clustering \citep{kaufman2009finding,wierzchon2018modern}. Our ultimate goal is to identify a small, finite partition of the set that distinguishes different classes of configurations, without a priori knowing the number of clusters that may be involved. In many clustering algorithms (K-means \citep[Ch.~3.1]{wierzchon2018modern}, Gaussian Mixtures \citep[Ch.~3.2]{wierzchon2018modern}, BIRCH \citep{zhang1996birch}, etc.), the number of clusters is a hyperparameter that ought to be passed to the algorithm. Some more complex algorithms do not require this (e.g. spectral \citep{ng2001spectral}\citep[Ch.~5]{wierzchon2018modern} or HDBSCAN \citep{campello2013density,mcinnes2017accelerated}), but do instead offer additional complications. To decide on the number of clusters, we treat the problem as one of optimisation with respect to hyperparameters of the clustering algorithms. We define the relevant number of clusters as that corresponding to the set of clustering algorithms and hyperparameters that best clusters the set. We choose the common Calinski-Harabasz measure as a measure of goodness \citep{zhao2012cluster,calinski1974dendrite}. This scalar measure compares the variance of elements within clusters to the variance between clusters, additionally penalising having more clusters. Such a measure is not perfect; in particular, it performs poorly for clusters that are nested or have very anisotropic shapes (what \cite[Ch.~4.3]{wierzchon2018modern} refer to as non-Voronoi clusters). We thus especialise ourselves to "simple" clusters. Table~\ref{tab:clustering_algorithms} summarises the used clustering algorithms and hyperparameters. This approach proves to be quite robust and suffice for the exploration needs in this paper.
\begin{table}
\centering

\begin{tabular}{p{3cm}p{4cm}p{4cm}c}
\textbf{Algorithm} & \textbf{Sklearn Class} & \textbf{Hyperparameters Scanned} & \textbf{Combinations} \\\hline\hline
K-Means & 
\texttt{MiniBatchKMeans} & 
\texttt{n\_clusters}, \texttt{init}, \texttt{max\_iter}, \texttt{batch\_size} &
180 \\
Ward & 
\texttt{AgglomerativeClustering} & 
\texttt{n\_clusters}, \texttt{linkage} & 
15 \\
SpectralClustering & 
\texttt{SpectralClustering} & 
\texttt{n\_clusters}, \texttt{eigen\_solver}, \texttt{affinity}, \texttt{gamma} & 
54 \\
GaussianMixture & 
\texttt{GaussianMixture} & 
\texttt{n\_components}, \texttt{covariance\_type}, \texttt{max\_iter}, \texttt{init\_params} & 
240 \\
BayesianGMM & 
\texttt{BayesianGaussianMixture} & 
\texttt{n\_components}, \texttt{covariance\_type}, \texttt{weight\_concentration\_prior}, \texttt{max\_iter}, \texttt{init\_params} &
960 \\
BIRCH & 
\texttt{Birch} & 
\texttt{n\_clusters}, \texttt{threshold}, \texttt{branching\_factor} & 
240 \\
AverageAgg. & 
\texttt{AgglomerativeClustering} & 
\texttt{n\_clusters}, \texttt{linkage}, \texttt{metric} &
60 \\
CompleteAgg. & 
\texttt{AgglomerativeClustering} & 
\texttt{n\_clusters}, \texttt{linkage}, \texttt{metric} &
45 \\
SingleAgg. & 
\texttt{AgglomerativeClustering} & 
\texttt{n\_clusters}, \texttt{linkage}, \texttt{metric} &
45 \\
AffinityPropagation & 
\texttt{AffinityPropagation} & 
\texttt{damping}, \texttt{max\_iter}, \texttt{convergence\_iter}, \texttt{preference} &
30 \\
\\
\multicolumn{3}{r}{\textbf{Total Parameter Combinations:}} & 1,869 \\
\end{tabular}
\caption{\textbf{Clustering algorithms and their hyperparameter search spaces}.  Clustering algorithms used (with their \texttt{scikit-learn} pairing \citep{scikit-learn}) and the hyperparameter scan performed to choose the number of clusters in the set analysed. Only algorithms that do not treat elements as noise have been considered here.}
\label{tab:clustering_algorithms}
\end{table}

Once a particular number of clusters and the assignment of each data-point to each cluster has been achieved, we can then assess the distribution of the $N_\mathrm{in}$ input features within each cluster in order to deem whether such configuration families are truly different in any meaningful way or not. If so, we can then register a representative configuration from each set (see Figure~\ref{fig:clust_N_2}), or parametrise the family (see Figure~\ref{fig:N_1_Acmhd}). This latter assessment can be made in a case-by-case basis, as the size of the set to evaluate is significantly reduced. Artificial additional clusters are not too much of a problem, so long as they only appear in reasonable numbers.
\par
To illustrate the achievements of the clustering and be able to assess the results further, it is convenient to have a way of representing the set on the plane, \textit{i.e.}, in 2D. For such representation we may use many of the existing dimensionality reduction approaches. In practice, we consider Multidimensional Scaling (MDS) \citep{borg2005modern}. There are multiple approaches to dimensionality reduction under MDS\footnote{In this work we use the default implementation in \texttt{scikit-learn.manifolds} \citep{scikit-learn} for MDS.}, attempting to faithfully preserve distances between points in different ways. No additional information is really necessary here for our purpose, given that we are using it as a simple representation vehicle, not for any quantitative end. Other dimensionality reduction techniques such as the simpler Principal Component Analysis (PCA) \citep{jolliffe2011principal} or more sophisticated Isomap \citep{tenenbaum2000global} could also be considered, but MDS is found to be most convenient in practice.

\section{Detailed contributions to $f_\mathcal{J}$}
\label{app:fJ}

In this Appendix we present some elements of the theoretical description of the particle precession $\omega_\alpha$, whose near-axis form was presented in Appendix~A of \cite{rodriguez2024maximum}. Precession as a function of $\alpha$ (the field line label) and $r$ (a normalised radial flux label) may be written as \citep[Appendix~A]{rodriguez2024maximum}
\begin{equation}
    \omega_\alpha(r,\alpha,\lambda)= \omega_{\alpha,-1}^{\mathrm{non-QI}}(\lambda)\frac{\cos\alpha}{r}+\omega_{\alpha,\mathrm{vac}}(\lambda)+\omega_{\alpha,0}^\mathrm{non-QI}(\lambda)\cos2\alpha, \label{eqn:precess_wa}
\end{equation}
where $\omega_{\alpha,-1}^{\mathrm{non-QI}}$ and $\omega_{\alpha,0}^\mathrm{non-QI}$ are driven by non-omnigenous behaviour (\textit{i.e.}, result in $\alpha$-dependent terms), Eqs.~(A19) and (A21) in \cite{rodriguez2024maximum}, and the omnigenous $\omega_{\alpha,\mathrm{vac}}$ is a purely second order quantity, Eq.~(A15) in \cite{rodriguez2024maximum}.
\par
The first term in Eq.~(\ref{eqn:precess_wa}), $\omega_{\alpha,-1}^\mathrm{non-QI}$, represents the net poloidal drift due to imperfections in the first order near-axis construction. The generally necessary departures from precise omnigeneity near $B_\mathrm{max}$ \citep{plunk2019direct,camacho-mata-2022,rodriguez2023higher} dominate precession sufficiently close to the magnetic axis, expecting $f_\mathcal{J}\approx 0.5$ there due to the field-line to field-line variation. In practice, though, this contribution quickly (in $r$) becomes overwhelmed by order $O(r^0)$ terms, given mainly the relatively small magnitude of the magnetic drift close to $B_\mathrm{max}$. 

Ignoring any QI-breaking contribution from first order, then, we may succinctly write the dominant non-omnigenous contribution to precession as \citep[Eq.~(A20)]{rodriguez2024maximum},

\begin{equation}
    \omega_{\alpha,0}^\mathrm{non-QI} = -\frac{2mv^2}{q \bar{B}}\oint F(\lambda,B_0)\left[B_{2c}-\frac{1}{4}\left(\frac{B_0^2d^2}{B_0'}\right)'\right]\mathrm{d}\varphi\Bigg/{\oint\frac{\mathrm{d}\varphi}{B_0\sqrt{1-\lambda B_0}}}, \label{eqn:2nd-order-deviation-wa}
\end{equation}
where $\oint$ denote bounce integrals along $B_0(\varphi)$. The integrand in square brackets (being zero) is precisely the stellarator symmetric QI condition at second order \citep[Eq.~(32c)]{rodriguez2023higher}. The field must therefore be sufficiently omnigenous so that $\omega_{\alpha,\mathrm{vac}}(\lambda) \geq \omega_{\alpha,0}^\mathrm{non-QI}(\lambda)$ if a large fraction $f_\mathcal{J}$ is to be achieved. We explicitly write $\omega_{\alpha,\mathrm{vac}}$ as, \citep[Eq.~(A15)]{rodriguez2024maximum},
\begin{equation}
    \omega_{\alpha,\mathrm{vac}} = \frac{2mv^2}{q \bar{B}}\oint F(\lambda,B_0)\left[B_{20}-\frac{1}{4}\left(\frac{B_0^2d^2}{B_0'}\right)'\right]\mathrm{d}\varphi\Bigg/\oint\frac{\mathrm{d}\varphi}{B_0\sqrt{1-\lambda B_0}}. \label{eqn:nae_w_alpha_appx}
\end{equation}
The radial gradient of the magnetic field strength dominates this term, \textit{i.e.,} $B_{20}$. The larger the gradient, the more positive $\omega_{\alpha,\mathrm{vac}}$, and thus the more maximum-$\mathcal{J}$ the field. This is the near-axis manifestation of the connection between MHD stability, Eq.~(\ref{eqn:mag_well_nae}), and maximum-$\mathcal{J}$. 

\section{Helicity of the field} \label{app:helicity}
The database presented in this paper has considered so-called \textit{half-helicity} QI near-axis fields only. The notion of axis helicity in stellarators was introduced by \cite{rodriguez2022phases}, and for QI fields in \cite{Camacho2023helicity}, later discussed in detail in \cite[Appendix~A]{rodriguez2024near}. These pieces of work restricted themselves to axes that can be described in cylindrical coordinates by smooth functions $R$ and $Z$, so that $\mathbf{r}_\mathrm{axis}(\phi)=R(\phi)\hat{\pmb R}(\phi)+Z(\phi)\hat{\pmb z}$, where $\phi$ is the cylindrical angle. Such a description is quite general and mostly sufficient, but it nevertheless excludes non-star shaped curves, knotted curves or curves with vertical segments. These exceptional cases, although rare, are not ruled out when constructing curves following \cite{plunk2025-geometric}, that is, by prescribing curvature and torsion directly (see Figure~\ref{fig:high_curv_solutions}). Hence, we must take additional care in defining helicity in the present context. 

At a high level, \textit{helicity} is a count of the number of times the normal of the magnetic axis rotates around itself in a toroidal transit. This may be related to the topological notion of \textit{self-linking number} \citep{pohl1968self,moffatt1992helicity,fuller1999geometric,fuller1971writhing,oberti2016torus,pfefferle2018non,rodriguez2022phases}: $S_L$ is the linking number ($L_k$) between our curve of interest and one generated by an infinitesimal displacement of the curve in the direction of the \textit{signed} binormal vector, $\mathbf{r}_b^\mathcal{\delta}(\phi)=\mathbf{r}_\mathrm{axis}(\phi)+\delta~\hat{\pmb \tau}(\phi)$, with $\delta>0$ small. This is an intrinsic property of the curve and its signed frame, and thus a `good' property of the axis. In the QI scenario, for curves of odd total order, the generated $\mathbf{r}_b^\delta$ is only closed upon two toroidal transits \citep{rodriguez2024near}, in which case we must divide $S_L/2$ to correct for that double transit. 

$S_L$ can be directly connected to the notion of helicity as a counting of poloidal turns, but it is customary to introduce some reference point, and in so doing, invoke the Calegareanu theorem for two neighbouring curves $C$ and $C^\star$,
\begin{equation}
    L_k(C,C^\star)=S_L(C) + \mathcal{N}. \label{eqn:calegareanu}
\end{equation}
Here $\mathcal{N}$ is defined as the number of rotations the unit span vector from $C$ to $C^\star$ makes respect to the signed frame of $C$, and thus it gives a direct connection to counting of turns. To exploit Eq.~(\ref{eqn:calegareanu}) we need some reference $C^\star$ that is not aligned with the frame, as otherwise we simply recover the definition of self-linking number (with $\mathcal{N}=0$). 

\begin{figure}
    \centering
    \includegraphics[width=0.7\linewidth]{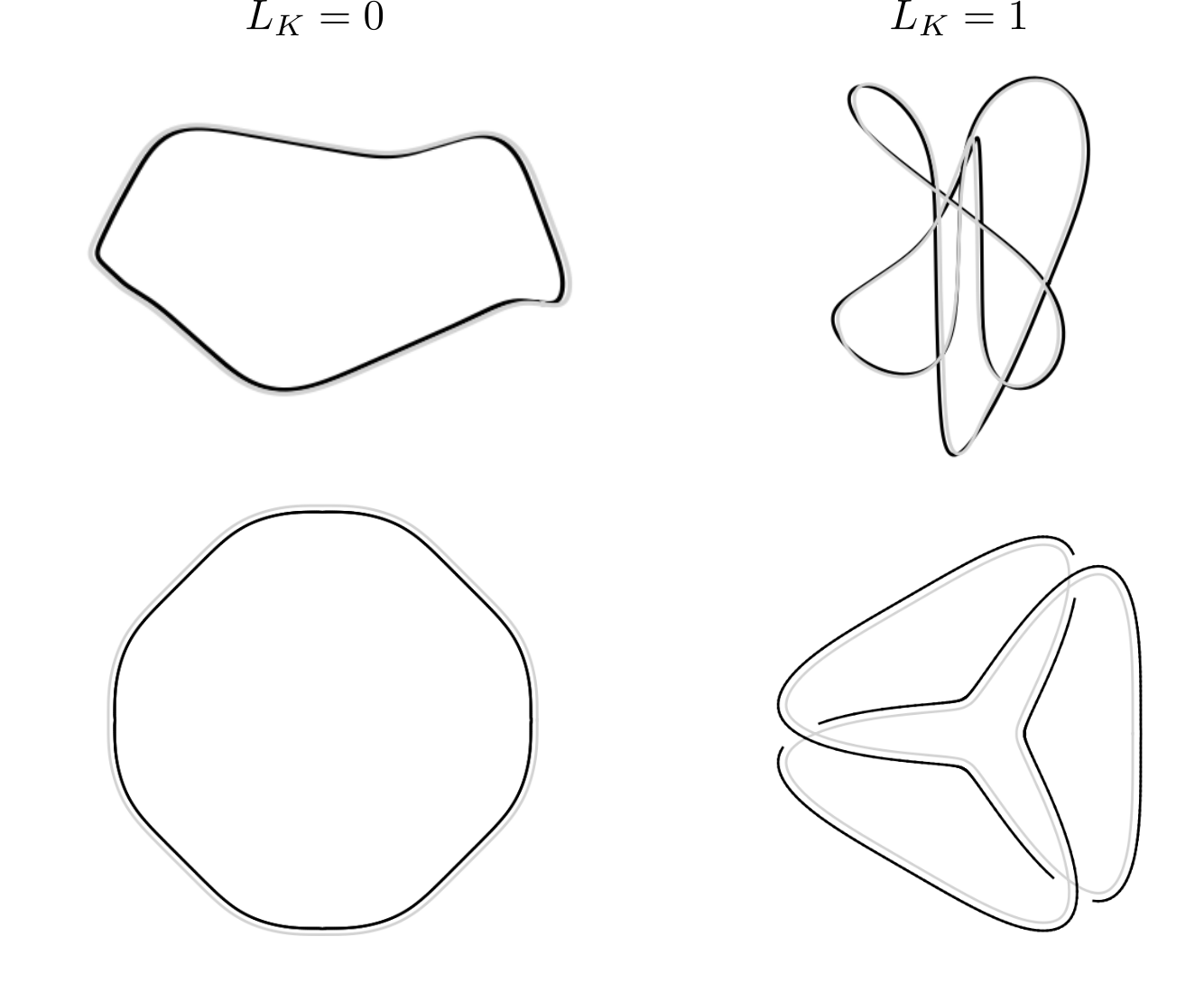}
    \caption{\textbf{Illustrating examples of different Kauffmann number curves.} 3D (top) and knot diagram (along the $z$-axis, bottom) for a $L_K=0$ and $L_K=1$ curves. The curve on the left has $L_K=0$ as it follows from the lack of crossings. The gray curve is generated by the blackboard framing, what we take to be the reference $\theta=0$ curve. The curve on the right corresponds to the tall trefoil in Figure~\ref{fig:high_curv_solutions}, and the knot diagram clearly shows a total non-vanishing writhe.}
    \label{fig:helicity_SL}
\end{figure}

In the standard picture of helicity, the one where the curve can be defined through functions $R$ and $Z$, the natural reference is what is known as the \textit{blackboard framing} of the curve along the $z$ axis. That is, define $C^\star$ to lie outside the original curve $C$ when looking down the $z$-axis, curve that we may take as a reference $\theta=0$ (see Figure~\ref{fig:helicity_SL}). Following Eq.~(\ref{eqn:calegareanu}), $\mathcal{N}$ will be the number of times the signed frame crosses $C^\star$ (with a sign). We must also compute the linking number $L_k(C,C^\star)$, but one may prove it to vanish for the non-exceptional $R,Z$-defined curves. The proof is immediate once we invoke the following proposition from \cite{chmutov2012introduction},
\begin{lemma}
    The self-linking number of a framed knot [the linking number between a curve $C$ and a $C^\star$ generated by the framing] given by a diagram $D$ with blackboard framing is equal to the total writhe of the diagram $D$.
\end{lemma}
The total writhe of a knot diagram vanishes if that projection of the curve onto the plane exhibits no crossings \citep[Sec.~1.3.2]{chmutov2012introduction}. And by definition, our projection onto the $x-y$ plane does not have such crossings. Hence, 
\begin{equation}
    \text{Helicity}=S_L,
\end{equation}
as originally pointed in \citep{rodriguez2022phases, thesis}. When the projection has crossings, then there is a linking related correction, which we may incorporate on to the definition 
(as in Eq.~(6.4) of \cite{thesis}),
\begin{equation}
    \text{Helicity}=S_L-L_K,
\end{equation}
where $L_K$ is the Kauffmann linking number (the linking with the blackboard framing). We can generally compute this number by constructing $C^\star$ deforming $C$ in the direction $\hat{\pmb{z}}\times\hat{\pmb{t}}$. This works even when the axis is knotted or turning back in the toroidal angle. Such measures follows more closely \cite{landreman2018a}.

This definition of helicity, although it matches the standard one, does however evidence that it is not fully intrinsic to the curve. It requires an external direction as well to count turns. When the curve has vertical portions this definition will clearly break. This makes this form of helicity not a fully satisfactory property, even though it matches the qualitative description given.  

\newpage
\section{Database statistics}
\begin{figure}
    \centering
    \includegraphics[width=\linewidth]{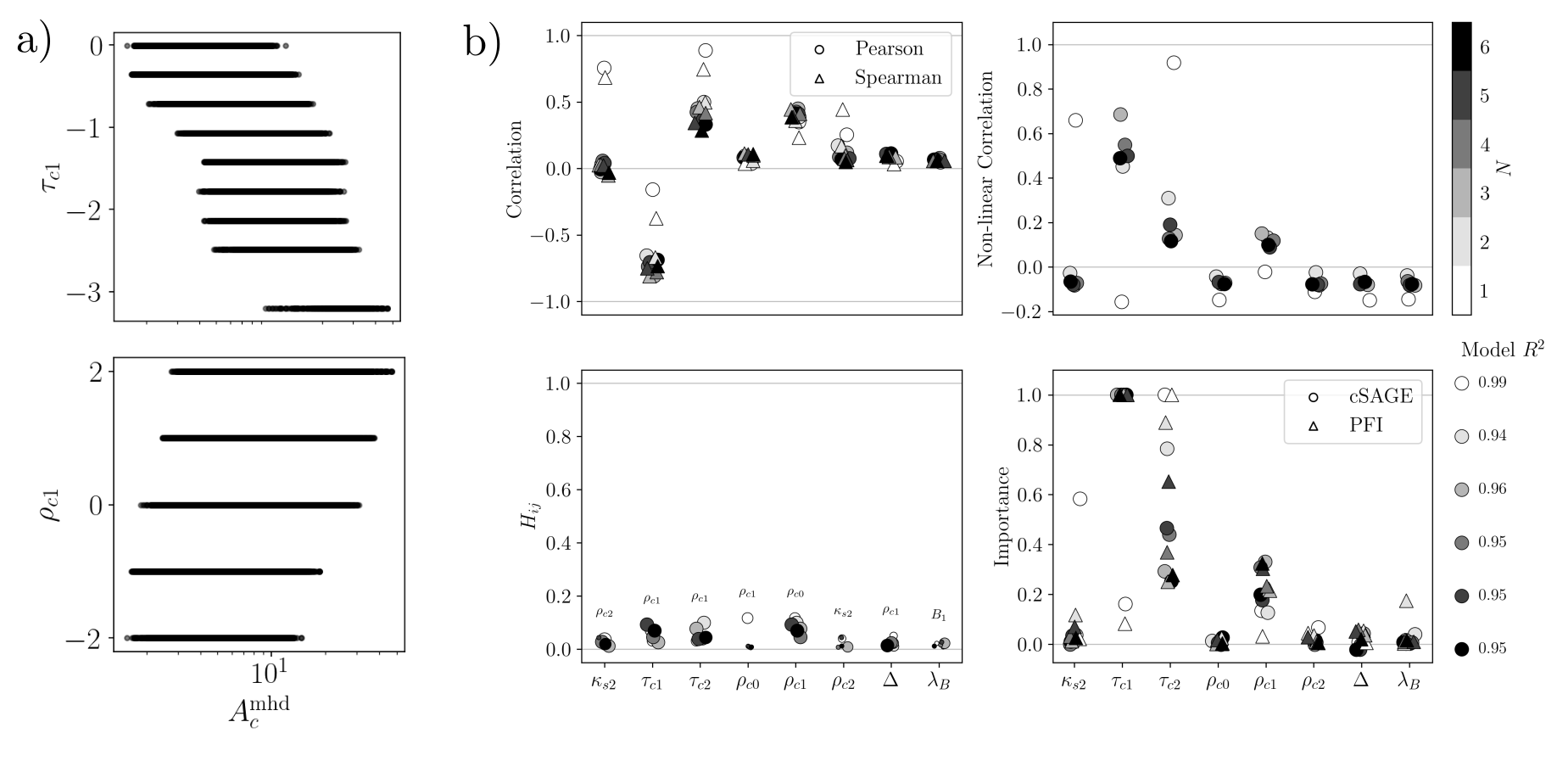}
    \caption{\textbf{Feature dependence statistical summary for $A_c^\mathrm{mhd}$.} (a) Scatter plots showing the univariate dependence of $A_c^\mathrm{mhd}$ on the features $\tau_{c1}$ and $\rho_{1c}$, for the $N=2$ subset. (b) Summary of key statistical measures describing the dependence of $A_c^\mathrm{mhd}$ on input features. (Top left) Linear correlation between the input features and $A_c^\mathrm{mhd}$. The different symbols indicate the Pearson and Spearman coefficients, while the colour distinguish different number of field periods. (Top right) Non-linear correlation coefficient, where the colour distinguish different number of field periods. (Bottom right) Relative feature importance with symbols representing the PFI and cSAGE measures, and color different number of field periods. The values listed to the right next to the scatter indicate the coefficient of determination of a multivariate SVM model fitted to $L_{\nabla\mathbf{B}}$ in terms of the input features. (Bottom left) Friedman H-statistic where the size of the scatter indicates the significance of the measure compared to a null reference distribution, and the colour represent different number of field periods. The labels indicate which other feature they are most closely linked to.}
    \label{fig:Acmhd_stats}
\end{figure}

\begin{figure}
    \centering
    \includegraphics[width=\linewidth]{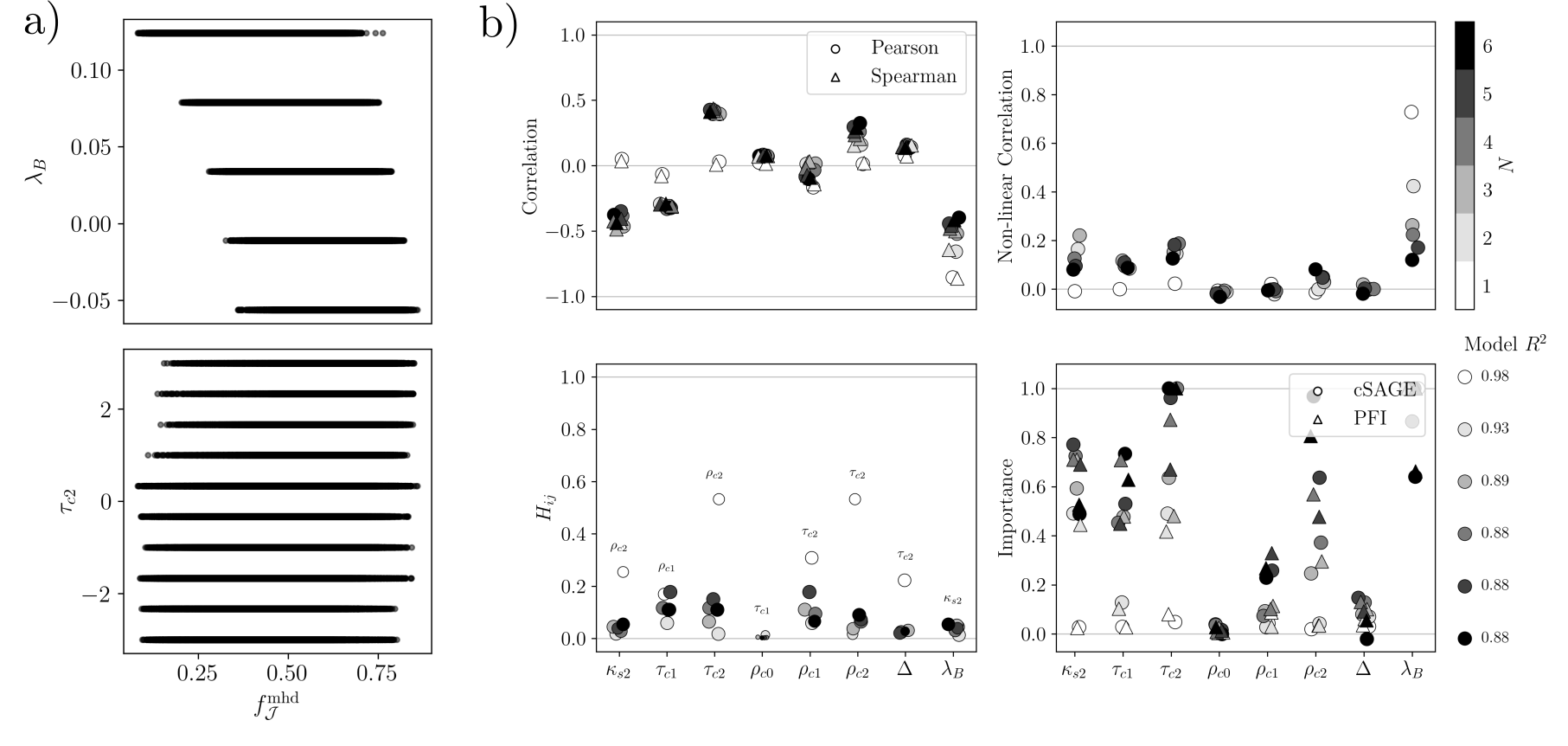}
    \caption{\textbf{Feature dependence statistical summary for $f_\mathcal{J}$.} }
    \label{fig:fmj_stats}
\end{figure}

\begin{figure}
    \centering
    \includegraphics[width=\linewidth]{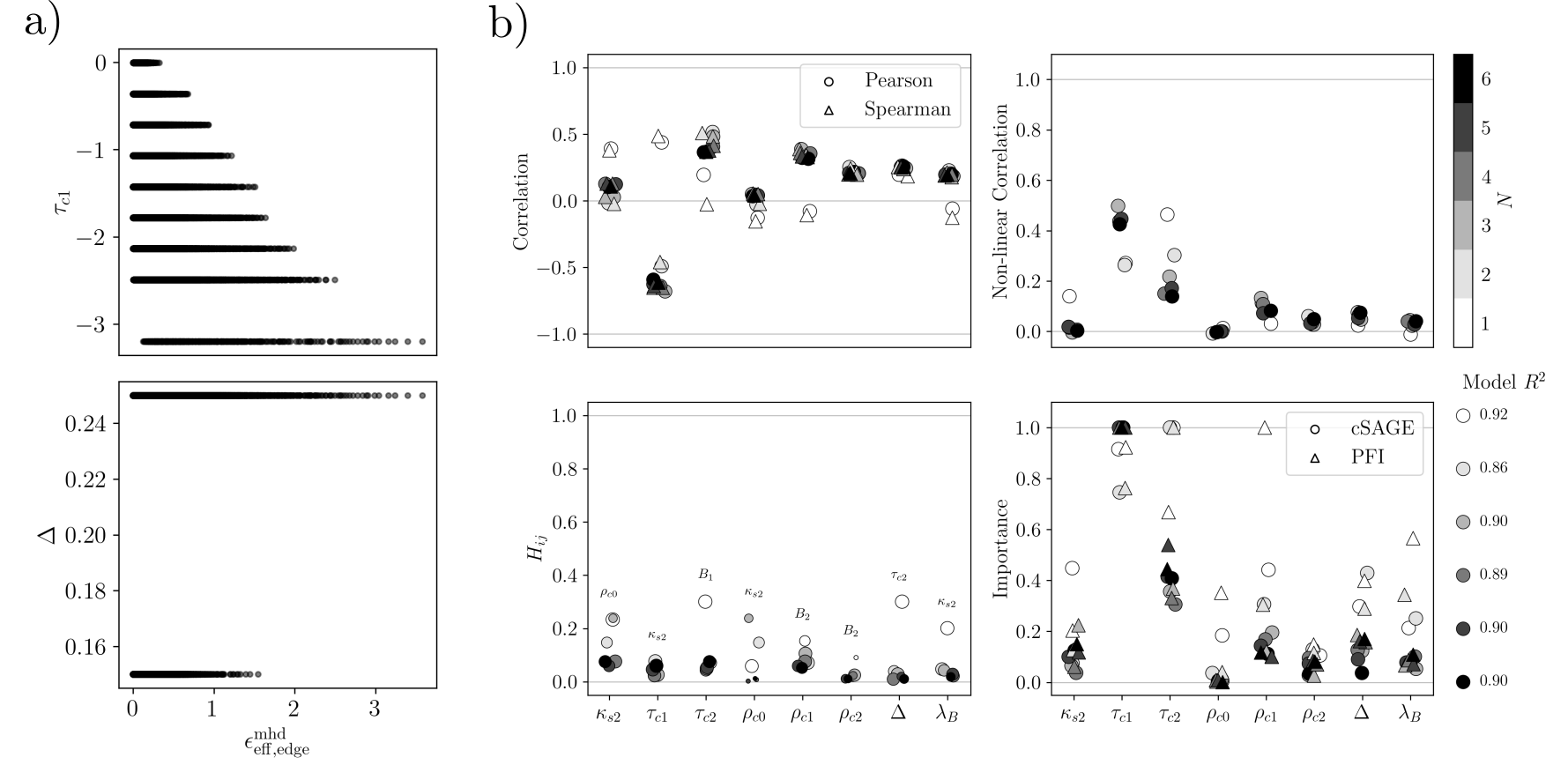}
    \caption{\textbf{Feature dependence statistical summary for $\epsilon_\mathrm{eff}^\mathrm{edge}$.} }
    \label{fig:epseff_stats}
\end{figure}

In this Appendix we collect the feature dependence statistical summary plots, Figures~\ref{fig:Acmhd_stats}-\ref{fig:epseff_stats}, for the various physics features discussed throughout the papers. We do so here instead of throughout the paper for clarity. We also include a Table summarising the mean, maximum and minimum values of different physics features throughout the database distinguishing between different $N$. The definitions of the features in the table can be found in the main text or in \cite{rodriguez2025near}.

\begin{table}
\centering
\begin{tabular}{l|cccccc}
\textbf{Feature} & $N=1$ & $N=2$ & $N=3$ & $N=4$ & $N=5$ & $N=6$ \\\hline
$L_{\nabla\mathbf{B}}$ & $0.35^{0.43}_{0.26}$ & $0.27^{0.42}_{0.13}$ & $0.20^{0.36}_{0.09}$ & $0.17^{0.33}_{0.08}$ & $0.14^{0.30}_{0.05}$ & $0.12^{0.26}_{0.04}$ \\
$A_c^\mathrm{mhd}$ & $4.6^{10.9}_{1.9}$ & $7.8^{40.1}_{1.6}$ & $15.3^{94.7}_{2.4}$ & $21.2^{132}_{3.9}$ & $28.4^{175}_{5.6}$ & $38.1^{217}_{6.7}$ \\
$\epsilon_\mathrm{eff,mhd}^\mathrm{edge}$ (\%) & $5.72^{65.6}_{0.11}$ & $11.6^{358}_{0.035}$ & $45.4^{1.13^3}_{0.043}$ & $68.0^{1.45^3}_{0.054}$ & $105^{2.03^3}_{0.070}$ & $158^{3.20^3}_{0.17}$ \\

$\mathcal{S}_\mathrm{max}$ & $94^{100^3}_{2}$ & $7.7^{20^4}_{0.5}$ & $3.5^{34^3}_{0.2}$ & $2.4^{5^3}_{0.1}$ & $3.6^{13^4}_{0.09}$ & $1.7^{5^3}_{0.07}$ \\
$\iota$ & $0.41^{0.83}_{-0.08}$ & $0.65^{1.43}_{0.15}$ & $0.94^{1.88}_{0.34}$ & $1.20^{2.50}_{0.46}$ & $1.45^{2.69}_{0.59}$ & $1.72^{3.59}_{0.73}$ \\
$\delta\!B$ & $0.048^{10.8}_{0.012}$ & $0.118^{0.613}_{0.015}$ & $0.26^{1.45}_{0.033}$ & $0.40^{1.63}_{0.062}$ & $0.59^{2.10}_{0.10}$ & $0.81^{3.03}_{0.14}$ \\
$\epsilon_{\mathrm{eff}}^{3/2,(0)}\times10^6$ & $4.20^{120}_{1.75_{-9}}$ & $4.85^{131}_{3.71_{-9}}$ & $5.81^{206}_{1.55_{-6}}$ & $7.19^{284}_{1.33_{-7}}$ & $9.27^{1540}_{1.48_{-3}}$ & $10.6^{2320}_{3.23_{-5}}$ \\
$f_\mathcal{J}^\mathrm{mhd}$ & $0.65^{0.87}_{0.36}$ & $0.52^{0.86}_{0.08}$ & $0.55^{0.91}_{0.04}$ & $0.55^{0.90}_{0.02}$ & $0.56^{0.89}_{0.00}$ & $0.57^{0.91}_{0.00}$ \\
$q_\mathrm{eff}$ & $0.72^{1.94}_{0.44}$ & $0.18^{0.47}_{0.08}$ & $0.09^{0.45}_{0.05}$ & $0.05^{0.16}_{0.03}$ & $0.03^{0.33}_{0.02}$ & $0.02^{0.26}_{0.01}$ \\
\end{tabular}
\caption{\textbf{Statistics of configuration features in the database.} The table summarises the statistics of the different features of configurations (rows) for different number of field periods (columns). The numbers denote the average value, with the maximum (superscript) and minimum (subscript), and the super or subscripts of these denoting $\times10^x$. }
\label{tab:summary_db}
\end{table}

\bibliographystyle{jpp}

\bibliography{jpp-instructions}

@book{Parra_Diaz_2024,
   title={Flexible Stellarator Physics Facility},
   url={http://dx.doi.org/10.2172/2372903},
   DOI={10.2172/2372903},
   institution={Office of Scientific and Technical Information (OSTI)},
   author={Parra Diaz, Felix and Baek, Seung Gyou and Churchill, Michael and Demers, Diane and Dudson, B. and Ferraro, Nathaniel and Geiger, Benedikt and Gerhardt, Stefan and Hammond, Kenneth and Hudson, Stuart and Jorge, R. and Kolemen, Egemen and Kriete, David and Kumar, S. and Landreman, M. and Lowe, C. and Maurer, David and Nespoli, Federico and Pablant, Novimir and Pueschel, M. and Punjabi, A. and Schwartz, Jacob and Swanson, C. and Wright, Adelle},
   year={2024},
   month=jun 
}

@article{warmer2024overview,
  title={Overview of European efforts and advances in Stellarator power plant studies},
  author={Warmer, Felix and Alguacil, J and Biek, D and Bogaarts, T and Bongiov{\`\i}, G and Bykov, V and Catal{\'a}n, JP and Duligal, RK and Fern{\'a}ndez-Berceruelo, I and Giambrone, S and others},
  journal={Fusion Engineering and Design},
  volume={202},
  pages={114386},
  year={2024},
  publisher={Elsevier}
}

@article{Hegna_2025,
  title={The Infinity Two fusion pilot plant baseline plasma physics design},
  author={Hegna, CC and Anderson, DT and Andrew, EC and Ayilaran, A and Bader, A and Bohm, TD and Mata, K Camacho and Canik, JM and Carbajal, L and Cerfon, A and others},
  journal={Journal of Plasma Physics},
  volume={91},
  number={3},
  pages={E76},
  year={2025},
  publisher={Cambridge University Press}
}

@article{LION_2025,
  title={Stellaris: A high-field quasi-isodynamic stellarator for a prototypical fusion power plant},
  author={Lion, J and Angl{\`e}s, J-C and Bonauer, L and Navarro, A Ba{\~n}{\'o}n and Ceron, SA Cadena and Davies, R and Drevlak, M and Foppiani, N and Geiger, J and Goodman, A and others},
  journal={Fusion Engineering and Design},
  volume={214},
  pages={114868},
  year={2025},
  publisher={Elsevier}
}

@article{Beidler_2001,
  title={The helias reactor HSR4/18},
  author={Beidler, CD and Harmeyer, E and Herrnegger, F and Igitkhanov, Yu and Kendl, A and Kisslinger, J and Kolesnichenko, Ya I and Lutsenko, VV and N{\"u}hrenberg, C and Sidorenko, I and others},
  journal={Nuclear Fusion},
  volume={41},
  number={12},
  pages={1759},
  year={2001},
  publisher={IOP Publishing}
}

@article{aeries-cs,
author = {F. Najmabadi and A. R. Raffray and S. I. Abdel-Khalik and L. Bromberg and L. Crosatti and L. El-Guebaly and P. R. Garabedian and A. A. Grossman and D. Henderson and A. Ibrahim and T. Ihli and T. B. Kaiser and B. Kiedrowski and L. P. Ku and J. F. Lyon and R. Maingi and S. Malang and C. Martin and T. K. Mau and B. Merrill and R. L. Moore and R. J. Peipert Jr. and D. A. Petti and D. L. Sadowski and M. Sawan and J. H. Schultz and R. Slaybaugh and K. T. Slattery and G. Sviatoslavsky and A. Turnbull and L. M. Waganer and X. R. Wang and J. B. Weathers and P. Wilson and J. C. Waldrop III and M. Yoda and M. Zarnstorffh},
title = {The ARIES-CS Compact Stellarator Fusion Power Plant},
journal = {Fusion Science and Technology},
volume = {54},
number = {3},
pages = {655--672},
year = {2008},
publisher = {Taylor \& Francis},
doi = {10.13182/FST54-655},
URL = {https://doi.org/10.13182/FST54-655},
eprint = {https://doi.org/10.13182/FST54-655}
}

@article{cadena2025constellaration,
  title={ConStellaration: A dataset of QI-like stellarator plasma boundaries and optimization benchmarks},
  author={Cadena, Santiago A and Merlo, Andrea and Laude, Emanuel and Bauer, Alexander and Agrawal, Atul and Pascu, Maria and Savtchouk, Marija and Guiraud, Enrico and Bonauer, Lukas and Hudson, Stuart and others},
  journal={arXiv preprint arXiv:2506.19583},
  year={2025}
}

@article{laia2025data,
  title={Data-driven approach to model the influence of magnetic geometry in the confinement of fusion devices},
  author={Laia, R and Jorge, R and Abreu, G},
  journal={Nuclear Fusion},
  volume={66},
  number={1},
  pages={016034},
  year={2025},
  publisher={IOP Publishing}
}

@article{plunk2025-geometric,
      title={A geometric approach to constructing quasi-isodynamic fields}, 
      author={G. G. Plunk and E. Rodríguez},
      year={2026},
  journal={Journal of Plasma Physics, {\it in press}},
      eprint={2508.12820},
      archivePrefix={arXiv},
      primaryClass={physics.plasm-ph},
      url={https://arxiv.org/abs/2508.12820}, 
}

@article{breiman2001random,
  title={Random forests},
  author={Breiman, Leo},
  journal={Machine learning},
  volume={45},
  number={1},
  pages={5--32},
  year={2001},
  publisher={Springer}
}

@article{shapley1953value,
  title={A value for n-person games},
  author={Shapley, Lloyd S and others},
  year={1953},
  publisher={Princeton University Press Princeton}
}

@article{zhang1996birch,
  title={BIRCH: an efficient data clustering method for very large databases},
  author={Zhang, Tian and Ramakrishnan, Raghu and Livny, Miron},
  journal={ACM sigmod record},
  volume={25},
  number={2},
  pages={103--114},
  year={1996},
  publisher={ACM New York, NY, USA}
}

@incollection{jolliffe2011principal,
  title={Principal component analysis},
  author={Jolliffe, Ian},
  booktitle={International encyclopedia of statistical science},
  pages={1094--1096},
  year={2011},
  publisher={Springer}
}

@inproceedings{mcinnes2017accelerated,
  title={Accelerated hierarchical density based clustering},
  author={McInnes, Leland and Healy, John},
  booktitle={2017 IEEE international conference on data mining workshops (ICDMW)},
  pages={33--42},
  year={2017},
  organization={IEEE}
}

@inproceedings{campello2013density,
  title={Density-based clustering based on hierarchical density estimates},
  author={Campello, Ricardo JGB and Moulavi, Davoud and Sander, J{\"o}rg},
  booktitle={Pacific-Asia conference on knowledge discovery and data mining},
  pages={160--172},
  year={2013},
  organization={Springer}
}

@article{ng2001spectral,
  title={On spectral clustering: Analysis and an algorithm},
  author={Ng, Andrew and Jordan, Michael and Weiss, Yair},
  journal={Advances in neural information processing systems},
  volume={14},
  year={2001}
}

@incollection{winter2002shapley,
title = {Chapter 53 The shapley value},
series = {Handbook of Game Theory with Economic Applications},
publisher = {Elsevier},
volume = {3},
pages = {2025-2054},
year = {2002},
issn = {1574-0005},
doi = {https://doi.org/10.1016/S1574-0005(02)03016-3},
url = {https://www.sciencedirect.com/science/article/pii/S1574000502030163},
author = {Eyal Winter}
}

@article{fisher2019all,
  title={All models are wrong, but many are useful: Learning a variable's importance by studying an entire class of prediction models simultaneously},
  author={Fisher, Aaron and Rudin, Cynthia and Dominici, Francesca},
  journal={Journal of Machine Learning Research},
  volume={20},
  number={177},
  pages={1--81},
  year={2019}
}

@book{molnar2020interpretable,
  title={Interpretable machine learning},
  author={Molnar, Christoph},
  year={2020},
  publisher={Lulu.com}
}

@misc{hastie2009elements,
  title={The elements of statistical learning},
  author={Hastie, Trevor and Tibshirani, Robert and Friedman, Jerome and others},
  year={2009},
  publisher={Springer series in statistics New-York}
}

@inproceedings{gori1997quasi,
  title={Quasi-isodynamic stellarators},
  author={Gori, S and Lotz, W and N{\"u}hrenberg, J},
  booktitle={Theory of Fusion Plasmas-Proceedings of the Joint Varenna-Laussane International Workshop},
  pages={335--342},
  year={1997}
}

@article{loizu2017equilibrium,
  title={Equilibrium beta-limits in classical stellarators},
  author={Loizu, Joaquim and Hudson, SR and N{\"u}hrenberg, C and Geiger, J and Helander, P},
  journal={Journal of Plasma Physics},
  volume={83},
  number={6},
  pages={715830601},
  year={2017},
  publisher={Cambridge University Press}
}

@article{shafranov1963,
  title={Equilibrium of a toroidal pinch in a magnetic field},
  author={Shafranov, VD},
  journal={Soviet Atomic Energy},
  volume={13},
  number={6},
  pages={1149--1158},
  year={1963},
  publisher={Springer}
}

@book{wierzchon2018modern,
  title={Modern algorithms of cluster analysis},
  author={Wierzcho{\'n}, S{\l}awomir T and K{\l}opotek, Mieczys{\l}aw A},
  year={2018},
  publisher={Springer}
}

@book{borg2005modern,
  title={Modern multidimensional scaling: Theory and applications},
  author={Borg, Ingwer and Groenen, Patrick JF},
  year={2005},
  publisher={Springer}
}

@article{pfefferle2018non,
  title={Non-planar elasticae as optimal curves for the magnetic axis of stellarators},
  author={Pfefferl{\'e}, David and Gunderson, Lee and Hudson, Stuart R and Noakes, Lyle},
  journal={Physics of Plasmas},
  volume={25},
  number={9},
  year={2018},
  publisher={AIP Publishing}
}

@article{giuliani2022single,
  title={Single-stage gradient-based stellarator coil design: optimization for near-axis quasi-symmetry},
  author={Giuliani, Andrew and Wechsung, Florian and Cerfon, Antoine and Stadler, Georg and Landreman, Matt},
  journal={Journal of Computational Physics},
  volume={459},
  pages={111147},
  year={2022},
  publisher={Elsevier}
}

@article{giuliani2024direct,
  title={Direct stellarator coil design using global optimization: application to a comprehensive exploration of quasi-axisymmetric devices},
  author={Giuliani, Andrew},
  journal={Journal of Plasma Physics},
  volume={90},
  number={3},
  pages={905900303},
  year={2024},
  publisher={Cambridge University Press}
}

@article{giuliani2025comprehensive,
  title={A comprehensive exploration of quasisymmetric stellarators and their coil sets},
  author={Giuliani, Andrew and Rodr{\'\i}guez, Eduardo and Spivak, Marina},
  journal={Journal of Plasma Physics},
  volume={91},
  number={5},
  pages={E128},
  year={2025},
  publisher={Cambridge University Press}
}

@article{scikit-learn,
  title={Scikit-learn: Machine Learning in {P}ython},
  author={Pedregosa, F. and Varoquaux, G. and Gramfort, A. and Michel, V.
          and Thirion, B. and Grisel, O. and Blondel, M. and Prettenhofer, P.
          and Weiss, R. and Dubourg, V. and Vanderplas, J. and Passos, A. and
          Cournapeau, D. and Brucher, M. and Perrot, M. and Duchesnay, E.},
  journal={Journal of Machine Learning Research},
  volume={12},
  pages={2825--2830},
  year={2011}
}

@article{goodman2024quasi,
  title={Quasi-isodynamic stellarators with low turbulence as fusion reactor candidates},
  author={Goodman, Alan G and Xanthopoulos, Pavlos and Plunk, Gabriel G and Smith, H{\aa}kan and N{\"u}hrenberg, Carolin and Beidler, Craig D and Henneberg, Sophia A and Roberg-Clark, Gareth and Drevlak, Michael and Helander, Per},
  journal={PRX Energy},
  volume={3},
  number={2},
  pages={023010},
  year={2024},
  publisher={APS}
}

@article{Kappel_2024,
doi = {10.1088/1361-6587/ad1a3e},
url = {https://dx.doi.org/10.1088/1361-6587/ad1a3e},
year = {2024},
month = {jan},
publisher = {IOP Publishing},
volume = {66},
number = {2},
pages = {025018},
author = {Kappel, John and Landreman, Matt and Malhotra, Dhairya},
title = {The magnetic gradient scale length explains why certain plasmas require close external magnetic coils},
journal = {Plasma Physics and Controlled Fusion}
}

@book{chmutov2012introduction,
  title={Introduction to Vassiliev knot invariants},
  author={Chmutov, Sergei and Duzhin, Sergei Vasilevich and Mostovoy, Jacob},
  year={2012},
  publisher={Cambridge University Press}
}

@book{kaufman2009finding,
  title={Finding groups in data: an introduction to cluster analysis},
  author={Kaufman, Leonard and Rousseeuw, Peter J},
  year={2009},
  publisher={John Wiley \& Sons}
}

@article{rodriguez2025near,
  title={Near-axis measures of quasi-isodynamic configurations},
  author={Rodriguez, Eduardo and Plunk, Gabriel G},
  journal={arXiv preprint arXiv:2505.02465},
  year={2025}
}

@article{goodman2023constructing, title={Constructing precisely quasi-isodynamic magnetic fields}, volume={89}, DOI={10.1017/S002237782300065X}, number={5}, journal={Journal of Plasma Physics}, publisher={Cambridge University Press}, author={Goodman, A.G. and Camacho Mata, K. and Henneberg, S.A. and Jorge, R. and Landreman, M. and Plunk, G.G. and Smith, H.M. and Mackenbach, R.J.J. and Beidler, C.D. and Helander, P. and et al.}, year={2023}, pages={905890504}}

@article{rodriguez2024near, title={Near-axis description of stellarator-symmetric quasi-isodynamic stellarators to second order}, volume={91}, DOI={10.1017/S0022377825000157}, number={2}, journal={Journal of Plasma Physics}, author={Rodríguez, E. and Plunk, G.G. and Jorge, R.}, year={2025}, pages={E59}}

@article{dudt2024magnetic,
  title={Magnetic fields with general omnigenity},
  author={Dudt, Daniel W and Goodman, Alan G and Conlin, Rory and Panici, Dario and Kolemen, Egemen},
  journal={Journal of Plasma Physics},
  volume={90},
  number={1},
  pages={905900120},
  year={2024},
  publisher={Cambridge University Press}
}

@article{plunk2024back,
  title={Back to the Figure-8 Stellarator},
  author={Plunk, GG and Drevlak, M and Rodriguez, E and Babin, R and Goodman, A and Hindenlang, F},
  journal={arXiv preprint arXiv:2411.16411},
  year={2024}
}

@article{rodriguez2024zonal,
  title={The zonal-flow residual does not tend to zero in the limit of small mirror ratio},
  author={Rodriguez, Eduardo and Plunk, Gabriel G},
  journal={arXiv preprint arXiv:2407.17824},
  year={2024}
}

@article{landreman2022mapping,
  title={Mapping the space of quasisymmetric stellarators using optimized near-axis expansion},
  author={Landreman, Matt},
  journal={Journal of Plasma Physics},
  volume={88},
  number={6},
  pages={905880616},
  year={2022},
  publisher={Cambridge University Press}
}

@article{boozer1998stellarator,
  title={What is a stellarator?},
  author={Boozer, Allen H},
  journal={Physics of Plasmas},
  volume={5},
  number={5},
  pages={1647--1655},
  year={1998},
  publisher={American Institute of Physics}
}

@book{wessonTok,
  author    = {Wesson, John},
  title     = {{Tokamaks; 4th ed.}},
  number    = {},
  publisher = {Oxford Univ. Press},
  address   = {Oxford},
  series    = {International series of monographs on physics},
  year      = {2011}
}

@article{mynick2006,
  author  = {Mynick,H. E. },
  title   = {Transport optimization in stellarators},
  journal = {Physics of Plasmas},
  volume  = {13},
  number  = {5},
  pages   = {058102},
  year    = {2006},
  doi     = {10.1063/1.2177643}
}

@article{nemov1999,
  author  = {Nemov,V. V.  and Kasilov,S. V.  and Kernbichler,W.  and Heyn,M. F. },
  title   = {Evaluation of $1/\nu$ neoclassical transport in stellarators},
  journal = {Physics of Plasmas},
  volume  = {6},
  number  = {12},
  pages   = {4622-4632},
  year    = {1999},
  doi     = {10.1063/1.873749}
}

@book{dodge2008concise,
  title={The concise encyclopedia of statistics},
  author={Dodge, Yadolah},
  year={2008},
  publisher={Springer Science \& Business Media}
}

@article{whitney1971direct,
  title={A direct method of nonparametric measurement selection},
  author={Whitney, A Wayne},
  journal={IEEE transactions on computers},
  volume={100},
  number={9},
  pages={1100--1103},
  year={1971},
  publisher={IEEE}
}

@article{friedman2008predictive,
  title={Predictive learning via rule ensembles},
  author={Friedman, Jerome H and Popescu, Bogdan E},
  year={2008}
}

@article{panici2025extending,
  title={Extending near-axis equilibria in DESC},
  author={Panici, Dario and Rodriguez, Eduardo and Conlin, Rory and Dudt, Daniel and Kolemen, Egemen},
  journal={arXiv preprint arXiv:2506.05170},
  year={2025}
}

@book{ericson2004real,
  title={Real-time collision detection},
  author={Ericson, Christer},
  year={2004},
  publisher={Crc Press}
}

@article{kay1986ray,
  title={Ray tracing complex scenes},
  author={Kay, Timothy L and Kajiya, James T},
  journal={ACM SIGGRAPH computer graphics},
  volume={20},
  number={4},
  pages={269--278},
  year={1986},
  publisher={ACM New York, NY, USA}
}

@article{covert2020understanding,
  title={Understanding global feature contributions with additive importance measures},
  author={Covert, Ian and Lundberg, Scott M and Lee, Su-In},
  journal={Advances in neural information processing systems},
  volume={33},
  pages={17212--17223},
  year={2020}
}

@book{liu2012feature,
  title={Feature selection for knowledge discovery and data mining},
  author={Liu, Huan and Motoda, Hiroshi},
  volume={454},
  year={2012},
  publisher={Springer science \& business media}
}

@book{shalev2014understanding,
  title={Understanding machine learning: From theory to algorithms},
  author={Shalev-Shwartz, Shai and Ben-David, Shai},
  year={2014},
  publisher={Cambridge university press}
}

@inproceedings{ewald2024guide,
  title={A guide to feature importance methods for scientific inference},
  author={Ewald, Fiona Katharina and Bothmann, Ludwig and Wright, Marvin N and Bischl, Bernd and Casalicchio, Giuseppe and K{\"o}nig, Gunnar},
  booktitle={World Conference on Explainable Artificial Intelligence},
  pages={440--464},
  year={2024},
  organization={Springer}
}

@article{kamath2021explainable,
  title={Explainable artificial intelligence: An introduction to interpretable machine learning},
  author={Kamath, Uday and Liu, John},
  year={2021},
  publisher={Springer Nature}
}

@book{scholkopf2002learning,
  title={Learning with kernels: support vector machines, regularization, optimization, and beyond},
  author={Sch{\"o}lkopf, Bernhard and Smola, Alexander J},
  year={2002},
  publisher={MIT press}
}

@article{rosenbluth1998poloidal,
  title={Poloidal flow driven by ion-temperature-gradient turbulence in tokamaks},
  author={Rosenbluth, MN and Hinton, FL},
  journal={Physical review letters},
  volume={80},
  number={4},
  pages={724},
  year={1998},
  publisher={APS}
}

@article{spitzer1958stellarator,
  title={The stellarator concept},
  author={Spitzer Jr, Lyman},
  journal={The Physics of Fluids},
  volume={1},
  number={4},
  pages={253--264},
  year={1958},
  publisher={American Institute of Physics}
}

@article{xiao2006short,
  title={Short wavelength effects on the collisionless neoclassical polarization and residual zonal flow level},
  author={Xiao, Yong and Catto, Peter J},
  journal={Physics of Plasmas},
  volume={13},
  number={10},
  year={2006}
}

@article{rodriguez2023constructing,
  title={Constructing the space of quasisymmetric stellarators through near-axis expansion},
  author={Rodr{\'\i}guez, E and Sengupta, W and Bhattacharjee, A},
  journal={Plasma Physics and Controlled Fusion},
  volume={65},
  number={9},
  pages={095004},
  year={2023},
  publisher={IOP Publishing}
}

@article{monreal2016residual,
  title={Residual zonal flows in tokamaks and stellarators at arbitrary wavelengths},
  author={Monreal, Pedro and Calvo, Iv{\'a}n and S{\'a}nchez, Edilberto and Parra, F{\'e}lix I and Bustos, Andr{\'e}s and K{\"o}nies, Axel and Kleiber, Ralf and G{\"o}rler, Tobias},
  journal={Plasma Physics and Controlled Fusion},
  volume={58},
  number={4},
  pages={045018},
  year={2016},
  publisher={IOP Publishing}
}

@article{landreman2021a,
  title     = {Figures of merit for stellarators near the magnetic axis},
  volume    = {87},
  number    = {1},
  journal   = {Journal of Plasma Physics},
  publisher = {Cambridge University Press},
  author    = {Landreman, M.},
  year      = {2021},
  pages     = {905870112}
}

@article{rodriguez2020necessary,
  title={Necessary and sufficient conditions for quasisymmetry},
  author={Rodr\'{i}guez, E. and Helander, P. and Bhattacharjee, A.},
  journal={Physics of Plasmas},
  volume={27},
  number={6},
  pages={062501},
  year={2020},
  publisher={AIP Publishing LLC}
}

@article{Helander_2009,
doi = {10.1088/0741-3335/51/5/055004},
url = {https://dx.doi.org/10.1088/0741-3335/51/5/055004},
year = {2009},
month = {feb},
publisher = {},
volume = {51},
number = {5},
pages = {055004},
author = {Helander, P. and Nührenberg, J.},
title = {Bootstrap current and neoclassical transport in quasi-isodynamic stellarators},
journal = {Plasma Physics and Controlled Fusion}
}

@article{Nührenberg_2010,
doi = {10.1088/0741-3335/52/12/124003},
url = {https://dx.doi.org/10.1088/0741-3335/52/12/124003},
year = {2010},
month = {nov},
publisher = {},
volume = {52},
number = {12},
pages = {124003},
author = {Jürgen Nührenberg},
title = {Development of quasi-isodynamic stellarators},
journal = {Plasma Physics and Controlled Fusion}
}

@article{plunk2019direct,
  title={Direct construction of optimized stellarator shapes. Part 3. Omnigenity near the magnetic axis},
  author={Plunk, G. G. and Landreman, M. and Helander, P.},
  journal={Journal of Plasma Physics},
  volume={85},
  number={6},
  pages={905850602},
  year={2019},
  publisher={Cambridge University Press}
}

@phdthesis{zhao2012cluster,
  title={Cluster validity in clustering methods},
  author={Zhao, Qinpei},
  year={2012},
  school={University of Eastern Finland}
}

@article{tenenbaum2000global,
  title={A global geometric framework for nonlinear dimensionality reduction},
  author={Tenenbaum, Joshua B and Silva, Vin de and Langford, John C},
  journal={science},
  volume={290},
  number={5500},
  pages={2319--2323},
  year={2000},
  publisher={American Association for the Advancement of Science}
}

@article{calinski1974dendrite,
  title={A dendrite method for cluster analysis},
  author={Cali{\'n}ski, Tadeusz and Harabasz, Jerzy},
  journal={Communications in Statistics-theory and Methods},
  volume={3},
  number={1},
  pages={1--27},
  year={1974},
  publisher={Taylor \& Francis}
}

@article{landreman2019,
  title     = {Constructing stellarators with quasisymmetry to high order},
  volume    = {85},
  doi       = {10.1017/S0022377819000783},
  number    = {6},
  journal   = {Journal of Plasma Physics},
  publisher = {Cambridge University Press},
  author    = {Landreman, M. and Sengupta, W.},
  year      = {2019},
  pages     = {815850601}
}

@article{rodriguez2023higher,
    author = {Rodríguez, E. and Plunk, G. G.},
    title = "{Higher order theory of quasi-isodynamicity near the magnetic axis of stellarators}",
    journal = {Physics of Plasmas},
    volume = {30},
    number = {6},
    pages = {062507},
    year = {2023},
    month = {06},
    issn = {1070-664X},
    doi = {10.1063/5.0150275},
    url = {https://doi.org/10.1063/5.0150275},
}

@article{jorge2020naeturb,
  title={The use of near-axis magnetic fields for stellarator turbulence simulations},
  author={Jorge, Rogerio and Landreman, Matt},
  journal={Plasma Physics and Controlled Fusion},
  volume={63},
  number={1},
  pages={014001},
  year={2020},
  publisher={IOP Publishing}
}

@article{camacho-mata-2022, title={Direct construction of stellarator-symmetric quasi-isodynamic magnetic configurations}, volume={88}, DOI={10.1017/S0022377822000812}, number={5}, journal={Journal of Plasma Physics}, publisher={Cambridge University Press}, author={Camacho Mata, K. and Plunk, G. G. and Jorge, R.}, year={2022}, pages={905880503}}

@book{Solovev1970,
address = {New York - London},
author = {Solov'ev, L. S. and Shafranov, V. D.},
publisher = {Consultants Bureau},
title = {{Reviews of Plasma Physics 5}},
year = {1970}
}

@article{lortz1976equilibrium,
  title={Equilibrium and stability of a three-dimensional toroidal MHD configuration near its magnetic axis},
  author={Lortz, D and N{\"u}hrenberg, J},
  journal={Zeitschrift f{\"u}r Naturforschung A},
  volume={31},
  number={11},
  pages={1277--1288},
  year={1976},
  publisher={Verlag der Zeitschrift f{\"u}r Naturforschung}
}

@article{mercier1964equilibrium,
  title={Equilibrium and stability of a toroidal magnetohydrodynamic system in the neighbourhood of a magnetic axis},
  author={Mercier, Claude},
  journal={Nuclear Fusion},
  volume={4},
  number={3},
  pages={213},
  year={1964},
  publisher={IOP Publishing}
}

@article{pohl1968self,
  title={The self-linking number of a closed space curve},
  author={Pohl, William F},
  journal={Journal of Mathematics and Mechanics},
  volume={17},
  number={10},
  pages={975--985},
  year={1968},
  publisher={JSTOR}
}

@article{frenet1852courbes,
  title={Sur les courbes {\`a} double courbure},
  author={Frenet, F},
  journal={Journal de math{\'e}matiques pures et appliqu{\'e}es},
  volume={17},
  pages={437--447},
  year={1852}
}

@article{boozer1983transport,
  title={Transport and isomorphic equilibria},
  author={Boozer, Allen H},
  journal={The Physics of Fluids},
  volume={26},
  number={2},
  pages={496--499},
  year={1983},
  publisher={American Institute of Physics}
}

@article{blank2004guiding,
  title={Guiding center motion},
  author={Blank, HJ de},
  journal={Fusion science and technology},
  volume={45},
  number={2T},
  pages={47--54},
  year={2004},
  publisher={Taylor \& Francis}
}

@article{nuhren1988,
title = "Quasi-helically symmetric toroidal stellarators",
journal = "Physics Letters A",
volume = "129",
number = "2",
pages = "113 - 117",
year = "1988",
author = "J. N{\"u}hrenberg and R. Zille"
}

@article{bernardin1986,
  author  = {Bernardin,M. P.  and Moses,R. W.  and Tataronis,J. A. },
  title   = {Isodynamical (omnigenous) equilibrium in symmetrically confined plasma           configurations},
  journal = {The Physics of Fluids},
  volume  = {29},
  number  = {8},
  pages   = {2605-2611},
  year    = {1986},
  doi     = {10.1063/1.865501}
}

@article{northrop1961guiding,
  title={The guiding center approximation to charged particle motion},
  author={Northrop, Theodore G},
  journal={Annals of Physics},
  volume={15},
  number={1},
  pages={79--101},
  year={1961},
  publisher={Elsevier}
}

@article{hall1975,
  author  = {Hall, Laurence S.  and McNamara, Brendan },
  title   = {Three‐dimensional equilibrium of the anisotropic, finite‐pressure guiding‐center plasma: Theory of the magnetic plasma},
  journal = {The Physics of Fluids},
  volume  = {18},
  number  = {5},
  pages   = {552-565},
  year    = {1975}
}

@article{littlejohn1983,
title={Variational principles of guiding centre motion},
volume={29},
DOI={10.1017/S002237780000060X},
number={1},
journal={Journal of Plasma Physics},
publisher={Cambridge University Press},
author={Littlejohn, Robert G.},
year={1983},
pages={111–125}}

@phdthesis{thesis, title={Quasisymmetry}, author={Rodr\'{i}guez, E.}, year={2022}}

@article{proll2012resilience,
  title={Resilience of quasi-isodynamic stellarators against trapped-particle instabilities},
  author={Proll, J. H. E. and Helander, P. and Connor, J. W. and Plunk, G. G.},
  journal={Physical Review Letters},
  volume={108},
  number={24},
  pages={245002},
  year={2012},
  publisher={APS}
}

@article{helander2013collisionless,
  title={Collisionless microinstabilities in stellarators. I. Analytical theory of trapped-particle modes},
  author={Helander, P. and Proll, J. H. E. and Plunk, G. G.},
  journal={Physics of Plasmas},
  volume={20},
  number={12},
  pages={122505},
  year={2013},
  publisher={American Institute of Physics}
}

@article{helander2014theory,
  title={Theory of plasma confinement in non-axisymmetric magnetic fields},
  author={Helander, P.},
  journal={Reports on Progress in Physics},
  volume={77},
  number={8},
  pages={087001},
  year={2014},
  publisher={IOP Publishing}
}

@article{rosenbluth1968,
author = {Rosenbluth, M. N. },
title = {Low‐Frequency Limit of Interchange Instability},
journal = {The Physics of Fluids},
volume = {11},
number = {4},
pages = {869-872},
year = {1968},
doi = {10.1063/1.1692009},
URL = {https://aip.scitation.org/doi/abs/10.1063/1.1692009},
eprint = {https://aip.scitation.org/doi/pdf/10.1063/1.1692009}}

@article{rodriguez2023mhd, title={Magnetohydrodynamic stability and the effects of shaping: a near-axis view for tokamaks and quasisymmetric stellarators}, volume={89}, DOI={10.1017/S0022377823000211}, number={2}, journal={Journal of Plasma Physics}, publisher={Cambridge University Press}, author={Rodríguez, E.}, year={2023}, pages={905890211}}

@article{zhu2025collisionless,
  title={Collisionless zonal-flow dynamics in quasisymmetric stellarators},
  author={Zhu, Hongxuan and Lin, Z and Bhattacharjee, A},
  journal={Journal of Plasma Physics},
  volume={91},
  number={1},
  pages={E28},
  year={2025},
  publisher={Cambridge University Press}
}

@book{animov2001differential,
  title={Differential geometry and topology of curves},
  author={Animov, Yu},
  year={2001},
  publisher={CRC press}
}

@article{garrenboozer1991a,
  author  = {Garren,D. A.  and Boozer,A. H. },
  title   = {Magnetic field strength of toroidal plasma equilibria},
  journal = {Physics of Fluids B: Plasma Physics},
  volume  = {3},
  number  = {10},
  pages   = {2805-2821},
  year    = {1991},
  doi     = {10.1063/1.859915}
}

@article{greene1997,
  title={A brief review of magnetic wells},
  author={Greene, J. M.},
  journal={Comments on Plasma Physics and Controlled Fusion},
  volume={17},
  pages={389--402},
  year={1997},
  publisher={Citeseer}
}

@article{Camacho2023helicity,
  title={Helicity of the magnetic axes of quasi-isodynamic stellarators},
  author={Camacho Mata, Katia and Plunk, Gabriel G},
  journal={Journal of Plasma Physics},
  volume={89},
  number={6},
  pages={905890609},
  year={2023},
  publisher={Cambridge University Press}
}

@article{oberti2016torus,
  title={On torus knots and unknots},
  author={Oberti, Chiara and Ricca, Renzo L},
  journal={Journal of Knot Theory and Its Ramifications},
  volume={25},
  number={06},
  pages={1650036},
  year={2016},
  publisher={World Scientific}
}

@article{moffatt1992helicity,
  title={Helicity and the C{\u{a}}lug{\u{a}}reanu invariant},
  author={Moffatt, Henry Keith and Ricca, Renzo L},
  journal={Proceedings of the Royal Society of London. Series A: Mathematical and Physical Sciences},
  volume={439},
  number={1906},
  pages={411--429},
  year={1992},
  publisher={The Royal Society London}
}

@book{fuller1999geometric,
  title={The geometric and topological structure of holonomic knots},
  author={Fuller Jr, Edgar Jackson},
  year={1999},
  publisher={University of Georgia}
}

@article{rodriguez2024maximum,
  title={The maximum-J property in quasi-isodynamic stellarators},
  author={Rodr\'{i}guez, Eduardo and Helander, Per and Goodman, AG},
  journal={Journal of Plasma Physics},
  volume={90},
  number={2},
  pages={905900212},
  year={2024},
  publisher={Cambridge University Press}
}

@article{fuller1971writhing,
  title={The writhing number of a space curve},
  author={Fuller, F Brock},
  journal={Proceedings of the National Academy of Sciences},
  volume={68},
  number={4},
  pages={815--819},
  year={1971},
  publisher={National Acad Sciences}
}

@article{rodriguez2022phases,
  title={Phases and phase-transitions in quasisymmetric configuration space},
  author={Rodr\'{i}guez, E. and Sengupta, W. and Bhattacharjee, A.},
  journal={Plasma Physics and Controlled Fusion},
  volume={64},
  number={10},
  pages={105006},
  year={2022},
  publisher={IOP Publishing}
}

@article{Cary1997,
    author = {Cary, J. R. and Shasharina, S. G.},
    title = "{Omnigenity and quasihelicity in helical plasma confinement systems}",
    journal = {Physics of Plasmas},
    volume = {4},
    number = {9},
    pages = {3323-3333},
    year = {1997},
    month = {09},
    issn = {1070-664X},
    doi = {10.1063/1.872473},
    url = {https://doi.org/10.1063/1.872473},
    eprint = {https://pubs.aip.org/aip/pop/article-pdf/4/9/3323/12664528/3323\_1\_online.pdf},
}

@article{mikhailov2002,
	doi = {10.1088/0029-5515/42/11/102},
	year = 2002,
	month = {oct},
	publisher = {{IOP} Publishing},
	volume = {42},
	number = {11},
	pages = {L23--L26},
	author = {M. I. Mikhailov and V. D. Shafranov and A. A. Subbotin and M. Y. Isaev and J. N\"{u}hrenberg and R. Zille and W. A. Cooper}
}

@article{skovoroda2005,
	doi = {10.1088/0741-3335/47/11/004},
	year = 2005,
	month = {oct},
	publisher = {{IOP} Publishing},
	volume = {47},
	number = {11},
	pages = {1911--1924},
	author = {A. A. Skovoroda},
	title = {3D toroidal geometry of currentless magnetic configurations with improved confinement},
	journal = {Plasma Physics and Controlled Fusion}
}

@article{parra2015less,
  title={Less constrained omnigeneous stellarators},
  author={Parra, Felix I and Calvo, Iv{\'a}n and Helander, Per and Landreman, Matt},
  journal={Nuclear Fusion},
  volume={55},
  number={3},
  pages={033005},
  year={2015},
  publisher={IOP Publishing}
}

@article{burby2020some,
  title={Some mathematics for quasi-symmetry},
  author={Burby, Joshua William and Kallinikos, Nikos and MacKay, Robert S},
  journal={Journal of Mathematical Physics},
  volume={61},
  number={9},
  year={2020},
  publisher={AIP Publishing}
}

@article{jorge2022c, title={A single-field-period quasi-isodynamic stellarator}, volume={88}, DOI={10.1017/S0022377822000873}, number={5}, journal={Journal of Plasma Physics}, publisher={Cambridge University Press}, author={Jorge, R. and Plunk, G.G. and Drevlak, M. and Landreman, M. and Lobsien, J.-F. and Camacho Mata, K. and Helander, P.}, year={2022}, pages={175880504}}

@article{landreman2020magnetic,
  title={Magnetic well and Mercier stability of stellarators near the magnetic axis},
  author={Landreman, M. and Jorge, R.},
  journal={Journal of Plasma Physics},
  volume={86},
  number={5},
  pages={905860510},
  year={2020},
  publisher={Cambridge University Press}
}

@article{sanchez2023quasi,
  title={A quasi-isodynamic configuration with good confinement of fast ions at low plasma $\beta$},
  author={S{\'a}nchez, E. and Velasco, J. L. and Calvo, I. and Mulas, S.},
  journal={Nuclear Fusion},
  volume={63},
  number={6},
  pages={066037},
  year={2023},
  publisher={IOP Publishing}
}

@article{landreman2018a,
  title     = {Direct construction of optimized stellarator shapes. Part 1. Theory in cylindrical coordinates},
  volume    = {84},
  doi       = {10.1017/S0022377818001289},
  number    = {6},
  journal   = {Journal of Plasma Physics},
  publisher = {Cambridge University Press},
  author    = {Landreman, M. and Sengupta, W.},
  year      = {2018},
  pages     = {905840616}
}

\end{document}